\documentclass[
  reprint,
  amsmath,amssymb,
  aps,
]{revtex4-2}

\usepackage{braket}
\usepackage{graphicx}
\usepackage{dcolumn}
\usepackage{diagbox}
\usepackage{bm}
\usepackage{multirow}
\usepackage{xcolor}

\newcommand{\C}[1]{\null}
\usepackage[colorlinks=true,linkcolor=blue,urlcolor=blue,citecolor=blue]{hyperref}

\begin{document}

\title{Magnetic quadrupole current generation and accumulation in \\ noncentrosymmetric systems}

\author{Yuuga Takasu}
\author{Satoru Hayami}
\affiliation{Graduate School of Science, Hokkaido University, Sapporo 060-0810, Japan}

\date{\today}

\begin{abstract}
Magnetization control via magnetic octupole injection has recently been proposed for a new class of centrosymmetric antiferromagnets, namely $d$-wave altermagnets, where the magnetic octupole is the lowest-rank magnetic multipole allowed by symmetry and serves as an alternative carrier to spin injection. 
In contrast, in noncentrosymmetric antiferromagnets, the magnetic quadrupole (MQ) constitutes the lowest-rank symmetry-allowed magnetic multipole, suggesting that MQ currents can provide an efficient route toward magnetization control through MQ injection. 
Here, we establish the symmetry conditions for MQ-current generation by constructing the multipole representation of the MQ conductivity tensor and show that MQ currents are generically allowed in noncentrosymmetric crystallographic point groups.
As a representative example, we demonstrate MQ-current generation in the linear-response regime associated with symmetry lowering from the centrosymmetric point group ($mmm$) to its noncentrosymmetric subgroup ($mm2$). 
Furthermore, we reveal MQ accumulation near sample edges, analogous to spin accumulation induced by the spin Hall effect. 
This edge accumulation provides direct evidence of MQ-current generation and constitutes a key prerequisite for realizing MQ injection and MQ-based magnetization control in noncentrosymmetric antiferromagnets.
\end{abstract}

\maketitle

\section{\label{Introduction}Introduction}

Spintronics based on antiferromagnets has attracted considerable attention as a promising route toward highly integrated, ultrafast, and secure information technologies owing to their robustness against external magnetic perturbations, negligible stray fields, and ultrafast spin dynamics~\cite{macdonald2011antiferromagnetic, jungwirth2016antiferromagnetic, Baltz_RevModPhys.90.015005}. 
Extensive efforts have been devoted to developing efficient techniques for the detection and manipulation of antiferromagnetic order.
Remarkable progress has been achieved through electrical readout methods based on anisotropic magnetoresistance~\cite{marti2014room,fina2014anisotropic,zhang2014planar,wong2014strain,moriyama2015sequential,kriegner2016multiple},
tunnelling anisotropic magnetoresistance~\cite{shick2010spin,park2011spin,wang2012room,wang2014evidence},
and anomalous Hall responses~\cite{chen2014anomalous,kubler2014non,nakatsuji2015large,kiyohara2016giant,nayak2016large, vsmejkal6crystal,PhysRevLett.130.036702,reichlova2024observation}, as well as electrical writing techniques based on
the N\'{e}el-order spin--orbit torque (SOT)~\cite{PhysRevLett.113.157201,doi:10.1126/science.aab1031}
and electric-field control of magnetic anisotropy~\cite{doi:10.1126/sciadv.adn0479}. 
Among them, spin injection and the resulting SOT play a central role in modern spintronics through the coupling $\bm{\mathrm{S}} \cdot \bm{\mathrm{M}}$ between the injected spin polarization $\bm{\mathrm{S}}$ and the local magnetization $\bm{\mathrm{M}}$~\cite{berger1984exchange,PhysRevLett.108.117201}.

Recently, altermagnets have attracted considerable attention as a new class of antiferromagnets exhibiting momentum-dependent spin splitting even in the absence of spin--orbit coupling while maintaining a vanishing net magnetization~\cite{noda2016momentum,okugawa2018weakly,ahn2019antiferromagnetism,naka2019spin,hayami2019momentum, vsmejkal2022beyond}.
Such unconventional antiferromagnetic states can be naturally understood within the framework of magnetic multipoles, which provide a unified description of symmetry and microscopic electronic degrees of freedom~\cite{Santini_RevModPhys.81.807,kuramoto2009multipole,suzuki2018first,kusunose2022generalization,hayami2024unified}. 
In particular, centrosymmetric $d$-wave altermagnets are characterized by magnetic octupoles (MOs)~\cite{mcclarty2024landau, Bhowal_PhysRevX.14.011019, oike2025thermodynamic, schiff2025collinear, sato2026quantum, sato2025orbital}, which constitute the lowest-rank magnetic multipoles coupled to the N\'{e}el vector~\cite{mcclarty2024landau}.
Based on this multipolar description, a mechanism for manipulating altermagnetic order through MO injection has recently been proposed~\cite{han2025harnessing,han2026deterministic}.
This mechanism relies on the coupling $\bm{O}_{ij}\cdot\bm{N}$ between injected MOs $\bm{O}_{ij}$ and the N\'{e}el vector $\bm{N}$, representing a multipolar analogue of the conventional SOT arising from $\bm{S}\cdot\bm{M}$.
Motivated by this idea, MO currents have been theoretically demonstrated in $4d$- and $5d$-transition metals~\cite{baek2025magnetic} and $d$-wave altermagnets~\cite{ko2025magnetic}, while their symmetry requirements have been systematically clarified within the multipole framework~\cite{takasu2026symmetry}.

The multipolar description also becomes important in noncentrosymmetric antiferromagnets, where magnetic quadrupoles (MQs), rather than magnetic octupoles, often constitute the lowest-rank magnetic multipoles coupled to the N\'eel vector.
Since MQs are odd under both spatial inversion ($\mathcal{P}$) and time-reversal ($\mathcal{T}$) operations~\cite{dubovik1990toroid, kopaev2009toroidal, Spaldin_0953-8984-20-43-434203}, they provide a microscopic basis for a variety of cross-correlated phenomena, including linear magnetoelectric effects~\cite{Fiebig0022-3727-38-8-R01, EdererPhysRevB.76.214404, tokura2014multiferroics, Yanagi_PhysRevB.97.020404, Hayami_PhysRevB.97.024414, thole2018magnetoelectric}, nonreciprocal excitations~\cite{Miyahara_JPSJ.81.023712, Miyahara_PhysRevB.89.195145, Hayami_doi:10.7566/JPSJ.85.053705, Okuma_PhysRevB.99.094401, Matsumoto_PhysRevB.101.224419}, nonreciprocal transports~\cite{tokura2018nonreciprocal, Watanabe_PhysRevX.11.011001, Suzuki_PhysRevB.105.075201, Yatsushiro_PhysRevB.105.155157, Hayami_PhysRevB.106.014420}, and nonlinear charge/spin Hall effects~\cite{Wang_PhysRevLett.127.277201, Liu_PhysRevLett.127.277202, Hayami_PhysRevB.106.024405, Kondo_PhysRevResearch.4.013186, Kirikoshi_PhysRevB.107.155109}.  
Indeed, MQ has been identified as a fundamental order parameter in a broad class of parity-breaking antiferromagnets, making MQs a central concept in modern multipolar magnetism~\cite{Gao_PhysRevB.98.060402, Shitade_PhysRevB.98.020407}.
These developments naturally raise the question of whether MQs can play a role analogous to spins and MOs in transport phenomena.
If MQ currents can be generated and injected into MQ-ordered antiferromagnets, they are expected to exert a torque through coupling to the MQ order parameter, providing a new route toward the electrical manipulation of antiferromagnetic states.
Furthermore, the nonequilibrium accumulation of MQs at sample boundaries, namely MQ accumulation, is anticipated to be the multipolar counterpart of spin accumulation~\cite{Fert_2002,kato2004observation,nikolic2005nonequilibrium,nomura2005edge, shitade2025intrinsic}, which serves as one of the most important experimental manifestations of spin transport and plays a central role in spin injection and SOT devices~\cite{johnson1985interfacial,johnson1988spin}.
Motivated by this analogy, establishing the conditions for MQ accumulation and its relation to MQ transport is an important step toward realizing MQ-based spintronic functionalities. 
Despite its potential importance, however, the symmetry conditions for MQ-current generation and the microscopic mechanism of MQ accumulation remain largely unexplored.

In this study, we investigate MQ transport phenomena from both symmetry and microscopic viewpoints.
By constructing the multipole representation of the MQ conductivity tensor~\cite{hayami2018classification,yatsushiro2021multipole}, we systematically classify the symmetry-allowed MQ conductivities under the 32 crystallographic point groups (CPGs).
Based on this classification, we identify the symmetry reduction necessary for generating MQ currents and construct a minimal tight-binding model realizing such conditions.
We then evaluate the MQ conductivity within linear response theory using the Kubo formula and demonstrate the emergence of MQ currents induced by an external electric field.
Furthermore, we numerically demonstrate MQ accumulation near sample edges, thereby establishing the multipolar analogue of spin accumulation.

This paper is organized as follows.
In Sec.~\ref{sec:Symmetry analysis of magnetic quadrupole conductivity}, we derive the multipole representation of
the MQ conductivity tensors and classify its symmetry-allowed components.
In Sec.~\ref{sec:Microscopic Modeling of Magnetic Quadrupole Current}, we construct a tight-binding model and evaluate the MQ conductivity using the Kubo formula.
We demonstrate the generation of MQ currents by applying an electric field.
In Sec.~\ref{sec:Edge Accumulation of Magnetic Quadrupoles Induced by MQ Hall Currents}, we investigate MQ accumulation in a system with edge.
Section~\ref{sec:Summary} provides a summary of the paper.
Appendix~\ref{app:Angle dependence of multipole} presents the angle dependence of the multipoles.
Appendix~\ref{app:Multipole representation of MQ conductivity} shows the multipole representation of the MQ conductivity tensor, and Appendix~\ref{app:Model analysis of spin accumulation} gives the result of spin accumulation.
\section{\label{sec:Symmetry analysis of magnetic quadrupole conductivity}Symmetry analysis of magnetic quadrupole conductivity}

In this section, we investigate the symmetry properties of the MQ conductivity tensor $\sigma_{i; j}^{n \alpha}$, which characterizes the generation of MQ currents by an external electric field.
The MQ conductivity is defined through the linear-response relation
\begin{equation}
  J_{i}^{n \alpha} = \sum_{j} \sigma_{i; j}^{n \alpha} E_{j},
  \label{eq:MQ conductivity}
\end{equation}
where $J_{i}^{n \alpha}$ and $E_{j}$ for $i, j, n, \alpha = x, y, z$ denote the MQ current and the electric field, respectively. The MQ current is defined in analogy with the spin current~\cite{sinova2004universal} as $J_{i}^{n \alpha} \equiv \{M_{n}^{\alpha}, v_{i}\}_{+}$, where $M_n^\alpha=r_n\otimes S_\alpha$ is the MQ operator.
The symbols $\{\cdot,\cdot\}_{+}$, $v_{i}$, $r_{n}$, and $S_{\alpha}$ represent the anticommutator, velocity, position vector, and spin angular momentum, respectively.

Since $\sigma_{i;j}^{n\alpha}$ is a rank-4 axial tensor, MQ-current generation requires broken spatial inversion symmetry.
Because the MQ current is even under spatial inversion whereas the electric field is odd, the MQ conductivity tensor $\sigma_{i;j}^{n\alpha}$ vanishes in centrosymmetric systems.
Therefore, finite MQ currents can arise only in noncentrosymmetric crystal structures.
In the following, we identify the CPGs that permit MQ-current generation and clarify the corresponding symmetry conditions.

To this end, we employ the multipole representation, which provides a unified framework for describing symmetry and microscopic electronic degrees of freedom~\cite{kusunose2022generalization,hayami2024unified}. 
In this framework, multipoles are classified according to their transformation properties under $\mathcal{P}$ and $\mathcal{T}$ operations into four categories: electric, magnetic, magnetic toroidal, and electric toroidal multipoles~\cite{dubovik1975multipole}. 
This classification is particularly useful for identifying symmetry-allowed response tensors and their nonvanishing components, including MQ conductivities discussed below.
By expressing the MQ conductivity tensor into multipole basis, one can directly determine the irreducible representation of each conductivity component and thereby classify the symmetry-allowed MQ conductivities in a given CPG.
Furthermore, the multipole representation establishes a direct connection between macroscopic transport coefficients and microscopic multipolar degrees of freedom, which serves as the basis for the model analysis in Sec.~\ref{sec:Microscopic Modeling of Magnetic Quadrupole Current}.

In the following, we first derive the multipole representation of the MQ conductivity tensor by decomposing it into Ohmic and Hall contributions in Sec.~\ref{sec:Correspondence to multipole}.
We then combine the obtained multipole representation with the classification of symmetry-allowed multipoles under the 32 CPGs and systematically identify the symmetry conditions required for finite MQ conductivities in Sec.~\ref{sec:Classification under 32 point groups}.

To facilitate the subsequent discussion, we first summarize the multipole notation used throughout this paper.
Depending on the crystal system, we use two types of multipole notation: $c$-multipole for the cubic point group and its subgroups, and $t$-multipole for the hexagonal point group and its subgroups.
Following Refs.~\cite{hayami2018classification,yatsushiro2021multipole}, the notations for $c$- (or $t$-) multipole $Z_{ij\cdots}$ are given by
$Z_{0}$ for the rank-0 monopole, $Z_{i} = \{Z_{x}$, $Z_{y}$, $Z_{z}\}$ for the rank-1 dipole, $Z_{ij} = \{Z_{u}$, $Z_{v}$, $Z_{yz}$, $Z_{zx}$, $Z_{xy}\}$ for the rank-2 quadrupole,
$Z_{ijk} = \{Z_{xyz}$, $Z^{\alpha}_{x}$, $Z^{\alpha}_{y}$, $Z^{\alpha}_{z}$, $Z^{\beta}_{x}$, $Z^{\beta}_{y}$, $Z^{\beta}_{z}\}$ ($= \{Z_{xyz}$, $Z_{z}^{\alpha}$, $Z_{z}^{\beta}$, $Z_{3u}$, $Z_{3v}$, $Z_{3a}$, $Z_{3b}\}$) for the rank-3 octupole,
and $Z_{ijkl} = \{Z_{4}$, $Z_{4u}$, $Z_{4v}$, $Z^{\alpha}_{4x}$, $Z^{\alpha}_{4y}$, $Z^{\alpha}_{4z}$, $Z^{\beta}_{4x}$, $Z^{\beta}_{4y}$, $Z^{\beta}_{4z}\}$ ($= \{Z_{40}$, $Z_{4a}$, $Z_{4b}$, $Z_{4u}^{\alpha}$, $Z_{4v}^{\alpha}$, $Z_{4u}^{\beta 1}$, $Z_{4v}^{\beta 1}$, $Z_{4u}^{\beta 2}$, $Z_{4v}^{\beta 2}\}$) for the rank-4 hexadecapole.
Each $Z_{ij\cdots}$ is obtained as a linear combination of the complex spherical-tensor expressions $Z_{lm}$ of rank $l$, where $m$ labels the tensor component.
Note that $t$-multipoles from rank 0 to rank 2 are identical to $c$-multipoles.
The angular dependence of multipoles are summarized in Table~\ref{tab:angle dependences of multipole} in Appendix~\ref{app:Angle dependence of multipole}.

\subsection{\label{sec:Correspondence to multipole}Correspondence to multipole}
To identify the symmetry properties of MQ transport, we express the MQ conductivity tensors $\sigma_{i; j}^{n \alpha}$ in the basis of multipoles.
In this representation, each tensor component is associated with a definite irreducible representation of the rotation group, enabling a direct correspondence between MQ transport responses and multipolar degrees of freedom.

To derive the multipole representations of $\sigma_{i; j}^{n \alpha}$, we introduce the tensor-product relation among multipoles:
\begin{equation}
  Z_{lm} = \sum_{m_{1}, m_{2}} C^{l_{1} l_{2}; m_{1} m_{2}}_{l_{1} l_{2}; l m} Z_{l_{1} m_{1}} \otimes Z_{l_{2} m_{2}},
  \label{eq:tensor-product relation between multipoles}
\end{equation}
where $C^{l_{1} l_{2}; q_{1} q_{2}}_{l_{1} l_{2}; l q}$ denotes the Clebsch--Gordan coefficient.
The angular momentum quantum numbers satisfy $l = |l_{1} - l_{2}|, |l_{1} - l_{2}| + 1, \cdots, l_{1} + l_{2}$ and $m = -l, -l + 1, \cdots, l$.
In the multipole notation, we also use the symbols $Z=X$ and $Y$ to distinguish polar and axial quantities, respectively.
The four types of multipoles and their corresponding $\mathcal P$- and $\mathcal T$-parities are summarized in Table~\ref{tab: 4-type multipoles}~\cite{dubovik1975multipole,kusunose2022generalization,hayami2024unified}.
The parity under $\mathcal P$ is determined by the multipole rank $l$, whereas the parity under $\mathcal T$ distinguishes electric and magnetic characters. 
Consequently, each multipole belongs to one of the four symmetry classes, which serve as fundamental building blocks for classifying response tensors.

\begin{table}[htbp]
  \caption{
    Four types of multipoles classified by their spatial-inversion ($\mathcal{P}$) and time-reversal ($\mathcal{T}$) parities.
    Polar (axial) multipoles transform as $\mathcal{P}\left(X_{lm}\right) = (-1)^{l} X_{lm}$ $\left[\mathcal{P}\left(Y_{lm}\right) = (-1)^{l+1} Y_{lm}\right]$.
    Combining the $\mathcal{P}$ and $\mathcal{T}$ characters yields four multipole classes:
    electric multipoles $Q_{lm}$,
    magnetic toroidal multipoles $T_{lm}$,
    electric toroidal multipoles $G_{lm}$,
    and magnetic multipoles $M_{lm}$.
  }
  \begin{tabular}{c||c|cc} \hline\hline
    \diagbox{$\mathcal{P}$}{$\mathcal{T}$}& & $+1$ & $-1$\\ \hline
    $(-1)^{l}$ & $X_{lm}$ & $Q_{lm}$ & $T_{lm}$ \rule[-3pt]{0pt}{12pt}\\
    $(-1)^{l+1}$ & $Y_{lm}$ & $G_{lm}$ & $M_{lm}$ \rule[-3pt]{0pt}{12pt}\\ \hline\hline
  \end{tabular}
  \label{tab: 4-type multipoles}
\end{table}

Since the MQ conductivity couples the electric field to the MQ current, its symmetry is determined by the transformation properties of the constituent quantities $J_i$, $E_j$, $r_n$, and $S_\alpha$.
Accordingly, $\sigma_{i;j}^{n\alpha}$ transforms as the tensor product $J_i\otimes E_j\otimes r_n\otimes S_\alpha$ under a symmetry operation $R$:
\begin{equation}
R[\sigma_{i;j}^{n\alpha}]
\leftrightarrow
R[J_i]
\otimes
R[E_j]
\otimes
R[r_n]
\otimes
R[S_\alpha],
\end{equation}
where $J_i=ev_i$ and $e$ denotes the electric charge.
Therefore, the multipole representation of $\sigma_{i;j}^{n\alpha}$ can be systematically constructed by sequentially combining these four vector quantities through Eq.~(\ref{eq:tensor-product relation between multipoles}). 

The decomposition naturally separates into two parts.
The first part, $J_i\otimes E_j$, describes the transport sector and is closely related to the ordinary electrical conductivity.
The second part, $r_n\otimes S_\alpha$, describes the MQ degree of freedom itself.
Their combination ultimately determines the symmetry of MQ transport.
We first investigate the correspondence between multipoles and $J_{i} \otimes E_{j}$, and then between multipoles and $r_{n} \otimes S_{\alpha}$.
Finally, we combine these two sectors to obtain the multipole representation of $\sigma_{i;j}^{n\alpha}$.

We begin with the transport sector $J_i\otimes E_j$, which is related to the electrical conductivity tensor $\sigma_{i;j}$ defined by $J_{i} = \sum_{j} \sigma_{i;j} E_{j}$.
Since both $J_{i}$ and $E_{j}$ are rank-1 tensors, their tensor product is decomposed into multipoles with ranks $l=0,1,2$:
\begin{align}
  \sigma_{i;j}
  &\leftrightarrow
  \sum_{lm} \sigma_{i;j}(Z_{lm}) J_{i} \otimes E_{j}, \\
  \sigma_{i;j}(Z_{lm})
  &\equiv
  \sum_{m_{i} m_{j}}
  C_{1, 1; l, m}^{1, 1; m_{i}, m_{j}}
  \bigg(\prod_{k = i, j} U_{k m_{k}}\bigg),
\end{align}
where $m = -l, -l + 1, \cdots, l$ and $m_{i, j} = -1, 0, 1$.
Here, $\sigma_{i;j}(Z_{lm})$ is a numerical coefficient specifying the transformation from the Cartesian tensor basis $J_i \otimes E_j$ to the multipole basis $Z_{lm}$. 
It is determined solely by the Clebsch--Gordan coefficients and the transformation matrix $U_{k m_k}$.
The matrix $U_{k m_{k}}$ transforms the rank-1 multipole basis $Z_{k}$ into the spherical-tensor basis $Z_{1m_{k}}$ through $Z_{1 m_{k}} = \sum_{k} U_{k m_{k}} Z_{k}$.

By construction, $\sum_{ij}\sigma_{i;j}(Z_{lm})J_i\otimes E_j$ transforms as the multipole $Z_{lm}$.
Since both $J_{i}$ and $E_{j}$ transform as polar vectors, the multipole representation of $\sigma_{i;j}$ is given by
\begin{equation}
  \begin{pmatrix}
    X_{0} - X_{u} + \sqrt{3} X_{v} & Y_{z} + \sqrt{3} X_{xy} & - Y_{y} + \sqrt{3} X_{zx} \\
    - Y_{z} + \sqrt{3} X_{xy} & X_{0} - X_{u} - \sqrt{3} X_{v} & Y_{x} + \sqrt{3} X_{yz} \\
    Y_{y} + \sqrt{3} X_{zx} & - Y_{x} + \sqrt{3} X_{yz} & X_{0} + 2 X_{u}
  \end{pmatrix},
  \label{eq:multipole representation of rank-2-polar tensor}
\end{equation}
where each element represents
\begin{equation}
  \begin{pmatrix}
    \sigma_{x; x} & \sigma_{x; y} & \sigma_{x; z} \\
    \sigma_{y; x} & \sigma_{y; y} & \sigma_{y; z} \\
    \sigma_{z; x} & \sigma_{z; y} & \sigma_{z; z}
  \end{pmatrix}.
\end{equation}
Note that the coefficients of the multipoles are represented by $\sigma_{i;j}(Z_{ij\cdots})$, which are linear combinations of $\sigma_{i;j}(Z_{lm})$.

The conductivity tensor can be decomposed into Ohmic and Hall parts: 
\begin{align}
  \sigma_{ij}^{(\mathrm{O})} &\equiv \frac{1}{2} \left(\sigma_{i;j} + \sigma_{j;i}\right), \\
  \sigma_{ij}^{(\mathrm{H})} &\equiv \frac{1}{2} \left(\sigma_{i;j} - \sigma_{j;i}\right),
\end{align}
where $\sigma_{ij}^{(\mathrm O)}$ and $\sigma_{ij}^{(\mathrm H)}$ are symmetric and antisymmetric under the exchange $i\leftrightarrow j$, respectively.
In the multipole representation, the Ohmic part is described by rank-0 and rank-2 polar multipoles, whereas the Hall part is described by rank-1 axial multipoles.
Consequently, the activation of a given multipole directly determines the symmetry-allowed MQ conductivity components.

We next consider the MQ sector $r_n\otimes S_\alpha$, which corresponds to the MQ operator $M_n^\alpha$.
The corresponding multipole representation is given by
\begin{equation}
  \begin{pmatrix}
    Y_{0} - Y_{u} + \sqrt{3} Y_{v} & X_{z} + \sqrt{3} Y_{xy} & - X_{y} + \sqrt{3} Y_{zx} \\
    - X_{z} + \sqrt{3} Y_{xy} & Y_{0} - Y_{u} - \sqrt{3} Y_{v} & X_{x} + \sqrt{3} Y_{yz} \\
    X_{y} + \sqrt{3} Y_{zx} & - X_{x} + \sqrt{3} Y_{yz} & Y_{0} + 2 Y_{u}
  \end{pmatrix},
  \label{eq:multipole representation of rank-2-axial tensor}
\end{equation}
where each element represents
\begin{equation*}
  \begin{pmatrix}
    M_{x}^{x} & M_{x}^{y} & M_{x}^{z} \\
    M_{y}^{x} & M_{y}^{y} & M_{y}^{z} \\
    M_{z}^{x} & M_{z}^{y} & M_{z}^{z}
  \end{pmatrix}.
\end{equation*}
Unlike the electrical conductivity tensor, the multipole representation of the MQ conductivity tensor is obtained by exchanging polar and axial multipoles, namely $X \leftrightarrow Y$.
This difference originates from the fact that the MQ operator consists of a polar vector $r_n$ and an axial vector $S_\alpha$, whereas the electrical conductivity is described by the polar electric-current operator alone.

Combining the transport sector $J_i\otimes E_j$ and the MQ sector $r_n\otimes S_\alpha$, we obtain the complete multipole representation of the MQ conductivity tensor $\sigma_{i;j}^{n\alpha}$:
\begin{widetext}
  \begin{align}
    \sigma_{i;j}^{n\alpha}
    \leftrightarrow &
    \sum_{lm} \sigma_{i;j}^{n\alpha}\left(Z_{lm}^{(l'l'')}\right) J_{i} \otimes E_{j} \otimes r_{n} \otimes S_{\alpha}, \\
    \sigma_{i;j}^{n\alpha}\left(Z_{lm}^{(l'l'')}\right)
    \equiv &
    \sum_{m' m''}
    \sum_{m_{i} m_{j}}
    \sum_{m_{n} m_{\alpha}}
    C_{l', l''; l, m}^{l', l''; m', m''}
    C_{1, 1; l', m'}^{1, 1; m_{i}, m_{j}}
    C_{1, 1; l'', m''}^{1, 1; m_{n}, m_{\alpha}}
    \bigg(\prod_{k = i, j, n, \alpha} U_{k m_{k}}\bigg),
    \label{eq:multipole representation of rank-4 tensor}
  \end{align}
\end{widetext}
where $l', l'' = 0, 1, 2$.
As in the case of ordinary electrical conductivity, the MQ conductivity tensor can be decomposed into Ohmic and Hall contributions:
\begin{align}
  \sigma_{ij}^{n\alpha(\mathrm{O})} &\equiv \frac{1}{2} \left(\sigma_{i;j}^{n\alpha} + \sigma_{j;i}^{n\alpha}\right), \\
  \sigma_{ij}^{n\alpha(\mathrm{H})} &\equiv \frac{1}{2} \left(\sigma_{i;j}^{n\alpha} - \sigma_{j;i}^{n\alpha}\right).
\end{align}
The former and latter are obtained by combining the MQ sector with the Ohmic sector ($\sigma_{ij}^{(\mathrm{O})}$) and Hall sector ($\sigma_{ij}^{(\mathrm{H})}$) of the electrical conductivity, respectively.
Equivalently, they are represented by $\sigma_{i;j}^{n\alpha} (Z_{lm}^{(l'(=0,2)l'')})$ and $\sigma_{i;j}^{n\alpha} (Z_{lm}^{(l'(=1)l'')})$.
Note that $Z_{lm}^{l'l''}$ with odd (even) rank $l=1,3$ ($l=0,2,4$) corresponds to polar multipoles $X$ (axial multipoles $Y$).

The obtained decomposition establishes a direct correspondence between MQ transport responses and multipolar degrees of freedom.
In Appendix~\ref{app:Multipole representation of MQ conductivity}, we provide the complete multipole representations of the Ohmic and Hall MQ conductivities for all $c$- and $t$-multipoles.

\subsection{\label{sec:Classification under 32 point groups}Classification under 32 point groups}

Having established the multipole representation of the MQ conductivity tensor, we now classify the symmetry-allowed MQ conductivities under the 32 CPGs.
The primary, secondary, and tertiary axes used throughout the classification are listed in Table~\ref{tab:primary axes}.
The classification is performed by combining the multipole representations derived in Sec.~\ref{sec:Correspondence to multipole} with the systematic classification of symmetry-allowed multipoles under CPGs~\cite{hayami2018classification,watanabe2018group,yatsushiro2021multipole}.
We here suppose the presence of the $\mathcal{T}$ symmetry, where $X$ and $Y$ corespond to the electric ($Q$) and electric toroidal ($G$) multipoles, respectively.
Since each component of the MQ conductivity tensor is associated with a definite multipole, finite MQ conductivities are obtained only when the corresponding multipole belongs to the totally symmetric representation of the crystal point group. 
This procedure enables a complete symmetry classification of MQ transport without relying on microscopic details of a specific material or model.
\begin{table}[htbp]
\caption{
  Definition of the primary, secondary, and tertiary axes for the seven crystal systems in Cartesian coordinates~\cite{yatsushiro2021multipole}.
}
\centering
\begin{tabular}{l|ccc} \hline\hline
 & Primary & Secondary & Tertiary \\ \hline
Cubic        & $\langle100\rangle$ & $\langle111\rangle$ & $\langle110\rangle$ \\
Tetragonal   & [001] & [100] & [110] \\
Orthorhombic & [100] & [010] & [001] \\
Monoclinic   & [010] &  &  \\
Triclinic    &  &  &  \\
Hexagonal    & [001] & [100] & [010] \\
Trigonal     & [001] & [010] &  \\ \hline\hline
\end{tabular}
\label{tab:primary axes}
\end{table}

The resulting symmetry-allowed MQ Ohmic and Hall conductivities are summarized in Tables~\ref{tab:symmetry allowed MQ current under Oh} and \ref{tab:symmetry allowed MQ current under D6h}.
To compactly represent the nonzero tensor components, we use the matrix form
\begin{equation}
  \left[\begin{array}{c|c|c}
    \{M_{n}^{\alpha}\}_{xx} & \{M_{n}^{\alpha}\}_{xy} & \{M_{n}^{\alpha}\}_{xz} \\ \hline
    \{M_{n}^{\alpha}\}_{yx} & \{M_{n}^{\alpha}\}_{yy} & \{M_{n}^{\alpha}\}_{yz} \\ \hline
    \{M_{n}^{\alpha}\}_{zx} & \{M_{n}^{\alpha}\}_{zy} & \{M_{n}^{\alpha}\}_{zz}
  \end{array}\right],
  \label{eq:nonzero MQ conductivity}
\end{equation}
where $\{M_{n}^{\alpha}\}_{ij}$ lists the nonzero tensor components of the MQ Ohmic/Hall conductivities $\sigma_{ij}^{n\alpha(\mathrm{O/H})}$ corresponding to an input electric field in the $j$ direction and an output MQ current in the $i$ direction.
For example, the symmetry-allowed Ohmic MQ conductivities $\sigma_{ij}^{n\alpha(\mathrm{O})}$ under the CPG $432$ are represented by
\begin{align}
  \left[\begin{array}{c|c|c} \bm{M}_{k}^{k} & M_{x}^{y}, M_{y}^{x} & M_{x}^{z}, M_{z}^{x} \\ \hline & \bm{M}_{k}^{k} & M_{y}^{z}, M_{z}^{y} \\ \hline & & \bm{M}_{k}^{k} \\\end{array}\right],
\end{align}
where the element $\{M_{n}^{\alpha}\}_{xx} = \bm{M}_{k}^{k} \equiv \{M_{x}^{x}, M_{y}^{y}, M_{z}^{z}\}$, for instance, indicate the nonzero components $\sigma_{xx}^{xx(\mathrm{O})}, \sigma_{xx}^{yy(\mathrm{O})}, \sigma_{xx}^{zz(\mathrm{O})}$.
Similarly, the symmetry-allowed Hall MQ conductivities $\sigma_{ij}^{n\alpha(\mathrm{H})}$ under the CPG $432$ are listed as follows:
\begin{align}
  \left[\begin{array}{c|c|c}- & M_{x}^{y}, M_{y}^{x} & M_{x}^{z}, M_{z}^{x} \\ \hline & - & M_{y}^{z}, M_{z}^{y} \\ \hline & & -\\\end{array}\right],
\end{align}
where the symbol ``--" denotes the absence of symmetry-allowed components.
Noted that only independent tensor components are listed; the remaining components follow from
$\sigma_{ij}^{\C{m}n\alpha(\mathrm{O})} = \sigma_{ji}^{\C{m}n\alpha(\mathrm{O})}$ and 
$\sigma_{ij}^{\C{m}n\alpha(\mathrm{H})} = -\sigma_{ji}^{\C{m}n\alpha(\mathrm{H})}$.
\begin{table*}
\centering
\caption{
  Symmetry-allowed $\mathcal{P}$-parity-odd multipoles and MQ Ohmic and Hall conductivity tensors for the CPG $m\bar{3}m$ and its subgroups.
  Only independent tensor components are listed; the remaining components follow from $\sigma_{ij}^{\C{m}n\alpha(\mathrm{O})} = \sigma_{ji}^{\C{m}n\alpha(\mathrm{O})}$ and $\sigma_{ij}^{\C{m}n\alpha(\mathrm{H})} = -\sigma_{ji}^{\C{m}n\alpha(\mathrm{H})}$.
  The symbol ``--" denotes the absence of symmetry-allowed components.
  For brevity, we introduce the shorthand notations $\bm{M}_{k}^{k} = \{M_{x}^{x}, M_{y}^{y}, M_{z}^{z}\}$ and $\mathrm{all} = \{\bm{M}_{k}^{k}, M_{y}^{z}, M_{z}^{x}, M_{x}^{y}, M_{z}^{y}, M_{x}^{z}, M_{y}^{x}\}$.
  Since the MQ conductivity tensor is odd under spatial inversion, all MQ conductivity components vanish in the centrosymmetric groups $m\bar{3}m$, $m\bar{3}$, $4/mmm$, $4/m$, $mmm$, $2/m$, and $\bar{1}$.
}
\scalebox{0.7}{
\begin{tabular}{c|*{5}{c}|c|c} \hline \hline
CPG & rank 0 & rank 1 & rank 2 & rank 3 & rank 4 & Ohmic & Hall \\ \hline
$432$ & \scalebox{1}{$G_{0}$ } & & & & \scalebox{1}{$G_{4}$ } & \scalebox{1}{$\left[\begin{array}{c|c|c} \bm{M}_{k}^{k} & M_{x}^{y}, M_{y}^{x} & M_{x}^{z}, M_{z}^{x} \\ \hline & \bm{M}_{k}^{k} & M_{y}^{z}, M_{z}^{y} \\ \hline & & \bm{M}_{k}^{k} \\\end{array}\right]$} & \scalebox{1}{$\left[\begin{array}{c|c|c}- & M_{x}^{y}, M_{y}^{x} & M_{x}^{z}, M_{z}^{x} \\ \hline & - & M_{y}^{z}, M_{z}^{y} \\ \hline & & -\\\end{array}\right]$} \\
$\bar{4}3m$ & & & & \scalebox{1}{$Q_{xyz}$ } & & \scalebox{1}{$\left[\begin{array}{c|c|c} M_{y}^{y}, M_{z}^{z} & M_{x}^{y}, M_{y}^{x} & M_{x}^{z}, M_{z}^{x} \\ \hline & M_{x}^{x}, M_{z}^{z} & M_{y}^{z}, M_{z}^{y} \\ \hline & & M_{x}^{x}, M_{y}^{y} \\\end{array}\right]$} & \scalebox{1}{$\left[\begin{array}{c|c|c}- & M_{x}^{y}, M_{y}^{x} & M_{x}^{z}, M_{z}^{x} \\ \hline & - & M_{y}^{z}, M_{z}^{y} \\ \hline & & -\\\end{array}\right]$} \\
$23$ & \scalebox{1}{$G_{0}$ } & & & \scalebox{1}{$Q_{xyz}$ } & \scalebox{1}{$G_{4}$ } & \scalebox{1}{$\left[\begin{array}{c|c|c} \bm{M}_{k}^{k} & M_{x}^{y}, M_{y}^{x} & M_{x}^{z}, M_{z}^{x} \\ \hline & \bm{M}_{k}^{k} & M_{y}^{z}, M_{z}^{y} \\ \hline & & \bm{M}_{k}^{k} \\\end{array}\right]$} & \scalebox{1}{$\left[\begin{array}{c|c|c}- & M_{x}^{y}, M_{y}^{x} & M_{x}^{z}, M_{z}^{x} \\ \hline & - & M_{y}^{z}, M_{z}^{y} \\ \hline & & -\\\end{array}\right]$} \\
$422$ & \scalebox{1}{$G_{0}$ } & & \scalebox{1}{$G_{u}$ } & & \scalebox{1}{$G_{4}$, $G_{4u}$ } & \scalebox{1}{$\left[\begin{array}{c|c|c} \bm{M}_{k}^{k} & M_{x}^{y}, M_{y}^{x} & M_{x}^{z}, M_{z}^{x} \\ \hline & \bm{M}_{k}^{k} & M_{y}^{z}, M_{z}^{y} \\ \hline & & \bm{M}_{k}^{k} \\\end{array}\right]$} & \scalebox{1}{$\left[\begin{array}{c|c|c}- & M_{x}^{y}, M_{y}^{x} & M_{x}^{z}, M_{z}^{x} \\ \hline & - & M_{y}^{z}, M_{z}^{y} \\ \hline & & -\\\end{array}\right]$} \\
$\bar{4}2m$ & & & \scalebox{1}{$G_{v}$ } & \scalebox{1}{$Q_{xyz}$ } & \scalebox{1}{$G_{4v}$ } & \scalebox{1}{$\left[\begin{array}{c|c|c} \bm{M}_{k}^{k} & M_{x}^{y}, M_{y}^{x} & M_{x}^{z}, M_{z}^{x} \\ \hline & \bm{M}_{k}^{k} & M_{y}^{z}, M_{z}^{y} \\ \hline & & M_{x}^{x}, M_{y}^{y} \\\end{array}\right]$} & \scalebox{1}{$\left[\begin{array}{c|c|c}- & M_{x}^{y}, M_{y}^{x} & M_{x}^{z}, M_{z}^{x} \\ \hline & - & M_{y}^{z}, M_{z}^{y} \\ \hline & & -\\\end{array}\right]$} \\
$4mm$ & & \scalebox{1}{$Q_{z}$ } & & \scalebox{1}{$Q^{\alpha}_{z}$ } & \scalebox{1}{$G^{\alpha}_{4z}$ } & \scalebox{1}{$\left[\begin{array}{c|c|c} M_{x}^{y}, M_{y}^{x} & M_{x}^{x}, M_{y}^{y} & M_{y}^{z}, M_{z}^{y} \\ \hline & M_{x}^{y}, M_{y}^{x} & M_{x}^{z}, M_{z}^{x} \\ \hline & & M_{x}^{y}, M_{y}^{x} \\\end{array}\right]$} & \scalebox{1}{$\left[\begin{array}{c|c|c}- & \bm{M}_{k}^{k} & M_{y}^{z}, M_{z}^{y} \\ \hline & - & M_{x}^{z}, M_{z}^{x} \\ \hline & & -\\\end{array}\right]$} \\
$4$ & \scalebox{1}{$G_{0}$ } & \scalebox{1}{$Q_{z}$ } & \scalebox{1}{$G_{u}$ } & \scalebox{1}{$Q^{\alpha}_{z}$ } & \scalebox{1}{$G_{4}$, $G_{4u}$, $G^{\alpha}_{4z}$ } & \scalebox{1}{$\left[\begin{array}{c|c|c} M_{x}^{y}, M_{y}^{x}, \bm{M}_{k}^{k} & M_{x}^{x}, M_{x}^{y}, M_{y}^{x}, M_{y}^{y} & M_{x}^{z}, M_{y}^{z}, M_{z}^{x}, M_{z}^{y} \\ \hline & M_{x}^{y}, M_{y}^{x}, \bm{M}_{k}^{k} & M_{x}^{z}, M_{y}^{z}, M_{z}^{x}, M_{z}^{y} \\ \hline & & M_{x}^{y}, M_{y}^{x}, \bm{M}_{k}^{k} \\\end{array}\right]$} & \scalebox{1}{$\left[\begin{array}{c|c|c}- & M_{x}^{y}, M_{y}^{x}, \bm{M}_{k}^{k} & M_{x}^{z}, M_{y}^{z}, M_{z}^{x}, M_{z}^{y} \\ \hline & - & M_{x}^{z}, M_{y}^{z}, M_{z}^{x}, M_{z}^{y} \\ \hline & & -\\\end{array}\right]$} \\
$\bar{4}$ & & & \scalebox{1}{$G_{v}$, $G_{xy}$ } & \scalebox{1}{$Q_{xyz}$, $Q^{\beta}_{z}$ } & \scalebox{1}{$G_{4v}$, $G^{\beta}_{4z}$ } & \scalebox{1}{$\left[\begin{array}{c|c|c} M_{x}^{y}, M_{y}^{x}, \bm{M}_{k}^{k} & M_{x}^{y}, M_{y}^{x}, \bm{M}_{k}^{k} & M_{x}^{z}, M_{y}^{z}, M_{z}^{x}, M_{z}^{y} \\ \hline & M_{x}^{y}, M_{y}^{x}, \bm{M}_{k}^{k} & M_{x}^{z}, M_{y}^{z}, M_{z}^{x}, M_{z}^{y} \\ \hline & & M_{x}^{x}, M_{x}^{y}, M_{y}^{x}, M_{y}^{y} \\\end{array}\right]$} & \scalebox{1}{$\left[\begin{array}{c|c|c}- & M_{x}^{x}, M_{x}^{y}, M_{y}^{x}, M_{y}^{y} & M_{x}^{z}, M_{y}^{z}, M_{z}^{x}, M_{z}^{y} \\ \hline & - & M_{x}^{z}, M_{y}^{z}, M_{z}^{x}, M_{z}^{y} \\ \hline & & -\\\end{array}\right]$} \\
$222$ & \scalebox{1}{$G_{0}$ } & & \scalebox{1}{$G_{u}$, $G_{v}$ } & \scalebox{1}{$Q_{xyz}$ } & \scalebox{1}{$G_{4}$, $G_{4u}$, $G_{4v}$ } & \scalebox{1}{$\left[\begin{array}{c|c|c} \bm{M}_{k}^{k} & M_{x}^{y}, M_{y}^{x} & M_{x}^{z}, M_{z}^{x} \\ \hline & \bm{M}_{k}^{k} & M_{y}^{z}, M_{z}^{y} \\ \hline & & \bm{M}_{k}^{k} \\\end{array}\right]$} & \scalebox{1}{$\left[\begin{array}{c|c|c}- & M_{x}^{y}, M_{y}^{x} & M_{x}^{z}, M_{z}^{x} \\ \hline & - & M_{y}^{z}, M_{z}^{y} \\ \hline & & -\\\end{array}\right]$} \\
$mm2$ & & \scalebox{1}{$Q_{z}$ } & \scalebox{1}{$G_{xy}$ } & \scalebox{1}{$Q^{\alpha}_{z}$, $Q^{\beta}_{z}$ } & \scalebox{1}{$G^{\alpha}_{4z}$, $G^{\beta}_{4z}$ } & \scalebox{1}{$\left[\begin{array}{c|c|c} M_{x}^{y}, M_{y}^{x} & \bm{M}_{k}^{k} & M_{y}^{z}, M_{z}^{y} \\ \hline & M_{x}^{y}, M_{y}^{x} & M_{x}^{z}, M_{z}^{x} \\ \hline & & M_{x}^{y}, M_{y}^{x} \\\end{array}\right]$} & \scalebox{1}{$\left[\begin{array}{c|c|c}- & \bm{M}_{k}^{k} & M_{y}^{z}, M_{z}^{y} \\ \hline & - & M_{x}^{z}, M_{z}^{x} \\ \hline & & -\\\end{array}\right]$} \\
$2$ & \scalebox{1}{$G_{0}$ } & \scalebox{1}{$Q_{y}$ } & \scalebox{1}{$G_{u}$, $G_{v}$, $G_{zx}$ } & \scalebox{1}{$Q_{xyz}$, $Q^{\alpha}_{y}$, $Q^{\beta}_{y}$ } & \scalebox{1}{$G_{4}$, $G_{4u}$, $G_{4v}$, $G^{\alpha}_{4y}$, $G^{\beta}_{4y}$ } & \scalebox{1}{$\left[\begin{array}{c|c|c} M_{x}^{z}, M_{z}^{x}, \bm{M}_{k}^{k} & M_{x}^{y}, M_{y}^{x}, M_{y}^{z}, M_{z}^{y} & M_{x}^{z}, M_{z}^{x}, \bm{M}_{k}^{k} \\ \hline & M_{x}^{z}, M_{z}^{x}, \bm{M}_{k}^{k} & M_{x}^{y}, M_{y}^{x}, M_{y}^{z}, M_{z}^{y} \\ \hline & & M_{x}^{z}, M_{z}^{x}, \bm{M}_{k}^{k} \\\end{array}\right]$} & \scalebox{1}{$\left[\begin{array}{c|c|c}- & M_{x}^{y}, M_{y}^{x}, M_{y}^{z}, M_{z}^{y} & M_{x}^{z}, M_{z}^{x}, \bm{M}_{k}^{k} \\ \hline & - & M_{x}^{y}, M_{y}^{x}, M_{y}^{z}, M_{z}^{y} \\ \hline & & -\\\end{array}\right]$} \\
$m$ & & \scalebox{1}{$Q_{x}$, $Q_{z}$ } & \scalebox{1}{$G_{xy}$, $G_{yz}$ } & \scalebox{1}{$Q^{\alpha}_{x}$, $Q^{\beta}_{x}$, $Q^{\alpha}_{z}$, $Q^{\beta}_{z}$ } & \scalebox{1}{$G^{\alpha}_{4x}$, $G^{\beta}_{4x}$, $G^{\alpha}_{4z}$, $G^{\beta}_{4z}$ } & \scalebox{1}{$\left[\begin{array}{c|c|c} M_{x}^{y}, M_{y}^{x}, M_{y}^{z}, M_{z}^{y} & M_{x}^{z}, M_{z}^{x}, \bm{M}_{k}^{k} & M_{x}^{y}, M_{y}^{x}, M_{y}^{z}, M_{z}^{y} \\ \hline & M_{x}^{y}, M_{y}^{x}, M_{y}^{z}, M_{z}^{y} & M_{x}^{z}, M_{z}^{x}, \bm{M}_{k}^{k} \\ \hline & & M_{x}^{y}, M_{y}^{x}, M_{y}^{z}, M_{z}^{y} \\\end{array}\right]$} & \scalebox{1}{$\left[\begin{array}{c|c|c}- & M_{x}^{z}, M_{z}^{x}, \bm{M}_{k}^{k} & M_{x}^{y}, M_{y}^{x}, M_{y}^{z}, M_{z}^{y} \\ \hline & - & M_{x}^{z}, M_{z}^{x}, \bm{M}_{k}^{k} \\ \hline & & -\\\end{array}\right]$} \\
$1$ & \multicolumn{5}{c|}{All $\mathcal{P}$-odd multipoles} & \scalebox{1}{$\left[\begin{array}{c|c|c}\mathrm{all} & \mathrm{all} & \mathrm{all} \\ \hline & \mathrm{all} & \mathrm{all} \\ \hline & & \mathrm{all}\\\end{array}\right]$} & \scalebox{1}{$\left[\begin{array}{c|c|c}- & \mathrm{all} & \mathrm{all} \\ \hline & - & \mathrm{all} \\ \hline & & -\\\end{array}\right]$} \\ \hline \hline
\end{tabular}
}
\label{tab:symmetry allowed MQ current under Oh}
\end{table*}
\begin{table*}
\centering
\caption{
  Symmetry-allowed $\mathcal{P}$-parity-odd multipoles and MQ Ohmic and Hall conductivity tensors for the CPG of hexagonal and trigonal systems.
  Since the MQ conductivity tensor is odd under spatial inversion, all MQ conductivity components are forbidden in the centrosymmetric groups $6/mmm$, $6/m$, $\bar{3}m$, and $\bar{3}$.
}
\scalebox{0.65}{
\begin{tabular}{c|*{5}{c}|c|c} \hline \hline
CPG & rank 0 & rank 1 & rank 2 & rank 3 & rank 4 & Ohmic & Hall \\ \hline
$622$ & \scalebox{1}{$G_{0}$ } & & \scalebox{1}{$G_{u}$ } & & \scalebox{1}{$G_{40}$ } & \scalebox{1}{$\left[\begin{array}{c|c|c} M_{k}^{k} & M_{x}^{y}, M_{y}^{x} & M_{x}^{z}, M_{z}^{x} \\ \hline & M_{k}^{k} & M_{y}^{z}, M_{z}^{y} \\ \hline & & M_{k}^{k} \\\end{array}\right]$} & \scalebox{1}{$\left[\begin{array}{c|c|c}- & M_{x}^{y}, M_{y}^{x} & M_{x}^{z}, M_{z}^{x} \\ \hline & - & M_{y}^{z}, M_{z}^{y} \\ \hline & & -\\\end{array}\right]$} \\
$\bar{6}m2$ & & & & \scalebox{1}{$Q_{3b}$ } & \scalebox{1}{$G_{4b}$ } & \scalebox{1}{$\left[\begin{array}{c|c|c} M_{x}^{z}, M_{z}^{x} & M_{y}^{z}, M_{z}^{y} & M_{x}^{x}, M_{y}^{y} \\ \hline & M_{x}^{z}, M_{z}^{x} & M_{x}^{y}, M_{y}^{x} \\ \hline & & -\\\end{array}\right]$} & \scalebox{1}{$\left[\begin{array}{c|c|c}- & - & M_{x}^{x}, M_{y}^{y} \\ \hline & - & M_{x}^{y}, M_{y}^{x} \\ \hline & & -\\\end{array}\right]$} \\
$6mm$ & & \scalebox{1}{$Q_{z}$ } & & \scalebox{1}{$Q^{\alpha}_{z}$ } & & \scalebox{1}{$\left[\begin{array}{c|c|c} M_{x}^{y}, M_{y}^{x} & M_{x}^{x}, M_{y}^{y} & M_{y}^{z}, M_{z}^{y} \\ \hline & M_{x}^{y}, M_{y}^{x} & M_{x}^{z}, M_{z}^{x} \\ \hline & & M_{x}^{y}, M_{y}^{x} \\\end{array}\right]$} & \scalebox{1}{$\left[\begin{array}{c|c|c}- & M_{k}^{k} & M_{y}^{z}, M_{z}^{y} \\ \hline & - & M_{x}^{z}, M_{z}^{x} \\ \hline & & -\\\end{array}\right]$} \\
$6$ & \scalebox{1}{$G_{0}$ } & \scalebox{1}{$Q_{z}$ } & \scalebox{1}{$G_{u}$ } & \scalebox{1}{$Q^{\alpha}_{z}$ } & \scalebox{1}{$G_{40}$ } & \scalebox{1}{$\left[\begin{array}{c|c|c} M_{x}^{y}, M_{y}^{x}, M_{k}^{k} & M_{x}^{x}, M_{x}^{y}, M_{y}^{x}, M_{y}^{y} & M_{x}^{z}, M_{y}^{z}, M_{z}^{x}, M_{z}^{y} \\ \hline & M_{x}^{y}, M_{y}^{x}, M_{k}^{k} & M_{x}^{z}, M_{y}^{z}, M_{z}^{x}, M_{z}^{y} \\ \hline & & M_{x}^{y}, M_{y}^{x}, M_{k}^{k} \\\end{array}\right]$} & \scalebox{1}{$\left[\begin{array}{c|c|c}- & M_{x}^{y}, M_{y}^{x}, M_{k}^{k} & M_{x}^{z}, M_{y}^{z}, M_{z}^{x}, M_{z}^{y} \\ \hline & - & M_{x}^{z}, M_{y}^{z}, M_{z}^{x}, M_{z}^{y} \\ \hline & & -\\\end{array}\right]$} \\
$\bar{6}$ & & & & \scalebox{1}{$Q_{3a}$, $Q_{3b}$ } & \scalebox{1}{$G_{4a}$, $G_{4b}$ } & \scalebox{1}{$\left[\begin{array}{c|c|c} M_{x}^{z}, M_{y}^{z}, M_{z}^{x}, M_{z}^{y} & M_{x}^{z}, M_{y}^{z}, M_{z}^{x}, M_{z}^{y} & M_{x}^{x}, M_{x}^{y}, M_{y}^{x}, M_{y}^{y} \\ \hline & M_{x}^{z}, M_{y}^{z}, M_{z}^{x}, M_{z}^{y} & M_{x}^{x}, M_{x}^{y}, M_{y}^{x}, M_{y}^{y} \\ \hline & & -\\\end{array}\right]$} & \scalebox{1}{$\left[\begin{array}{c|c|c}- & - & M_{x}^{x}, M_{x}^{y}, M_{y}^{x}, M_{y}^{y} \\ \hline & - & M_{x}^{x}, M_{x}^{y}, M_{y}^{x}, M_{y}^{y} \\ \hline & & -\\\end{array}\right]$} \\
$32$ & \scalebox{1}{$G_{0}$ } & & \scalebox{1}{$G_{u}$ } & \scalebox{1}{$Q_{3b}$ } & \scalebox{1}{$G_{40}$, $G_{4b}$ } & \scalebox{1}{$\left[\begin{array}{c|c|c} M_{x}^{z}, M_{z}^{x}, M_{k}^{k} & M_{x}^{y}, M_{y}^{x}, M_{y}^{z}, M_{z}^{y} & M_{x}^{x}, M_{x}^{z}, M_{y}^{y}, M_{z}^{x} \\ \hline & M_{x}^{z}, M_{z}^{x}, M_{k}^{k} & M_{x}^{y}, M_{y}^{x}, M_{y}^{z}, M_{z}^{y} \\ \hline & & M_{k}^{k} \\\end{array}\right]$} & \scalebox{1}{$\left[\begin{array}{c|c|c}- & M_{x}^{y}, M_{y}^{x} & M_{x}^{x}, M_{x}^{z}, M_{y}^{y}, M_{z}^{x} \\ \hline & - & M_{x}^{y}, M_{y}^{x}, M_{y}^{z}, M_{z}^{y} \\ \hline & & -\\\end{array}\right]$} \\
$3m$ & & \scalebox{1}{$Q_{z}$ } & & \scalebox{1}{$Q_{3a}$, $Q^{\alpha}_{z}$ } & \scalebox{1}{$G_{4a}$ } & \scalebox{1}{$\left[\begin{array}{c|c|c} M_{x}^{y}, M_{y}^{x}, M_{y}^{z}, M_{z}^{y} & M_{x}^{x}, M_{x}^{z}, M_{y}^{y}, M_{z}^{x} & M_{x}^{y}, M_{y}^{x}, M_{y}^{z}, M_{z}^{y} \\ \hline & M_{x}^{y}, M_{y}^{x}, M_{y}^{z}, M_{z}^{y} & M_{x}^{x}, M_{x}^{z}, M_{y}^{y}, M_{z}^{x} \\ \hline & & M_{x}^{y}, M_{y}^{x} \\\end{array}\right]$} & \scalebox{1}{$\left[\begin{array}{c|c|c}- & M_{k}^{k} & M_{x}^{y}, M_{y}^{x}, M_{y}^{z}, M_{z}^{y} \\ \hline & - & M_{x}^{x}, M_{x}^{z}, M_{y}^{y}, M_{z}^{x} \\ \hline & & -\\\end{array}\right]$} \\
$3$ & \scalebox{1}{$G_{0}$ } & \scalebox{1}{$Q_{z}$ } & \scalebox{1}{$G_{u}$ } & \scalebox{1}{$Q_{3a}$, $Q_{3b}$, $Q^{\alpha}_{z}$ } & \scalebox{1}{$G_{40}$, $G_{4a}$, $G_{4b}$ } & \scalebox{1}{$\left[\begin{array}{c|c|c} \mathrm{all} & M_{x}^{x}, M_{x}^{y}, M_{x}^{z}, M_{y}^{x}, M_{y}^{y}, M_{y}^{z}, M_{z}^{x}, M_{z}^{y} & M_{x}^{x}, M_{x}^{y}, M_{x}^{z}, M_{y}^{x}, M_{y}^{y}, M_{y}^{z}, M_{z}^{x}, M_{z}^{y} \\ \hline & \mathrm{all} & M_{x}^{x}, M_{x}^{y}, M_{x}^{z}, M_{y}^{x}, M_{y}^{y}, M_{y}^{z}, M_{z}^{x}, M_{z}^{y} \\ \hline & & M_{x}^{y}, M_{y}^{x}, M_{k}^{k} \\\end{array}\right]$} & \scalebox{1}{$\left[\begin{array}{c|c|c}- & M_{x}^{y}, M_{y}^{x}, M_{k}^{k} & M_{x}^{x}, M_{x}^{y}, M_{x}^{z}, M_{y}^{x}, M_{y}^{y}, M_{y}^{z}, M_{z}^{x}, M_{z}^{y} \\ \hline & - & M_{x}^{x}, M_{x}^{y}, M_{x}^{z}, M_{y}^{x}, M_{y}^{y}, M_{y}^{z}, M_{z}^{x}, M_{z}^{y} \\ \hline & & -\\\end{array}\right]$} \\ \hline \hline
\end{tabular}
}
\label{tab:symmetry allowed MQ current under D6h}
\end{table*}

Several general features emerge from the classification. 
First, all MQ conductivities vanish in centrosymmetric crystals, reflecting the odd spatial-inversion parity of the rank-4 axial tensor $\sigma_{ij}^{n\alpha(\mathrm O/\mathrm H)}$. 
Thus, finite MQ-current generation is an intrinsic transport property of parity-broken systems. 
In this respect, MQ transport is closely related to other cross-correlated responses in noncentrosymmetric systems, such as the Edelstein effects \cite{EDELSTEIN1990233} and nonlinaer Hall transport \cite{PhysRevLett.115.216806,du2021nonlinear}.

Second, the allowed MQ conductivities are directly tied to symmetry-allowed odd-parity multipoles. 
Once inversion symmetry is broken, the activated odd-parity multipoles determine which MQ-current components can appear. 
For example, the symmetry reduction from $mmm$ to $mm2$ activates the odd-parity 
electric dipole $Q_{z}$, 
electric toroidal quadrupole $G_{xy}$,
electric octupole $Q^{\alpha}_{z}$, $Q^{\beta}_{z}$, and 
electric toroidal hexadecapole $G^{\alpha}_{4z}$, $G^{\beta}_{4z}$,
which is forbidden in the centrosymmetric phase.
Simultaneously, finite MQ conductivity components become symmetry-allowed, indicating that MQ transport can emerge directly from parity-breaking multipolar degrees of freedom.
Similarly, the reduction from $6/mmm$ to $622$ activates odd-parity
electric toroidal monopole $G_{0}$, 
electric toroidal quadrupole $G_{u}$, and 
electric toroidal hexadecapole $G_{40}$, 
leading to a distinct set of MQ conductivity components. 
These examples demonstrate that different odd-parity multipoles leave characteristic fingerprints in MQ transport responses, establishing a one-to-one correspondence between parity-breaking multipolar order parameters and MQ transport phenomena.

The classification also reveals a clear distinction between the Ohmic and Hall MQ conductivities. 
As shown in Sec.~\ref{sec:Correspondence to multipole}, the Ohmic sector originates from the rank-$0$ and rank-$2$ parts of the transport tensor, whereas the Hall sector originates from the rank-$1$ part. 
Consequently, the two responses generally obey different symmetry constraints and need not coexist within the same point group. 
For example, only the Hall part $\sigma_{xy}^{zz\mathrm{(H)}}$ is symmetry allowed, whereas the Ohmic part $\sigma_{xy}^{zz\mathrm{(O)}}$ is not under the CPG $4mm$.
From a physical viewpoint, this distinction reflects the different multipolar origins of symmetric and antisymmetric MQ transport.

Finally, the number of independent MQ conductivity components increases as the crystal symmetry is lowered. 
High-symmetry noncentrosymmetric point groups strongly restrict the possible MQ-current components, whereas lower-symmetry groups allow a richer set of Ohmic and Hall responses. 
This observation provides a symmetry-based guideline for engineering MQ transport: one can identify candidate systems by searching for symmetry-lowering distortions or antiferromagnetic orders that activate the relevant odd-parity multipoles.

The classification tables therefore provide more than a catalog of nonzero tensor components. 
They establish a direct correspondence among crystal symmetry, multipolar order parameters, and MQ transport responses, offering practical guidelines for identifying candidate materials and constructing minimal microscopic models. 
In the following section, we demonstrate this connection explicitly by introducing a model in which the symmetry is lowered from $mmm$ to $mm2$, leading to finite MQ currents and MQ accumulation.
\section{\label{sec:Microscopic Modeling of Magnetic Quadrupole Current}Microscopic Modeling of Magnetic Quadrupole Current}
In Sec.~\ref{sec:Classification under 32 point groups}, we established a symmetry-based correspondence between odd-parity multipoles and MQ transport responses.
The classification revealed that finite MQ conductivities emerge only when the corresponding odd-parity multipoles become active.
To demonstrate this connection explicitly, we construct a minimal microscopic model that realizes the symmetry reduction from the centrosymmetric point group $mmm$ to the noncentrosymmetric point group $mm2$.

Among the odd-parity multipoles discussed in Sec.~\ref{sec:Classification under 32 point groups}, we focus on the electric dipole $Q_{z}$, which belongs to the $\mathrm{B}_{1u}$ irreducible representation under $mmm$.
When inversion symmetry is broken, $Q_{z}$ acquires a finite expectation value and enters the totally symmetric representation of $mm2$.
According to the symmetry analysis, this activation of $Q_{z}$ allows finite MQ Hall conductivity components that are forbidden in the centrosymmetric phase.
In the following, we verify this prediction by directly calculating the MQ Hall conductivity tensor within a microscopic tight-binding model.
The model furthermore serves as the basis for the MQ accumulation analysis presented in Sec.~\ref{sec:Edge Accumulation of Magnetic Quadrupoles Induced by MQ Hall Currents}.

\subsection{\label{sec:Tight-binding model}Tight-binding model}
\begin{figure}[htbp]
  \centering
  \includegraphics[width=0.8\columnwidth]{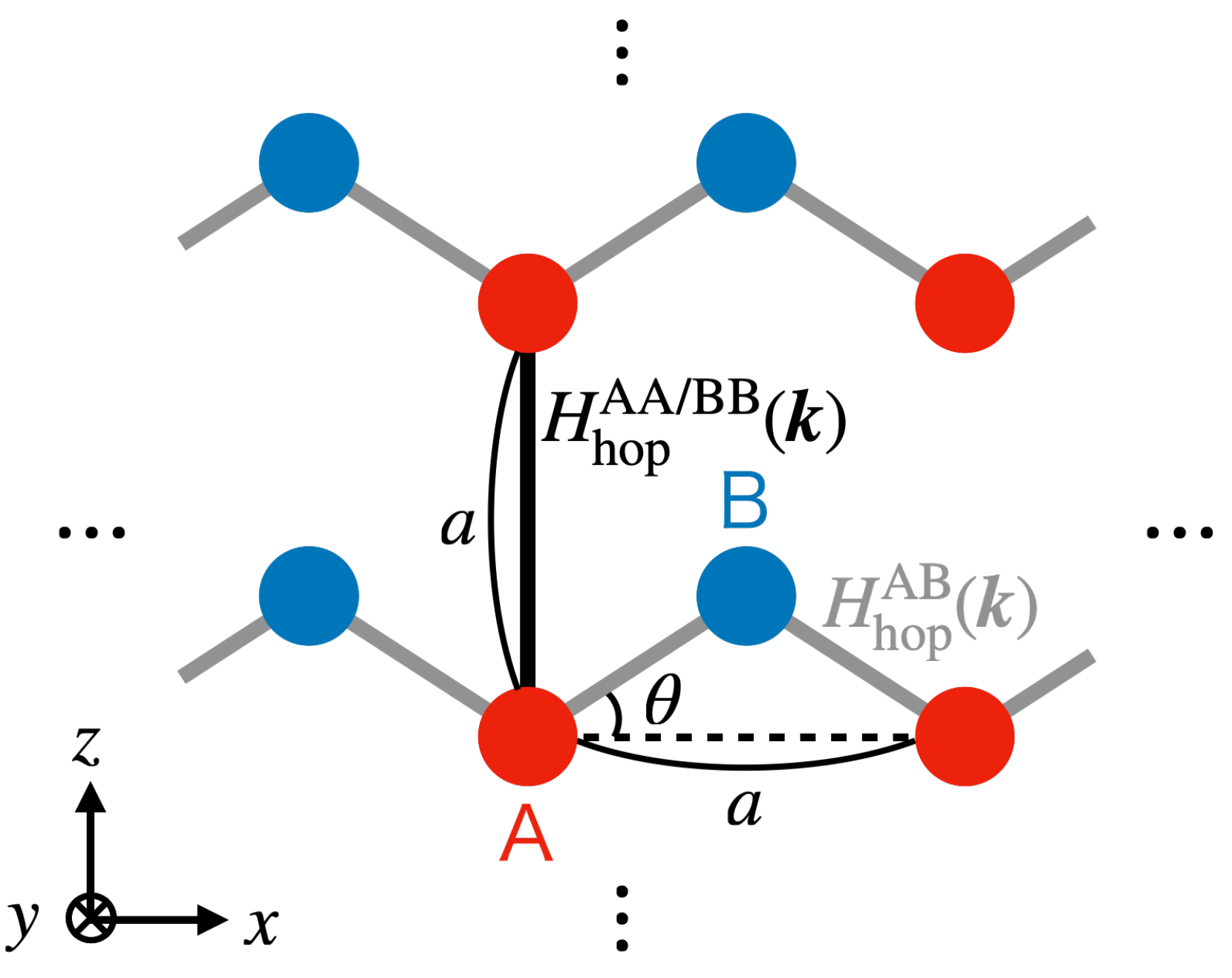}
  \caption{
    Crystal structure of the two-dimensional coupled zigzag chains in the $xz$ plane.
    The black (gray) bonds denote the intra-sublattice hopping Hamiltonians $H_{\mathrm{hop}}^{\mathrm{AA/BB}}(\bm{k})$ [inter-sublattice hopping $H_{\mathrm{hop}}^{\mathrm{AB}}(\bm{k})$], which together preserve the centrosymmetric point-group symmetry $mmm$.
  }
  \label{fig:hopping for MQ current}
\end{figure}

To capture the essential ingredients of MQ transport within a minimal setting, we consider a two-dimensional coupled zigzag-chain structure shown in Fig.~\ref{fig:hopping for MQ current}.
Zigzag-chain systems have been widely studied as minimal models of odd-parity electronic orders, where staggered magnetic, orbital, or multipolar order parameters spontaneously break spatial inversion symmetry~\cite{Yanase_JPSJ.83.014703,Hayami_doi:10.7566/JPSJ.84.064717,hayami2016emergent}.
Such parity breaking gives rise to various cross-correlated phenomena, including magnetoelectric effects~\cite{Hayami_PhysRevB.90.081115,cysne2021orbital,venderbos2025berry,venderbos2025topological,kanda2026revisiting} and nonreciprocal transport~\cite{Suzuki_PhysRevB.105.075201,Yatsushiro_PhysRevB.105.155157}.
The local inversion asymmetry of the zigzag geometry further induces sublattice-dependent electronic structures and antisymmetric spin--orbit coupling, making it an ideal platform for investigating MQ transport associated with odd-parity multipoles.

We consider one $s$ orbital and one $p_z$ orbital at each site.
This minimal basis supports the odd-parity electric dipole $Q_{z}$ through atomic $s$--$p_z$ hybridization and therefore provides the simplest microscopic model for studying MQ transport induced by parity-breaking multipoles.
The Hamiltonian is given by
\begin{equation}
  \mathcal{H} = \sum_{\bm{k}} \sum_{\rho, \rho'} \sum_{\tau, \tau'} \sum_{\sigma, \sigma'} \braket{\rho\tau\sigma | H(\bm{k}) | \rho'\tau'\sigma'} c_{\bm{k} \rho \tau \sigma}^{\dagger} c_{\bm{k} \rho' \tau' \sigma'},
  \label{eq:Hamiltonian}
\end{equation}
where $c_{\bm{k} \rho \tau \sigma}^{\dagger}$ and $c_{\bm{k} \rho \tau \sigma}$ creates (annihilates) an electron with wave vector $\bm{k} = (k_{x}, k_{z})$, sublattice index $\rho = \mathrm{A}, \mathrm{B}$, orbital index $\tau = s, p_{z}$, and spin index $\sigma = \uparrow, \downarrow$.
The Hamiltonian matrix $H(\bm{k})$ consists of four parts:
\begin{equation}
  H(\bm{k}) = H_{\mathrm{hop}}(\bm{k}) + H_{\mathrm{ASOC}}(\bm{k}) + \Delta_{sp} + q Q_{z}.
\end{equation}
The first term describes electron hopping, the second term represents antisymmetric spin--orbit coupling (ASOC), the third term corresponds to the energy difference between the $s$ and $p_z$ orbitals, and the last term introduces the odd-parity electric dipole that lowers the crystal symmetry.

The hopping Hamiltonian consists of intra-sublattice and inter-sublattice contributions,
\begin{widetext}
  \begin{align}
    &H_{\mathrm{hop}}(\bm{k}) = H_{\mathrm{hop}}^{\mathrm{AA/BB}}(\bm{k}) + H_{\mathrm{hop}}^{\mathrm{AB}}(\bm{k}),\\
    &H_{\mathrm{hop}}^{\mathrm{AA/BB}}(\bm{k})
    =
    \rho_{0} \otimes \begin{pmatrix}
      2 \alpha V_{ss\sigma} \cos{\left(a k_{z} \right)} & 2 i \alpha V_{sp\sigma} \sin{\left(a k_{z} \right)} \\
      - 2 i \alpha V_{sp\sigma} \sin{\left(a k_{z} \right)} & 2 \alpha V_{pp\sigma} \cos{\left(a k_{z} \right)}
    \end{pmatrix} \otimes \sigma_{0}, \\
    &H_{\mathrm{hop}}^{\mathrm{AB}}(\bm{k}) = \left(\rho_{x} + i \rho_{y}\right) \otimes
    \begin{pmatrix}V_{ss\sigma} e^{- \frac{i a k_{z} \tan{\theta}}{2}} \cos{\left(\frac{a k_{x}}{2} \right)} & - V_{sp\sigma} e^{- \frac{i a k_{z} \tan{\theta}}{2}} \sin{\theta} \cos{\left(\frac{a k_{x}}{2} \right)}\\V_{sp\sigma} e^{- \frac{i a k_{z} \tan{\theta}}{2}} \sin{\theta} \cos{\left(\frac{a k_{x}}{2} \right)} & \left(V_{pp\sigma} \sin^{2}{\theta} + V_{pp\pi} \cos^{2}{\theta}\right) e^{- \frac{i a k_{z} \tan{\theta}}{2}} \cos{\left(\frac{a k_{x}}{2} \right)}\end{pmatrix}
    \otimes \sigma_{0} + \mathrm{h.c.},
    \label{eq:hopping/AB}
  \end{align}
\end{widetext}
which are constructed from the Slater--Koster parametrization~\cite{slater1954simplified}.
The basis is taken as $\{\ket{\mathrm{A}, s}, \ket{\mathrm{A}, p_{z}}, \ket{\mathrm{B},s}, \ket{\mathrm{B},p_{z}}\}$, and the Slater--Koster parameters $V_{ss\sigma}$ for $s$ orbitals, $V_{pp\sigma}$ and $V_{pp\pi}$ for $p_z$ orbitals, and $V_{sp\sigma}$ for the off-diagonal $s$--$p_z$ hybridization are used.
Here, $a$ is the lattice constant, $\theta$ is the angle between the bond direction and the $x$-axis, $\rho_{i}$ and $\sigma_{i}$ for $i = 0, x, y, z$ are the Pauli matrices for the sublattice and spin spaces, respectively, and $\alpha$ is a scaling factor for the relative hopping strength between A--A (B--B) and A--B bonds.

The second term, $H_{\mathrm{ASOC}}(\bm{k})$, represents the sublattice-dependent ASOC, given by
\begin{align}
  H_{\mathrm{ASOC}}(\bm{k}) = \rho_{z} \otimes \begin{pmatrix}
    -2 \alpha_{s} \sin{\left(k_{x} a\right)} & 0 \\
    0 & -2\alpha_{p} \sin{\left(k_{x} a\right)} \\
  \end{pmatrix}
  \otimes \sigma_{y},
  \label{eq:ASOC}
\end{align}
where $\alpha_{s}$ and $\alpha_{p}$ denote the ASOC strengths for the $s$ and $p_{z}$ orbitals, respectively.
The staggered factor $\rho_z$ reflects the local inversion asymmetry of the zigzag chain and gives rise to opposite ASOC fields on the two sublattices.
The third term represents the energy splitting between the $s$ and $p_{z}$ orbitals:
\begin{equation}
  \Delta_{sp} = \rho_{0} \otimes \begin{pmatrix}
    0 & 0\\
    0 & \Delta\\
  \end{pmatrix}
  \otimes \sigma_{0},
  \label{eq:energy gap}
\end{equation}
where $\Delta_{sp}$ denotes the energy difference between the $s$ and $p_z$ orbitals.
These three terms preserve the centrosymmetric point-group symmetry $mmm$ and therefore cannot generate MQ conductivities.

To describe the symmetry reduction from $mmm$ to $mm2$ and induce MQ transport, we introduce an additional on-site $s$--$p_z$ hybridization term, as the fourth term in the Hamiltonian:
\begin{equation}
  q Q_{z}
  = \rho_{0} \otimes \begin{pmatrix}
    0 & q \\
    q & 0 \\
  \end{pmatrix}
  \otimes \sigma_{0},
  \label{eq:Qz}
\end{equation}
which represents the odd-parity electric dipole $Q_{z}$.
The $Q_{z}$ term mixes orbitals with opposite spatial parity and activates an odd-parity electric dipole degree of freedom. 
Since $Q_{z}$ belongs to the $\mathrm{B}_{1u}$ irreducible representation of $mmm$, a finite $q$ breaks inversion symmetry and lowers the crystal symmetry to $mm2$.

\subsection{Kubo formalism and symmetry properties of MQ conductivity}\label{sec:Kubo formalism and symmetry properties of MQ conductivity}

We evaluate the MQ conductivity tensor $\sigma_{i; j}^{n\alpha}$ within the framework of linear-response theory.
The response tensor of the operator $O$ induced by the electric fields is calculated using the Kubo formula,
\begin{align}
  \sigma_{j}(O)
  =
  -i\frac{e\hbar}{V} \sum_{\xi_{1} \xi_{2}}
  \frac{f(\epsilon_{\xi_{1}}) - f(\epsilon_{\xi_{2}})}{\epsilon_{\xi_{1}} - \epsilon_{\xi_{2}}} 
  \frac{\langle \xi_{1} | O | \xi_{2} \rangle \langle \xi_{2} | v_{j} | \xi_{1} \rangle}{\epsilon_{\xi_{1}} - \epsilon_{\xi_{2}} + i \delta},
  \label{eq:Kubo formula}
\end{align}
and we evaluate the MQ conductivity tensor as
\begin{align}
  \sigma_{i; j}^{n\alpha} &= \sigma_{j}(J_{i}^{n\alpha}),
\end{align}
where $V$ is the system volume and $\delta$ is a phenomenological broadening factor.
The index $\xi_{i}$ represents the combined band and momentum indices while $\epsilon_{\xi_{i}}$ and $\ket{\xi_{i}}$ denote the corresponding eigenenergy and eigenvector, respectively.
The Fermi--Dirac distribution function is given by
\begin{equation}
  f(\epsilon_{\xi_{i}}) = \frac{1}{e^{\left(\epsilon_{\xi_{i}} - \mu\right)/k_{\mathrm{B}} T} + 1},
\end{equation}
where $\mu$, $k_{\mathrm{B}}$, and $T$ are the chemical potential, the Boltzmann constant, and the temperature, respectively.
The velocity operator $v_{i}$ is defined as
\begin{equation}
  v_{i} = \frac{1}{\hbar} \frac{\partial H(\bm{k})}{\partial k_{i}}.
\end{equation}
Note that $\sigma_{j}(O)$ is a real quantity.

For later discussion, it is useful to decompose Eq.~(\ref{eq:Kubo formula}) into dissipative and dissipationless contributions.
The dissipative contribution is given by
\begin{align}
  \nonumber
  \sigma_{j}^{(\mathrm{J})}(O)
  =
  -\frac{e\hbar\delta}{V} \sum_{\xi_{1} \xi_{2}}
  \frac{f(\epsilon_{\xi_{1}}) - f(\epsilon_{\xi_{2}})}{\epsilon_{\xi_{1}} - \epsilon_{\xi_{2}}}
  \frac{\langle \xi_{1} | O | \xi_{2} \rangle \langle \xi_{2} | v_{j} | \xi_{1} \rangle}{\left(\epsilon_{\xi_{1}} - \epsilon_{\xi_{2}}\right)^{2} + \delta^{2}},
\end{align}
and the dissipationless contribution is given by
\begin{align}
  \nonumber
  \sigma_{j}^{(\mathrm{E})}(O)
  =
  -i\frac{e\hbar}{V} \sum_{\xi_{1} \xi_{2}}^{\epsilon_{\xi_{1}} \neq \epsilon_{\xi_{2}}}
  \frac{f(\epsilon_{\xi_{1}}) - f(\epsilon_{\xi_{2}})}{\left(\epsilon_{\xi_{1}} - \epsilon_{\xi_{2}}\right)^{2} + \delta^{2}}
  \langle \xi_{1} | O | \xi_{2} \rangle \langle \xi_{2} | v_{j} | \xi_{1} \rangle,
\end{align}
which correspond to the intraband and interband contributions, respectively.
Their behaviors under $\mathcal{T}$ symmetry are opposite:
\begin{align}
  \mathcal{T} \left(\sigma^{(\mathrm{J})}(O)\right) &= t_{O} t_{v_{j}} \sigma^{(\mathrm{J})}(O), \\
  \mathcal{T} \left(\sigma^{(\mathrm{E})}(O)\right) &= -t_{O} t_{v_{j}} \sigma^{(\mathrm{E})}(O),
\end{align}
where $t_{O}$ denotes the $\mathcal{T}$ parity of the operator $O$, defined by $\mathcal{T}(O) = t_{O} O$.

This distinction is particularly important for MQ transport.
Since the MQ-current operator $J_i^{n\alpha}\propto r_nS_\alpha v_i$ is even under time reversal, the dissipative and dissipationless MQ conductivities obey different symmetry constraints.
In the presence of $\mathcal T$ symmetry, $\sigma^{(\mathrm J)}(O)$ [$\sigma^{(\mathrm E)}(O)$] vanishes when $t_O t_{v_j}=-1$ ($+1$). 
This means that electric-type multipoles with $\mathcal{T}$ even contribute to $\sigma^{(\mathrm{E})}(O)$ $[\sigma^{(\mathrm{J})}(O)]$, while magnetic-type multipoles with $\mathcal{T}$ odd contribute to $\sigma^{(\mathrm{J})}(O)$ $[\sigma^{(\mathrm{E})}(O)]$ if the $\mathcal{T}$ parity of the operator $O$ is even (odd).
Consequently, the allowed MQ conductivity components depend not only on crystal symmetry but also on the $\mathcal{T}$ character of the response function.
This provides an additional symmetry criterion for identifying MQ transport phenomena in realistic materials.

\subsection{Numerical results of MQ Hall transport\label{sec:Numerical results of MQ Hall transport}}
We now numerically demonstrate the emergence of MQ Hall transport induced by the symmetry-lowering parameter $q$.
In the $mm2$ phase, the symmetry-allowed components are listed in the form of Eq.~(\ref{eq:nonzero MQ conductivity}) as shown in Table.~\ref{tab:symmetry allowed MQ current under Oh}:
\begin{align}
  \left[\begin{array}{c|c|c}- & \bm{M}_{k}^{k} & M_{y}^{z}, M_{z}^{y} \\ \hline & - & M_{x}^{z}, M_{z}^{x} \\ \hline & & -\\\end{array}\right].
\end{align}
Among them, we focus on the MQ Hall conductivity $\sigma_{zx}^{zy(\mathrm H)}$, which describes the generation of an MQ current carrying the $zy$ component in response to an electric field in the $x$ direction.
To avoid the gauge ambiguity associated with the position operator $r_n$ in periodic systems, we adopt the atomic-limit MQ operator
\begin{equation}
  \tilde{M}_{z}^{y}
  \equiv
  Q_{z}\otimes\sigma_y.
\end{equation}
Using this operator, we evaluate
\begin{align}
  \sigma_{zx}^{zy(\mathrm{H/E})} = \frac{\sigma_{x}(J_{z}^{n\alpha} = \{\tilde{M}_{z}^{y}, v_{z}\}_{+}) - \sigma_{z}(J_{x}^{n\alpha} = \{\tilde{M}_{z}^{y}, v_{x}\}_{+})}{2},
\end{align}
where only the dissipationless component survives in the presence of $\mathcal{T}$ symmetry.
Here, $\tilde{M}_{z}^{y}$ contains the magnetic quadrupole component $M_{yz}\propto r_yS_z+r_zS_y$ and the 
magnetic-toroidal dipole component $T_{x} \propto r_{y} S_{z} - r_{z} S_{y}$, whose expectation values are equivalent in the present atomic-limit representation.
In the following numerical calculations, we set $e = \hbar = k_{\mathrm{B}} = a = 1$ and use the parameter set listed in Table~\ref{tab:parameter}.
\begin{table}[htbp]
  \centering
  \caption{
    Parameter values used in Figs.~\ref{fig:q dependence of MQ Hall conductivity}, \ref{fig:M_z^y under mmm}, \ref{fig:M_z^y under mm2}, and \ref{fig:S_y under mm2}.
  }
  \begin{tabular}{cccc} \hline\hline
    Parameter & Value & Parameter & Value \\ \hline
    $V_{ss\sigma}$ & $-1$ & $\alpha$ & $0.5$ \\
    $V_{sp\sigma}$ & $-0.5$ & $\Delta_{sp}$ & $20$ \\
    $V_{pp\sigma}$ & $-0.6$ & $\theta$ & $\pi/4$ \\
    $V_{pp\pi}$ & $0.3$ & $T$ & $10^{-3}$ \\
    $\alpha_{s}$ & $1$ & $\mu$ & $0$ \\
    $\alpha_{p}$ & $0.5$ & $\delta$ & $10^{-2}$ \\ \hline\hline
  \end{tabular}
  \label{tab:parameter}
\end{table}

\begin{figure}[htbp]
  \centering
  \includegraphics[width=0.7\columnwidth]{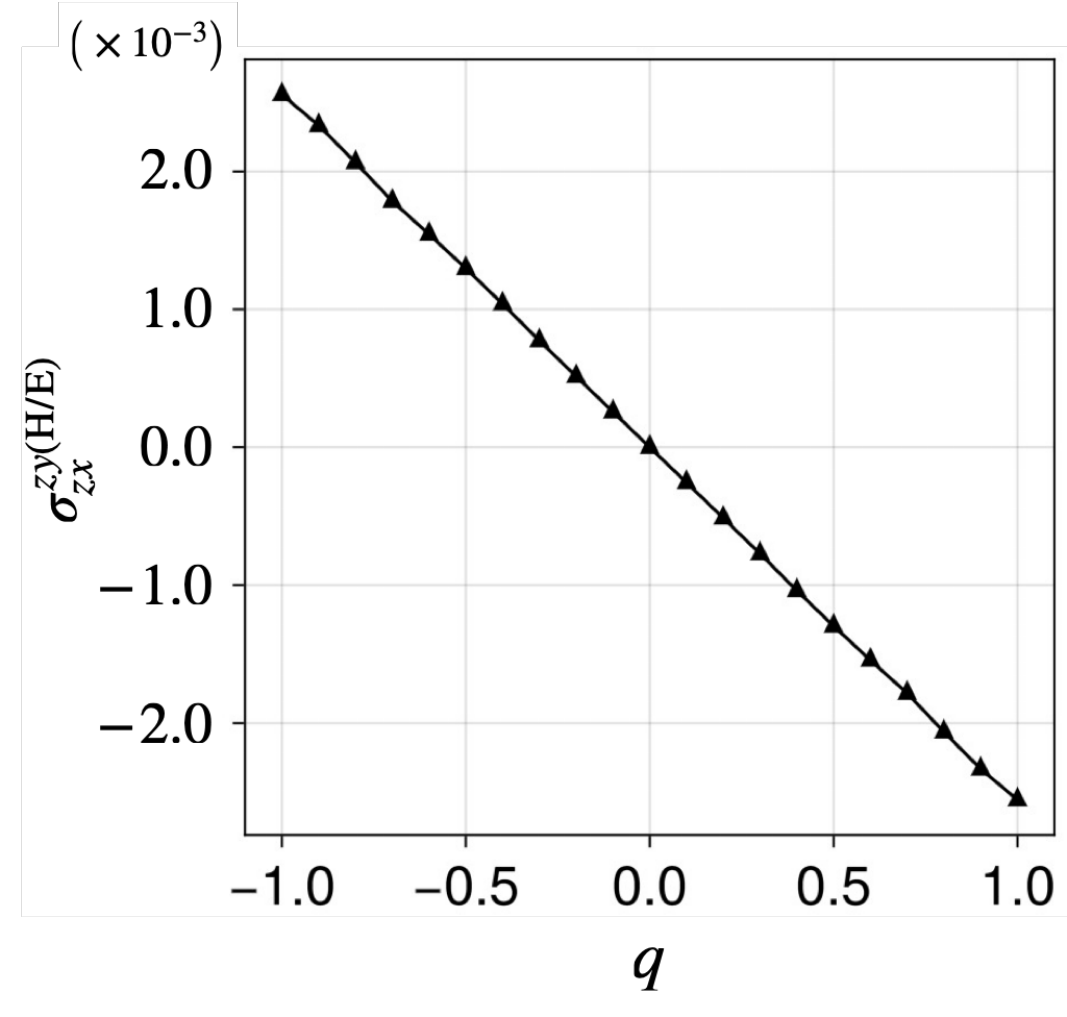}
  \caption{
    Dependence of the dissipationless MQ Hall conductivity $\sigma_{zx}^{zy(\mathrm{H/E})}$ on the symmetry-lowering parameter $q$.
    The finite response for $q\neq0$ demonstrates MQ-current generation induced by the symmetry reduction from $mmm$ to $mm2$.
    We set $V=2^{16}$.
  }
  \label{fig:q dependence of MQ Hall conductivity}
\end{figure}

Figure~\ref{fig:q dependence of MQ Hall conductivity} shows the $q$ dependence of the dissipationless MQ Hall conductivity $\sigma_{zx}^{zy(\mathrm{H/E})}$.
The conductivity vanishes at $q=0$, consistent with the centrosymmetric $mmm$ symmetry, where MQ conductivities are forbidden.
Once $q$ becomes finite, inversion symmetry is broken and the electric dipole $Q_{z}$ is activated, leading to a finite MQ Hall response.
This result directly verifies the symmetry prediction that MQ currents can be generated by activating an odd-parity multipole through the reduction from $mmm$ to $mm2$.

We also find that the dissipative contribution $\sigma_{zx}^{zy(\mathrm{H/J})}$ vanishes in the present model.
This behavior is consistent with the $\mathcal{T}$ selection rule discussed in Sec.~\ref{sec:Kubo formalism and symmetry properties of MQ conductivity}.
Since the MQ-current operator is even under time reversal whereas the velocity operator is odd, one obtains $t_{J_i^{n\alpha}}t_{v_j}=-1$.
Thus, in the $\mathcal T$-symmetric Hamiltonian considered here, the dissipative contribution is forbidden, while the dissipationless contribution is allowed.
The finite value of $\sigma_{zx}^{zy(\mathrm{H/E})}$ therefore represents a dissipationless MQ Hall current generated by parity breaking.
\section{\label{sec:Edge Accumulation of Magnetic Quadrupoles Induced by MQ Hall Currents}Edge Accumulation of Magnetic Quadrupoles}

In Sec.~\ref{sec:Microscopic Modeling of Magnetic Quadrupole Current}, we demonstrated that the symmetry reduction from $mmm$ to $mm2$ activates the MQ Hall conductivity $\sigma_{zx}^{zy(\mathrm{H})}$.
This result indicates that the parity-broken system supports a transverse MQ transport response to an applied electric field.
A natural question is whether the same symmetry lowering also gives rise to an observable real-space signature near sample boundaries.

In conventional spintronics, spin accumulation near sample edges has often been discussed in connection with spin Hall currents~\cite{Fert_2002,kato2004observation,nikolic2005nonequilibrium,nomura2005edge}.
However, in the presence of spin--orbit coupling, spin current is not uniquely defined, and spin accumulation cannot always be understood solely in terms of spin-current flow~\cite{PhysRevB.68.241315, PhysRevLett.96.076604}.
This observation suggests that the accumulation itself, rather than the current alone, should be treated as an experimentally relevant quantity~\cite{PhysRevB.98.174422, PhysRevB.105.L201202, PhysRevB.106.045203, gxpm-2gkq}.

Motivated by this viewpoint, we examine the spatial distribution of the induced MQ moment in a finite-size system.
We show that, once the symmetry is lowered to $mm2$, an antisymmetric MQ polarization appears near the sample edges.
This edge MQ accumulation provides a real-space signature associated with parity-broken MQ transport, in analogy with spin accumulation in spin Hall systems, while avoiding a direct identification of the accumulation solely with the flow of a uniquely defined MQ current.

\begin{figure}[htbp]
  \centering
  \includegraphics[width=0.8\columnwidth]{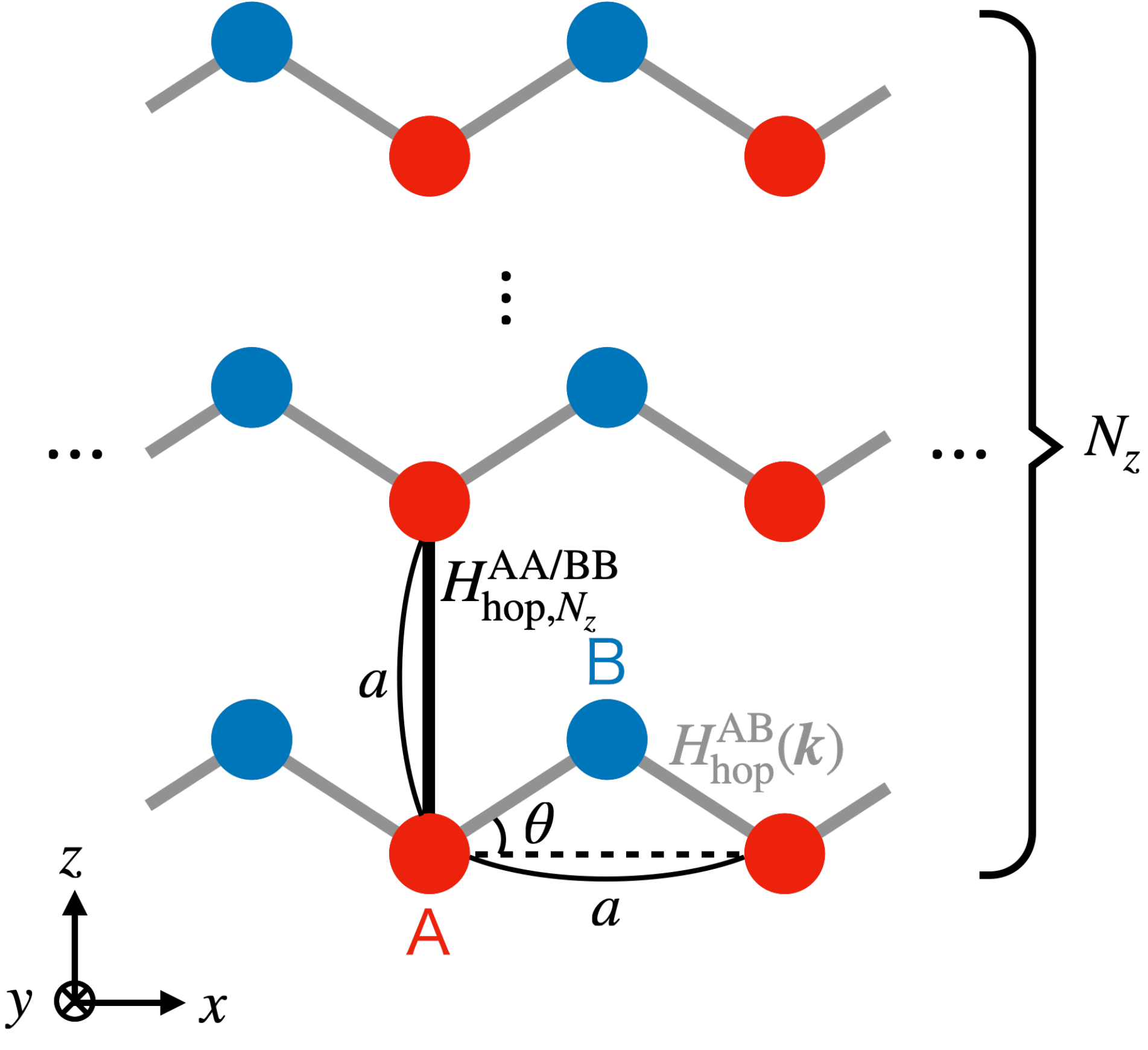}
  \caption{
    Crystal structure of two-dimensional coupled zigzag chains having the edges along $z$ direction.
    Black and gray lines represent the hopping processes described by
    $H_{\mathrm{hop},N_z}^{\mathrm{AA/BB}}$
    and
    $H_{\mathrm{hop}}^{\mathrm{AB}}(\bm{k})$,
    respectively.
    $N_z$ stands for the number of $x$-directional zigzag chains. 
  }
  \label{fig:hopping for MQ accumulation}
\end{figure}

\subsection{\label{sec:Finite-size tight-binding model}Finite-size tight-binding model}
To investigate the MQ accumulation at system boundaries, we introduce a finite-size version of the model in Eq.~(\ref{eq:Hamiltonian}).
We consider coupled zigzag chains consisting of $N_z$ layers along the $z$ direction, while retaining translational symmetry along the $x$ direction, as illustrated in Fig.~\ref{fig:hopping for MQ accumulation}.
The Hamiltonian is given by 
\begin{widetext}
  \begin{align}
    \mathcal{H} = \sum_{k_{x}} \sum_{\kappa, \kappa'} \sum_{\rho, \rho'} \sum_{\tau, \tau'} \sum_{\sigma, \sigma'} \braket{\kappa\rho\tau\sigma | H(k_{x}) | \kappa'\rho'\tau'\sigma'} c_{k_{x} \kappa \rho \tau \sigma}^{\dagger} c_{k_{x} \kappa' \rho' \tau' \sigma'},
    \label{eq:Hamiltonian with edge}
  \end{align}
\end{widetext}
where $c_{k_{x} \kappa \rho \tau \sigma}^{\dagger}$ and $c_{k_{x} \kappa \rho \tau \sigma}$ denote the fermionic creation and annihilation operators of $\kappa$th layer.

The Hamiltonian matrix is expressed as
\begin{widetext}
  \begin{align}
    H^{N_{z}}(k_{x}) = I_{N_{z}} \otimes\left[H_{\mathrm{hop}}^{\mathrm{AB}}(\bm{k}) + H_{\mathrm{ASOC}}(\bm{k}) + \Delta_{sp} + q Q_{z}\right]_{\bm{k} = (k_{x}, 0)}
    + H_{\mathrm{hop}, N_{z}}^{\mathrm{AA/BB}},
  \end{align}
\end{widetext}
where $I_{N_{z}}$ represents $N_{z}$-dimensional identity matrix.
The first four terms correspond to the bulk Hamiltonian introduced in Sec.~\ref{sec:Tight-binding model}, namely the symmetry-allowed A--B hopping, the ASOC term, the orbital-energy splitting, and the electric dipole order parameter. 
The last term describes interlayer hopping Hamiltonian between A-A (B-B) site along the $z$-axis, which is given by
\begin{align}
  \scalebox{0.9}{$\braket{\kappa, \rho\tau\sigma | H_{\mathrm{hop}, N_{z}}^{\mathrm{AA/BB}} | \kappa+1, \rho'\tau'\sigma'}
  = \rho_{0} \otimes \begin{pmatrix}
    \alpha V_{sp\sigma} & -\alpha V_{sp\sigma} \\
    \alpha V_{sp\sigma} & \alpha V_{pp\sigma}
  \end{pmatrix} \otimes \sigma_{0}$}.
\end{align}
The open boundary condition along the $z$ direction allows us to examine how MQ moments accumulate near the edges when an electric field is applied.

\subsection{\label{sec:Symmetry analysis of electric-field-induced MQ moments}Symmetry analysis of electric-field-induced magnetic quadrupole moments}
Before presenting numerical results, we discuss the symmetry properties of the induced MQ moment.
As discussed in Sec.~\ref{sec:Correspondence to multipole}, the response associated with $\tilde{M}_{z}^{y}\leftrightarrow r_z\otimes S_y$ under an external electric field $E_x$ is classified as
\begin{align}
  r_{z} \otimes S_{y} \otimes E_{x} \leftrightarrow X_{0} \oplus 2 X_{u} \oplus 2 X_{v} \oplus Y_{xyz},
\end{align}
where the coefficients of multipoles stand for the independent multipole numbers.
Since all of these multipoles belong to irreducible representations allowed in the $mmm$ point group
~\cite{hayami2018classification,watanabe2018group,yatsushiro2021multipole}, an electric-field-induced MQ moment is symmetry allowed both in the $mmm$ and $mm2$ phases.
Therefore, the existence of a finite MQ polarization itself does not necessarily indicate the presence of MQ Hall transport.

The crucial distinction lies in its spatial symmetry.
Under $mmm$, the system preserves the twofold rotation symmetry $C_{2y}$,
and thus the layer-resolved MQ moment satisfies
\begin{align}
  \tilde{M}_z^y(C_{2y}\kappa)
  =
  \tilde{M}_z^y(\kappa).
\end{align}
Consequently, any MQ polarization generated without MQ Hall transport must be spatially symmetric with respect to the sample center.
When the symmetry is reduced to $mm2$, the MQ Hall conductivity becomes finite and produces an additional contribution associated with the transverse flow of MQ moments.
This contribution is expected to appear as a $C_{2y}$-odd component of the spatial MQ distribution.
The extraction of this antisymmetric component therefore enables us to identify the MQ accumulation arising specifically from the MQ Hall current.

\subsection{\label{sec:Numerical results of MQ accumulation}Numerical results of MQ accumulation}
To characterize the spatial distribution of MQ moments, we calculate the layer-resolved conductivity
\begin{align}
  \sigma_x^{zy}(\kappa)
  \equiv
  \sigma_x[\tilde{M}_z^y(\kappa)],
\end{align}
where the operator $\tilde{M}_z^y(\kappa)$ is defined by
\begin{align}
  \braket{\kappa, \rho\tau\sigma | \tilde{M}_{z}^{y}(\kappa) | \kappa, \rho'\tau'\sigma'} = \tilde{M}_{z}^{y}.
  \label{eq:Qzy tilde}
\end{align}
As discussed in Sec.~\ref{sec:Kubo formalism and symmetry properties of MQ conductivity}, only the dissipative coutribution $\sigma_{x}^{zy(\mathrm{J})}(\kappa)$ is symmetry allowed due to the condition $t_{\tilde{M}_{z}^{y}(\kappa)} = -1$.

\begin{figure}[htbp]
  \centering
  \includegraphics[width=0.7\columnwidth]{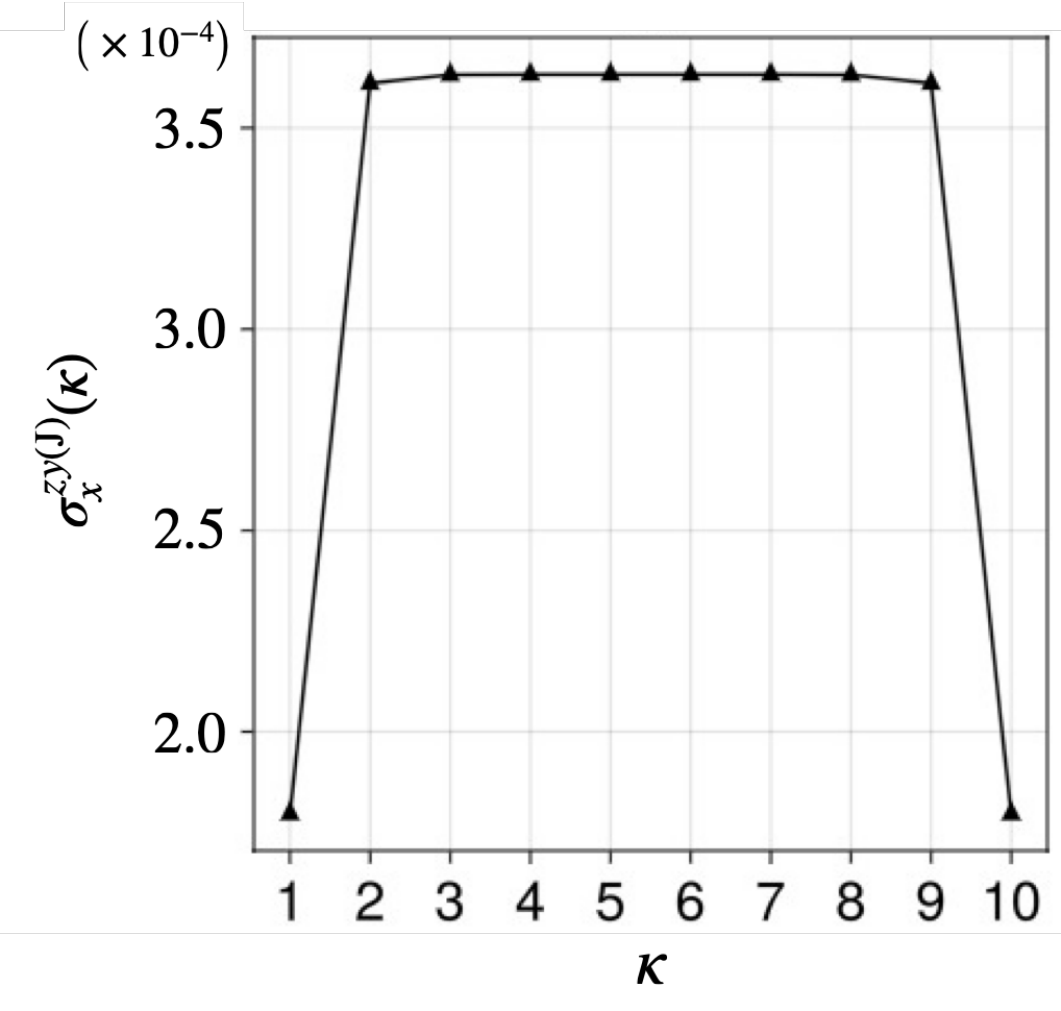}
  \caption{
    Layer dependence of $\sigma_x^{zy(\mathrm J)}(\kappa)$ for $q=0$, $N_z=10$, and $V=2^{15}N_z$.
  }
  \label{fig:M_z^y under mmm}
\end{figure}

We first consider the centrosymmetric case ($q=0$), where the system belongs to the $mmm$ point group.
Figure~\ref{fig:M_z^y under mmm} shows the spatial profile of $\sigma_x^{zy(\mathrm J)}(\kappa)$.
Although a finite MQ response is induced by the electric field, the distribution remains symmetric with respect to the sample center, satisfying
\begin{align}
  \sigma_x^{zy(\mathrm J)}(C_{2y}\kappa)
  =
  \sigma_x^{zy(\mathrm J)}(\kappa).
\end{align}
This result is consistent with the symmetry argument above and indicates that no MQ Hall transport exists in the $mmm$ phase.

\begin{figure*}[htbp]
  \centering
  \includegraphics[width=1.6\columnwidth]{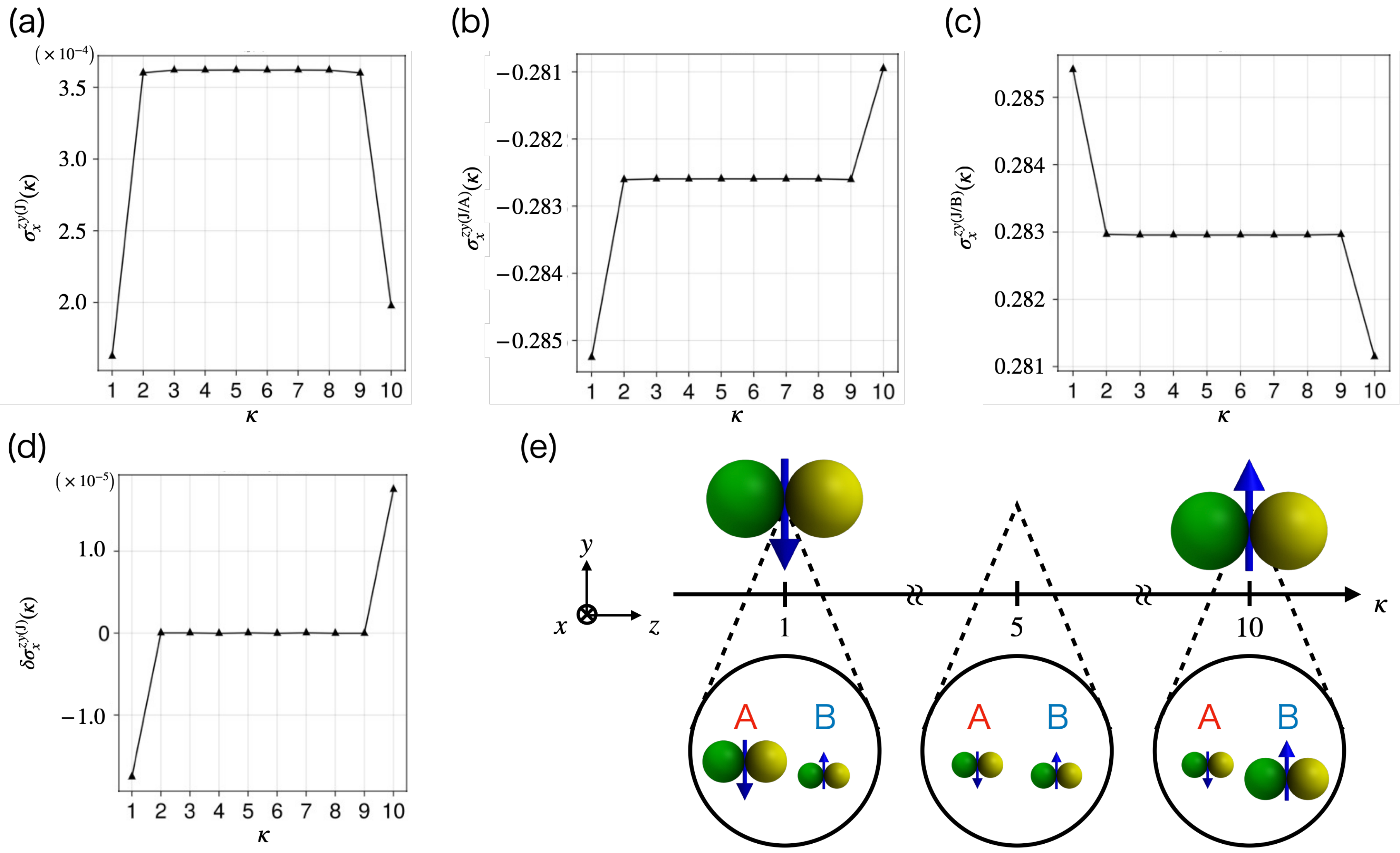}
  \caption{
    (a) Layer dependence of the MQ conductivity $\sigma_{x}^{zy(\mathrm{J})}(\kappa)$ for $q = 1$, $N_{z} = 10$, and $V = 2^{15} N_{z}$.
    (b), (c) Layer dependence of the sublattice-resolved contributions, $\sigma_{x}^{zy(\mathrm{J/A})}(\kappa)$ and $\sigma_{x}^{zy(\mathrm{J/B})}(\kappa)$, obtained from Eq.~(\ref{eq:Qzy tilde}) by restricting $\rho, \rho' = \mathrm{A}$ and $\mathrm{B}$, respectively.
    (d) $C_{2y}$-odd component of the MQ conductivity, $\left[\sigma_{x}^{zy(\mathrm{J})}(\kappa)-\sigma_{x}^{zy(\mathrm{J})}(C_{2y}\kappa)\right]\big/2$, which isolates the contribution associated with the MQ Hall current.
    (e) Schematic illustration of the antisymmetric MQ accumulation.
    The opposite signs near the two edges indicate the accumulation of MQ moments (blue arrows) carried by the MQ Hall current, while the circles represent the corresponding sublattice-resolved conductivities.
  }
  \label{fig:M_z^y under mm2}
\end{figure*}

We next turn to the symmetry-broken case ($q=1$), where the MQ Hall conductivity becomes finite.
The resulting MQ distribution is shown in Fig.~\ref{fig:M_z^y under mm2}(a). 
Unlike the $mmm$ case, the profile is no longer invariant under $C_{2y}$.
To isolate the contribution arising from MQ Hall transport, we extract the antisymmetric component
\begin{align}
  \delta \sigma_x^{zy(\mathrm J)}(\kappa)
  =
  \frac{
  \sigma_x^{zy(\mathrm J)}(\kappa)
  -
  \sigma_x^{zy(\mathrm J)}(C_{2y}\kappa)
  }{2},
\end{align}
which is shown in Fig.~\ref{fig:M_z^y under mm2}(d).

A pronounced enhancement of $\delta \sigma_x^{zy(\mathrm J)}(\kappa)$ appears near the two opposite edges with opposite signs, while it remains strongly suppressed in the bulk region.
This behavior demonstrates the emergence of antisymmetric MQ accumulation near the sample boundaries. 
The observed spatial profile accompanies the finite MQ Hall conductivity $\sigma_{zx}^{zy(\mathrm H)}$ activated by the symmetry reduction from $mmm$ to $mm2$, providing a real-space manifestation of the MQ Hall response.

Figures~\ref{fig:M_z^y under mm2}(b) and (c) further reveal that the accumulated MQ moments originate from opposite sublattice contributions.
Although the individual A- and B-sublattice components exhibit different spatial profiles, their $C_{2y}$-odd parts cooperate to produce the net edge accumulation shown in Fig.~\ref{fig:M_z^y under mm2}(d).
The resulting MQ polarization is schematically illustrated in Fig.~\ref{fig:M_z^y under mm2}(e).

The obtained edge profile is closely analogous to spin accumulation induced by the spin Hall effect.
In both phenomena, an applied electric field generates a transverse current carrying an internal degree of freedom.
The obtained antisymmetric edge profile closely resembles the spin accumulation observed in spin Hall systems, providing a real-space manifestation of MQ transport.
The present results therefore establish edge MQ accumulation as a characteristic real-space manifestation accompanying the MQ Hall response, providing an experimentally accessible signature of MQ transport.

From a symmetry viewpoint, the accumulation provides a direct consequence of the activation of the odd-parity electric dipole $Q_{z}$.
While the centrosymmetric $mmm$ phase only allows the $C_{2y}$-even MQ response, inversion-symmetry breaking generates an additional $C_{2y}$-odd component associated with the MQ Hall current.
The emergence of edge MQ accumulation can therefore be regarded as a real-space indicator of parity-breaking MQ transport.

These findings further suggest a possible route toward the experimental detection of MQ transport phenomena.
Since accumulated MQ moments correspond to spatially localized multipolar polarization near the sample boundaries, they may manifest themselves through edge-sensitive probes that couple to multipolar magnetic textures.
The MQ accumulation identified here thus provides a physically intuitive and experimentally relevant signature of MQ Hall transport.
\section{\label{sec:Summary}Summary}
In this work, we have established a comprehensive theoretical framework for MQ transport phenomena.
By deriving the multipole representation of the MQ conductivity tensor, we systematically classified the symmetry conditions for MQ transport under all 32 crystallographic point groups.
The classification reveals that MQ currents are governed by symmetry constraints distinct from those of conventional charge, spin, and orbital transport, and provides a unified symmetry-based criterion for identifying candidate materials and electronic states exhibiting MQ transport responses.

Based on this symmetry analysis, we investigated a microscopic two-dimensional coupled zigzag-chain model. 
We demonstrated that the reduction of crystal symmetry from the centrosymmetric $mmm$ phase to the noncentrosymmetric $mm2$ phase activates the MQ Hall conductivity $\sigma_{zx}^{zy(\mathrm H)}$, generating a transverse flow of magnetic quadrupoles under an applied electric field. 
Furthermore, by constructing a finite-size tight-binding model with open boundaries, we showed that the emergence of an antisymmetric accumulation of MQ moments near the sample edges in the presence of the MQ Hall response.
The obtained edge profile is closely analogous to spin accumulation in spin Hall systems and provides a real-space manifestation of MQ transport.
These results suggest that edge MQ accumulation can serve as an experimentally accessible signature of MQ transport in noncentrosymmetric systems.

The present results establish a direct connection between symmetry breaking, MQ Hall transport, and real-space MQ accumulation.
In particular, the emergence of edge MQ accumulation serves as a physically intuitive fingerprint of MQ transport that is absent in the higher-symmetry phase with spatial inversion symmetry.
This finding opens a route toward the experimental detection of MQ currents through probes sensitive to spatially inhomogeneous multipolar distributions. 
For example, the accumulated MQ moments may manifest themselves through characteristic edge-dependent signals in resonant x-ray scattering experiments~\cite{mannix2005order}, as well as in other spectroscopic and imaging techniques that couple to odd-parity multipolar magnetic textures.
Our work therefore provides a foundation for exploring odd-parity magnetic multipole transport phenomena and their experimental realization in materials.

\begin{acknowledgments}
  This research was supported by JSPS KAKENHI (Grant Nos. JP22H00101 and JP23H04869), JST CREST (Grant No.~JPMJCR23O4), and JST FOREST (Grant No.~JPMJFR2366).
\end{acknowledgments}

\appendix
\begin{widetext}
  \section{\label{app:Angle dependence of multipole}Angle dependence of multipole}
In this appendix, we present the angular dependence of $c$- and $t$-multipoles in Table~\ref{tab:angle dependences of multipole}. 
These angular functions constitute the basis of the multipole representation used throughout this work. 
It should be noted that, up to rank two, $t$-multipoles share the same angular dependence as the corresponding $c$-multipoles.
\begin{table*}[htbp]
  \caption{
    Angular basis functions of $c$- and $t$-multipoles up to rank 4.
    The multipoles are normalized as $\int_{S} d\Omega Z_{i}^{*} Z_{j} = \delta_{ij}$, where $S$ is the unit sphere, $\Omega$ is the solid angle, and $\delta_{ij}$ is the Kronecker delta.
    Here, $r^{2} = x^{2} + y^{2} + z^{2}$, while ``$\mathrm{cyclic}$'' denotes the cyclic permutation $x \to y \to z \to x$.
  }
  \scalebox{0.95}{\begin{minipage}{0.5\linewidth}
    \centering
    \begin{tabular}{ccc} \hline \hline
      rank & \emph{c}-multipole & angle dependence \\ \hline
      $0$ & $Z_{0}$ & $\frac{1}{2 \sqrt{\pi}}$ \rule[-8pt]{0pt}{20pt} \\ \hline
      $1$ & $Z_{x}, Z_{y}, Z_{z}$ & $\frac{\sqrt{3} x}{2 \sqrt{\pi} r}, \frac{\sqrt{3} y}{2 \sqrt{\pi} r}, \frac{\sqrt{3} z}{2 \sqrt{\pi} r}$ \rule[-8pt]{0pt}{20pt} \\ \hline
      $2$ & $Z_{u}$ & $\frac{\sqrt{5} \left(3 z^{2} - r^{2}\right)}{4 \sqrt{\pi} r^{2}}$ \rule[-8pt]{0pt}{20pt} \\
      & $Z_{v}$ & $\frac{\sqrt{15} \left(x^{2} - y^{2}\right)}{4 \sqrt{\pi} r^{2}}$ \rule[-8pt]{0pt}{20pt} \\
      & $Z_{yz}, Z_{zx}, Z_{xy}$ & $\frac{\sqrt{15} y z}{2 \sqrt{\pi} r^{2}}, \text{cyclic}$ \rule[-8pt]{0pt}{20pt} \\ \hline
      $3$ & $Z_{xyz}$ & $\frac{\sqrt{105} x y z}{2 \sqrt{\pi} r^{3}}$ \rule[-8pt]{0pt}{20pt} \\
      & $Z_{x}^{\alpha}, Z_{y}^{\alpha}, Z_{z}^{\alpha}$ & $\frac{\sqrt{7} x \left(5 x^{2} - 3 r^{2}\right)}{4 \sqrt{\pi} r^{3}}, \text{cyclic}$ \rule[-8pt]{0pt}{20pt} \\
      & $Z_{x}^{\beta}, Z_{y}^{\beta}, Z_{z}^{\beta}$ & $\frac{\sqrt{105} x \left(y^{2} - z^{2}\right)}{4 \sqrt{\pi} r^{3}}, \text{cyclic}$ \rule[-8pt]{0pt}{20pt} \\ \hline
      $4$ & $Z_{4}$ & $\frac{\sqrt{21} \left(x^{4} - 3 x^{2} y^{2} - 3 x^{2} z^{2} + y^{4} - 3 y^{2} z^{2} + z^{4}\right)}{4 \sqrt{\pi} r^{4}}$ \rule[-8pt]{0pt}{20pt} \\
      & $Z_{4u}$ & $\frac{\sqrt{15} \left(x^{4} - 12 x^{2} y^{2} + 6 x^{2} z^{2} + y^{4} + 6 y^{2} z^{2} - 2 z^{4}\right)}{8 \sqrt{\pi} r^{4}}$ \rule[-8pt]{0pt}{20pt} \\
      & $Z_{4v}$ & $- \frac{3 \sqrt{5} \left(x - y\right) \left(x + y\right) \left(x^{2} + y^{2} - 6 z^{2}\right)}{8 \sqrt{\pi} r^{4}}$ \rule[-8pt]{0pt}{20pt} \\
      & $Z_{4x}^{\alpha}, Z_{4y}^{\alpha}, Z_{4z}^{\alpha}$ & $\frac{3 \sqrt{35} y z \left(y^{2} - z^{2}\right)}{4 \sqrt{\pi} r^{4}}, \text{cyclic}$ \rule[-8pt]{0pt}{20pt} \\
      & $Z_{4x}^{\beta}, Z_{4y}^{\beta}, Z_{4z}^{\beta}$ & $\frac{3 \sqrt{5} y z \left(7 x^{2} - r^{2}\right)}{4 \sqrt{\pi} r^{4}}, \text{cyclic}$ \rule[-8pt]{0pt}{20pt} \\ \hline \hline
    \end{tabular}
  \end{minipage}}
  \scalebox{0.95}{\begin{minipage}{0.5\linewidth}
    \begin{tabular}{ccc} \hline \hline
      rank & \emph{t}-multipole & angle dependence \\ \hline
      $3$ & $Z_{z}^{\alpha}$ & $\frac{\sqrt{7} z \left(- 3 r^{2} + 5 z^{2}\right)}{4 \sqrt{\pi} r^{3}}$ \rule[-8pt]{0pt}{20pt} \\
      & $Z_{3a}$ & $\frac{\sqrt{70} x \left(x^{2} - 3 y^{2}\right)}{8 \sqrt{\pi} r^{3}}$ \rule[-8pt]{0pt}{20pt} \\
      & $Z_{3b}$ & $\frac{\sqrt{70} y \left(3 x^{2} - y^{2}\right)}{8 \sqrt{\pi} r^{3}}$ \rule[-8pt]{0pt}{20pt} \\
      & $Z_{3u}, Z_{3v}$ & $\frac{\sqrt{42} x \left(- r^{2} + 5 z^{2}\right)}{8 \sqrt{\pi} r^{3}}, \frac{\sqrt{42} y \left(- r^{2} + 5 z^{2}\right)}{8 \sqrt{\pi} r^{3}}$ \rule[-8pt]{0pt}{20pt} \\
      & $Z_{z}^{\beta}, Z_{xyz}$ & $\frac{\sqrt{105} z \left(x^{2} - y^{2}\right)}{4 \sqrt{\pi} r^{3}}, \frac{\sqrt{105} x y z}{2 \sqrt{\pi} r^{3}}$ \rule[-8pt]{0pt}{20pt} \\ \hline
      $4$ & $Z_{40}$ & $\frac{3 \left(3 r^{4} - 30 r^{2} z^{2} + 35 z^{4}\right)}{16 \sqrt{\pi} r^{4}}$ \rule[-8pt]{0pt}{20pt} \\
      & $Z_{4a}$ & $\frac{3 \sqrt{70} y z \left(3 x^{2} - y^{2}\right)}{8 \sqrt{\pi} r^{4}}$ \rule[-8pt]{0pt}{20pt} \\
      & $Z_{4b}$ & $\frac{3 \sqrt{70} x z \left(x^{2} - 3 y^{2}\right)}{8 \sqrt{\pi} r^{4}}$ \rule[-8pt]{0pt}{20pt} \\
      & $Z_{4u}^{\alpha}, Z_{4v}^{\alpha}$ & $\frac{3 \sqrt{10} x z \left(- 3 r^{2} + 7 z^{2}\right)}{8 \sqrt{\pi} r^{4}}, \frac{3 \sqrt{10} y z \left(- 3 r^{2} + 7 z^{2}\right)}{8 \sqrt{\pi} r^{4}}$ \rule[-8pt]{0pt}{20pt} \\
      & $Z_{4u}^{\beta 1}, Z_{4v}^{\beta 1}$ & $\frac{3 \sqrt{35} \left(x^{4} - 6 x^{2} y^{2} + y^{4}\right)}{16 \sqrt{\pi} r^{4}}, \frac{3 \sqrt{35} x y \left(x^{2} - y^{2}\right)}{4 \sqrt{\pi} r^{4}}$ \rule[-8pt]{0pt}{20pt} \\
      & $Z_{4u}^{\beta 2}, Z_{4v}^{\beta 2}$ & $\frac{3 \sqrt{5} \left(x^{2} - y^{2}\right) \left(- r^{2} + 7 z^{2}\right)}{8 \sqrt{\pi} r^{4}}, \frac{3 \sqrt{5} x y \left(- r^{2} + 7 z^{2}\right)}{4 \sqrt{\pi} r^{4}}$ \rule[-8pt]{0pt}{20pt} \\ \hline \hline
    \end{tabular}
  \end{minipage}}
  \label{tab:angle dependences of multipole}
\end{table*}
  \section{\label{app:Multipole representation of MQ conductivity}Multipole representation of MQ conductivity}
In this appendix, we present the multipole representation of the MQ conductivity tensors $\sigma_{ij}^{n\alpha(\mathrm O,\mathrm H)}$ up to rank 4.
Both \emph{c}- and \emph{t}-multipole bases are included.
For compactness, we arrange the tensor components in matrix form as
\begin{align}
  \nonumber
  & \sigma_{ij}^{n\alpha(\mathrm{O/L})} = \\
  & \begin{pmatrix}
    \sigma^{xx}_{xx} & \sigma^{xy}_{xx} & \sigma^{xz}_{xx} & \sigma^{yx}_{xx} & \sigma^{yy}_{xx} & \sigma^{yz}_{xx} & \sigma^{zx}_{xx} & \sigma^{zy}_{xx} & \sigma^{zz}_{xx}\\
    \sigma^{xx}_{yy} & \sigma^{xy}_{yy} & \sigma^{xz}_{yy} & \sigma^{yx}_{yy} & \sigma^{yy}_{yy} & \sigma^{yz}_{yy} & \sigma^{zx}_{yy} & \sigma^{zy}_{yy} & \sigma^{zz}_{yy}\\
    \sigma^{xx}_{zz} & \sigma^{xy}_{zz} & \sigma^{xz}_{zz} & \sigma^{yx}_{zz} & \sigma^{yy}_{zz} & \sigma^{yz}_{zz} & \sigma^{zx}_{zz} & \sigma^{zy}_{zz} & \sigma^{zz}_{zz}\\
  \end{pmatrix}^{\top},
\end{align}
\begin{align}
  \nonumber
  & \sigma_{ij}^{n\alpha(\mathrm{O/T})} = \\
  & \begin{pmatrix}
    \frac{\sigma^{xx}_{yz} + \sigma^{xx}_{zy}}{2} & \frac{\sigma^{xy}_{yz} + \sigma^{xy}_{zy}}{2} & \frac{\sigma^{xz}_{yz} + \sigma^{xz}_{zy}}{2} & \frac{\sigma^{yx}_{yz} + \sigma^{yx}_{zy}}{2} & \frac{\sigma^{yy}_{yz} + \sigma^{yy}_{zy}}{2} & \frac{\sigma^{yz}_{yz} + \sigma^{yz}_{zy}}{2} & \frac{\sigma^{zx}_{yz} + \sigma^{zx}_{zy}}{2} & \frac{\sigma^{zy}_{yz} + \sigma^{zy}_{zy}}{2} & \frac{\sigma^{zz}_{yz} + \sigma^{zz}_{zy}}{2}\\
    \frac{\sigma^{xx}_{xz} + \sigma^{xx}_{zx}}{2} & \frac{\sigma^{xy}_{xz} + \sigma^{xy}_{zx}}{2} & \frac{\sigma^{xz}_{xz} + \sigma^{xz}_{zx}}{2} & \frac{\sigma^{yx}_{xz} + \sigma^{yx}_{zx}}{2} & \frac{\sigma^{yy}_{xz} + \sigma^{yy}_{zx}}{2} & \frac{\sigma^{yz}_{xz} + \sigma^{yz}_{zx}}{2} & \frac{\sigma^{zx}_{xz} + \sigma^{zx}_{zx}}{2} & \frac{\sigma^{zy}_{xz} + \sigma^{zy}_{zx}}{2} & \frac{\sigma^{zz}_{xz} + \sigma^{zz}_{zx}}{2}\\
    \frac{\sigma^{xx}_{xy} + \sigma^{xx}_{yx}}{2} & \frac{\sigma^{xy}_{xy} + \sigma^{xy}_{yx}}{2} & \frac{\sigma^{xz}_{xy} + \sigma^{xz}_{yx}}{2} & \frac{\sigma^{yx}_{xy} + \sigma^{yx}_{yx}}{2} & \frac{\sigma^{yy}_{xy} + \sigma^{yy}_{yx}}{2} & \frac{\sigma^{yz}_{xy} + \sigma^{yz}_{yx}}{2} & \frac{\sigma^{zx}_{xy} + \sigma^{zx}_{yx}}{2} & \frac{\sigma^{zy}_{xy} + \sigma^{zy}_{yx}}{2} & \frac{\sigma^{zz}_{xy} + \sigma^{zz}_{yx}}{2}
  \end{pmatrix}^{\top},
\end{align}
\begin{align}
  \nonumber
  & \sigma_{ij}^{n\alpha(\mathrm{H})} = \\
  & \begin{pmatrix}
    \frac{\sigma^{xx}_{yz} - \sigma^{xx}_{zy}}{2} & \frac{\sigma^{xy}_{yz} - \sigma^{xy}_{zy}}{2} & \frac{\sigma^{xz}_{yz} - \sigma^{xz}_{zy}}{2} & \frac{\sigma^{yx}_{yz} - \sigma^{yx}_{zy}}{2} & \frac{\sigma^{yy}_{yz} - \sigma^{yy}_{zy}}{2} & \frac{\sigma^{yz}_{yz} - \sigma^{yz}_{zy}}{2} & \frac{\sigma^{zx}_{yz} - \sigma^{zx}_{zy}}{2} & \frac{\sigma^{zy}_{yz} - \sigma^{zy}_{zy}}{2} & \frac{\sigma^{zz}_{yz} - \sigma^{zz}_{zy}}{2}\\
    \frac{- \sigma^{xx}_{xz} + \sigma^{xx}_{zx}}{2} & \frac{- \sigma^{xy}_{xz} + \sigma^{xy}_{zx}}{2} & \frac{- \sigma^{xz}_{xz} + \sigma^{xz}_{zx}}{2} & \frac{- \sigma^{yx}_{xz} + \sigma^{yx}_{zx}}{2} & \frac{- \sigma^{yy}_{xz} + \sigma^{yy}_{zx}}{2} & \frac{- \sigma^{yz}_{xz} + \sigma^{yz}_{zx}}{2} & \frac{- \sigma^{zx}_{xz} + \sigma^{zx}_{zx}}{2} & \frac{- \sigma^{zy}_{xz} + \sigma^{zy}_{zx}}{2} & \frac{- \sigma^{zz}_{xz} + \sigma^{zz}_{zx}}{2}\\
    \frac{\sigma^{xx}_{xy} - \sigma^{xx}_{yx}}{2} & \frac{\sigma^{xy}_{xy} - \sigma^{xy}_{yx}}{2} & \frac{\sigma^{xz}_{xy} - \sigma^{xz}_{yx}}{2} & \frac{\sigma^{yx}_{xy} - \sigma^{yx}_{yx}}{2} & \frac{\sigma^{yy}_{xy} - \sigma^{yy}_{yx}}{2} & \frac{\sigma^{yz}_{xy} - \sigma^{yz}_{yx}}{2} & \frac{\sigma^{zx}_{xy} - \sigma^{zx}_{yx}}{2} & \frac{\sigma^{zy}_{xy} - \sigma^{zy}_{yx}}{2} & \frac{\sigma^{zz}_{xy} - \sigma^{zz}_{yx}}{2}\\
  \end{pmatrix}^{\top}.
\end{align}
Here, $\sigma_{ij}^{n\alpha(\mathrm{O})}$ is decomposed into the longitudinal part $\sigma_{ij}^{n\alpha(\mathrm{O/L})}$ with $i=j$ and the transverse part $\sigma_{ij}^{n\alpha(\mathrm{O/T})}$ with $i\neq j$, whereas $\sigma_{ij}^{n\alpha(\mathrm H)}$ denotes the Hall part, which is antisymmetric with respect to the exchange of the current and field directions.
In the following expressions, we use the simplified labels $(1), (2), \cdots$ instead of $(l'l'')$ in Eq.~(\ref{eq:multipole representation of rank-4 tensor}) for notation simplicity.
We also present the normalized MQ conductivity tensors $\tilde{\sigma}_{ij}^{n\alpha(\mathrm O,\mathrm H)}$ to improve readability.
Although these normalized tensors do not follow the normalization convention in Eq.~(\ref{eq:multipole representation of rank-4 tensor}), they transform according to the same irreducible representations as the corresponding $\sigma_{ij}^{n\alpha(\mathrm O,\mathrm H)}$ components under arbitrary point groups.
Finally, the multipole representation of rank-4 polar response tensors, such as the magnetic-toroidal quadrupole conductivity, can be obtained from the present expressions by exchanging $X\leftrightarrow Y$.

\subsection{$c$-multipole}
\subsubsection{rank $l=0$}
\begin{align}
  \nonumber
  & \sigma_{ij}^{n\alpha(\mathrm{O/L})} = \\
  & \scalebox{1}{$\begin{pmatrix}Y^{(1)}_{0} + 4 Y^{(2)}_{0} & Y^{(1)}_{0} - 2 Y^{(2)}_{0} & Y^{(1)}_{0} - 2 Y^{(2)}_{0}\\0 & 0 & 0\\0 & 0 & 0\\0 & 0 & 0\\Y^{(1)}_{0} - 2 Y^{(2)}_{0} & Y^{(1)}_{0} + 4 Y^{(2)}_{0} & Y^{(1)}_{0} - 2 Y^{(2)}_{0}\\0 & 0 & 0\\0 & 0 & 0\\0 & 0 & 0\\Y^{(1)}_{0} - 2 Y^{(2)}_{0} & Y^{(1)}_{0} - 2 Y^{(2)}_{0} & Y^{(1)}_{0} + 4 Y^{(2)}_{0}\end{pmatrix}$},
\end{align}
\begin{align}
  \nonumber
  & \sigma_{ij}^{n\alpha(\mathrm{O/T})} = \\
  & \scalebox{1}{$\begin{pmatrix}0 & 0 & 0\\0 & 0 & 3 Y^{(2)}_{0}\\0 & 3 Y^{(2)}_{0} & 0\\0 & 0 & 3 Y^{(2)}_{0}\\0 & 0 & 0\\3 Y^{(2)}_{0} & 0 & 0\\0 & 3 Y^{(2)}_{0} & 0\\3 Y^{(2)}_{0} & 0 & 0\\0 & 0 & 0\end{pmatrix}$},
\end{align}
\begin{align}
  \nonumber
  & \sigma_{ij}^{n\alpha(\mathrm{H})} = \\
  & \scalebox{1}{$\begin{pmatrix}0 & 0 & 0\\0 & 0 & Y^{(1)}_{0}\\0 & - Y^{(1)}_{0} & 0\\0 & 0 & - Y^{(1)}_{0}\\0 & 0 & 0\\Y^{(1)}_{0} & 0 & 0\\0 & Y^{(1)}_{0} & 0\\- Y^{(1)}_{0} & 0 & 0\\0 & 0 & 0\end{pmatrix}$}.
\end{align}

\subsubsection{rank $l=1$}
\begin{align}
  \nonumber
  & \sigma_{ij}^{n\alpha(\mathrm{O/L})} = \\
  & \scalebox{1}{$\begin{pmatrix}0 & 0 & 0\\X^{(1)}_{z} - 2 X^{(2)}_{z} + 2 X^{(3)}_{z} & X^{(1)}_{z} - 2 X^{(2)}_{z} - 2 X^{(3)}_{z} & X^{(1)}_{z} + 4 X^{(2)}_{z}\\- X^{(1)}_{y} + 2 X^{(2)}_{y} - 2 X^{(3)}_{y} & - X^{(1)}_{y} - 4 X^{(2)}_{y} & - X^{(1)}_{y} + 2 X^{(2)}_{y} + 2 X^{(3)}_{y}\\- X^{(1)}_{z} + 2 X^{(2)}_{z} + 2 X^{(3)}_{z} & - X^{(1)}_{z} + 2 X^{(2)}_{z} - 2 X^{(3)}_{z} & - X^{(1)}_{z} - 4 X^{(2)}_{z}\\0 & 0 & 0\\X^{(1)}_{x} + 4 X^{(2)}_{x} & X^{(1)}_{x} - 2 X^{(2)}_{x} + 2 X^{(3)}_{x} & X^{(1)}_{x} - 2 X^{(2)}_{x} - 2 X^{(3)}_{x}\\X^{(1)}_{y} - 2 X^{(2)}_{y} - 2 X^{(3)}_{y} & X^{(1)}_{y} + 4 X^{(2)}_{y} & X^{(1)}_{y} - 2 X^{(2)}_{y} + 2 X^{(3)}_{y}\\- X^{(1)}_{x} - 4 X^{(2)}_{x} & - X^{(1)}_{x} + 2 X^{(2)}_{x} + 2 X^{(3)}_{x} & - X^{(1)}_{x} + 2 X^{(2)}_{x} - 2 X^{(3)}_{x}\\0 & 0 & 0\end{pmatrix}$},
\end{align}
\begin{align}
  \nonumber
  & \sigma_{ij}^{n\alpha(\mathrm{O/T})} = \\
  & \scalebox{1}{$\begin{pmatrix}0 & 2 X^{(3)}_{y} & - 2 X^{(3)}_{z}\\3 X^{(2)}_{y} + X^{(3)}_{y} & 3 X^{(2)}_{x} - X^{(3)}_{x} & 0\\- 3 X^{(2)}_{z} - X^{(3)}_{z} & 0 & - 3 X^{(2)}_{x} + X^{(3)}_{x}\\- 3 X^{(2)}_{y} + X^{(3)}_{y} & - 3 X^{(2)}_{x} - X^{(3)}_{x} & 0\\- 2 X^{(3)}_{x} & 0 & 2 X^{(3)}_{z}\\0 & 3 X^{(2)}_{z} + X^{(3)}_{z} & 3 X^{(2)}_{y} - X^{(3)}_{y}\\3 X^{(2)}_{z} - X^{(3)}_{z} & 0 & 3 X^{(2)}_{x} + X^{(3)}_{x}\\0 & - 3 X^{(2)}_{z} + X^{(3)}_{z} & - 3 X^{(2)}_{y} - X^{(3)}_{y}\\2 X^{(3)}_{x} & - 2 X^{(3)}_{y} & 0\end{pmatrix}$},
\end{align}
\begin{align}
  \nonumber
  & \sigma_{ij}^{n\alpha(\mathrm{H})} = \\
  & \scalebox{1}{$\begin{pmatrix}X^{(1)}_{x} + 4 X^{(3)}_{x} & X^{(1)}_{y} - 2 X^{(3)}_{y} & X^{(1)}_{z} - 2 X^{(3)}_{z}\\- X^{(2)}_{y} + 3 X^{(3)}_{y} & X^{(2)}_{x} + 3 X^{(3)}_{x} & 0\\- X^{(2)}_{z} + 3 X^{(3)}_{z} & 0 & X^{(2)}_{x} + 3 X^{(3)}_{x}\\X^{(2)}_{y} + 3 X^{(3)}_{y} & - X^{(2)}_{x} + 3 X^{(3)}_{x} & 0\\X^{(1)}_{x} - 2 X^{(3)}_{x} & X^{(1)}_{y} + 4 X^{(3)}_{y} & X^{(1)}_{z} - 2 X^{(3)}_{z}\\0 & - X^{(2)}_{z} + 3 X^{(3)}_{z} & X^{(2)}_{y} + 3 X^{(3)}_{y}\\X^{(2)}_{z} + 3 X^{(3)}_{z} & 0 & - X^{(2)}_{x} + 3 X^{(3)}_{x}\\0 & X^{(2)}_{z} + 3 X^{(3)}_{z} & - X^{(2)}_{y} + 3 X^{(3)}_{y}\\X^{(1)}_{x} - 2 X^{(3)}_{x} & X^{(1)}_{y} - 2 X^{(3)}_{y} & X^{(1)}_{z} + 4 X^{(3)}_{z}\end{pmatrix}$}.
\end{align}

\subsubsection{rank $l=2$}
\begin{align}
  \nonumber
  & \sigma_{ij}^{n\alpha(\mathrm{O/L})} = \\
  & \scalebox{0.6}{$\begin{pmatrix}\sqrt{2} Y^{(1)}_{u} + \sqrt{10} Y^{(3)}_{u} + 4 \sqrt{70} Y^{(4)}_{u} - \sqrt{6} Y^{(1)}_{v} - \sqrt{30} Y^{(3)}_{v} - 4 \sqrt{210} Y^{(4)}_{v} & \sqrt{2} Y^{(1)}_{u} + \sqrt{10} Y^{(3)}_{u} - 8 \sqrt{70} Y^{(4)}_{u} + \sqrt{6} Y^{(1)}_{v} - \sqrt{30} Y^{(3)}_{v} & - 2 \sqrt{2} Y^{(1)}_{u} + \sqrt{10} Y^{(3)}_{u} + 4 \sqrt{70} Y^{(4)}_{u} - \sqrt{30} Y^{(3)}_{v} + 4 \sqrt{210} Y^{(4)}_{v}\\2 \sqrt{2} Y^{(2)}_{xy} - \sqrt{30} Y^{(3)}_{xy} - 2 \sqrt{210} Y^{(4)}_{xy} & - 2 \sqrt{2} Y^{(2)}_{xy} - \sqrt{30} Y^{(3)}_{xy} - 2 \sqrt{210} Y^{(4)}_{xy} & - \sqrt{30} Y^{(3)}_{xy} + 4 \sqrt{210} Y^{(4)}_{xy}\\2 \sqrt{2} Y^{(2)}_{zx} - \sqrt{30} Y^{(3)}_{zx} - 2 \sqrt{210} Y^{(4)}_{zx} & - \sqrt{30} Y^{(3)}_{zx} + 4 \sqrt{210} Y^{(4)}_{zx} & - 2 \sqrt{2} Y^{(2)}_{zx} - \sqrt{30} Y^{(3)}_{zx} - 2 \sqrt{210} Y^{(4)}_{zx}\\- 2 \sqrt{2} Y^{(2)}_{xy} - \sqrt{30} Y^{(3)}_{xy} - 2 \sqrt{210} Y^{(4)}_{xy} & 2 \sqrt{2} Y^{(2)}_{xy} - \sqrt{30} Y^{(3)}_{xy} - 2 \sqrt{210} Y^{(4)}_{xy} & - \sqrt{30} Y^{(3)}_{xy} + 4 \sqrt{210} Y^{(4)}_{xy}\\\sqrt{2} Y^{(1)}_{u} + \sqrt{10} Y^{(3)}_{u} - 8 \sqrt{70} Y^{(4)}_{u} - \sqrt{6} Y^{(1)}_{v} + \sqrt{30} Y^{(3)}_{v} & \sqrt{2} Y^{(1)}_{u} + \sqrt{10} Y^{(3)}_{u} + 4 \sqrt{70} Y^{(4)}_{u} + \sqrt{6} Y^{(1)}_{v} + \sqrt{30} Y^{(3)}_{v} + 4 \sqrt{210} Y^{(4)}_{v} & - 2 \sqrt{2} Y^{(1)}_{u} + \sqrt{10} Y^{(3)}_{u} + 4 \sqrt{70} Y^{(4)}_{u} + \sqrt{30} Y^{(3)}_{v} - 4 \sqrt{210} Y^{(4)}_{v}\\- \sqrt{30} Y^{(3)}_{yz} + 4 \sqrt{210} Y^{(4)}_{yz} & 2 \sqrt{2} Y^{(2)}_{yz} - \sqrt{30} Y^{(3)}_{yz} - 2 \sqrt{210} Y^{(4)}_{yz} & - 2 \sqrt{2} Y^{(2)}_{yz} - \sqrt{30} Y^{(3)}_{yz} - 2 \sqrt{210} Y^{(4)}_{yz}\\- 2 \sqrt{2} Y^{(2)}_{zx} - \sqrt{30} Y^{(3)}_{zx} - 2 \sqrt{210} Y^{(4)}_{zx} & - \sqrt{30} Y^{(3)}_{zx} + 4 \sqrt{210} Y^{(4)}_{zx} & 2 \sqrt{2} Y^{(2)}_{zx} - \sqrt{30} Y^{(3)}_{zx} - 2 \sqrt{210} Y^{(4)}_{zx}\\- \sqrt{30} Y^{(3)}_{yz} + 4 \sqrt{210} Y^{(4)}_{yz} & - 2 \sqrt{2} Y^{(2)}_{yz} - \sqrt{30} Y^{(3)}_{yz} - 2 \sqrt{210} Y^{(4)}_{yz} & 2 \sqrt{2} Y^{(2)}_{yz} - \sqrt{30} Y^{(3)}_{yz} - 2 \sqrt{210} Y^{(4)}_{yz}\\\sqrt{2} Y^{(1)}_{u} - 2 \sqrt{10} Y^{(3)}_{u} + 4 \sqrt{70} Y^{(4)}_{u} - \sqrt{6} Y^{(1)}_{v} + 4 \sqrt{210} Y^{(4)}_{v} & \sqrt{2} Y^{(1)}_{u} - 2 \sqrt{10} Y^{(3)}_{u} + 4 \sqrt{70} Y^{(4)}_{u} + \sqrt{6} Y^{(1)}_{v} - 4 \sqrt{210} Y^{(4)}_{v} & - 2 \sqrt{2} Y^{(1)}_{u} - 2 \sqrt{10} Y^{(3)}_{u} - 8 \sqrt{70} Y^{(4)}_{u}\end{pmatrix}$},
\end{align}
\begin{align}
  \nonumber
  & \sigma_{ij}^{n\alpha(\mathrm{O/T})} = \\
  & \scalebox{1}{$\begin{pmatrix}- \sqrt{6} Y^{(1)}_{yz} + 4 \sqrt{210} Y^{(4)}_{yz} & - \sqrt{6} Y^{(1)}_{zx} - 2 \sqrt{210} Y^{(4)}_{zx} & - \sqrt{6} Y^{(1)}_{xy} - 2 \sqrt{210} Y^{(4)}_{xy}\\- \sqrt{2} Y^{(2)}_{zx} - 3 \sqrt{210} Y^{(4)}_{zx} & \sqrt{2} Y^{(2)}_{yz} - 3 \sqrt{210} Y^{(4)}_{yz} & 6 \sqrt{70} Y^{(4)}_{u} - 2 \sqrt{2} Y^{(2)}_{v}\\- \sqrt{2} Y^{(2)}_{xy} - 3 \sqrt{210} Y^{(4)}_{xy} & \sqrt{6} Y^{(2)}_{u} - 3 \sqrt{70} Y^{(4)}_{u} - \sqrt{2} Y^{(2)}_{v} - 3 \sqrt{210} Y^{(4)}_{v} & \sqrt{2} Y^{(2)}_{yz} - 3 \sqrt{210} Y^{(4)}_{yz}\\\sqrt{2} Y^{(2)}_{zx} - 3 \sqrt{210} Y^{(4)}_{zx} & - \sqrt{2} Y^{(2)}_{yz} - 3 \sqrt{210} Y^{(4)}_{yz} & 6 \sqrt{70} Y^{(4)}_{u} + 2 \sqrt{2} Y^{(2)}_{v}\\- \sqrt{6} Y^{(1)}_{yz} - 2 \sqrt{210} Y^{(4)}_{yz} & - \sqrt{6} Y^{(1)}_{zx} + 4 \sqrt{210} Y^{(4)}_{zx} & - \sqrt{6} Y^{(1)}_{xy} - 2 \sqrt{210} Y^{(4)}_{xy}\\\sqrt{6} Y^{(2)}_{u} - 3 \sqrt{70} Y^{(4)}_{u} + \sqrt{2} Y^{(2)}_{v} + 3 \sqrt{210} Y^{(4)}_{v} & - \sqrt{2} Y^{(2)}_{xy} - 3 \sqrt{210} Y^{(4)}_{xy} & \sqrt{2} Y^{(2)}_{zx} - 3 \sqrt{210} Y^{(4)}_{zx}\\\sqrt{2} Y^{(2)}_{xy} - 3 \sqrt{210} Y^{(4)}_{xy} & - \sqrt{6} Y^{(2)}_{u} - 3 \sqrt{70} Y^{(4)}_{u} + \sqrt{2} Y^{(2)}_{v} - 3 \sqrt{210} Y^{(4)}_{v} & - \sqrt{2} Y^{(2)}_{yz} - 3 \sqrt{210} Y^{(4)}_{yz}\\- \sqrt{6} Y^{(2)}_{u} - 3 \sqrt{70} Y^{(4)}_{u} - \sqrt{2} Y^{(2)}_{v} + 3 \sqrt{210} Y^{(4)}_{v} & \sqrt{2} Y^{(2)}_{xy} - 3 \sqrt{210} Y^{(4)}_{xy} & - \sqrt{2} Y^{(2)}_{zx} - 3 \sqrt{210} Y^{(4)}_{zx}\\- \sqrt{6} Y^{(1)}_{yz} - 2 \sqrt{210} Y^{(4)}_{yz} & - \sqrt{6} Y^{(1)}_{zx} - 2 \sqrt{210} Y^{(4)}_{zx} & - \sqrt{6} Y^{(1)}_{xy} + 4 \sqrt{210} Y^{(4)}_{xy}\end{pmatrix}$},
\end{align}
\begin{align}
  \nonumber
  & \sigma_{ij}^{n\alpha(\mathrm{H})} = \\
  & \scalebox{1}{$\begin{pmatrix}0 & 2 \sqrt{30} Y^{(2)}_{zx} & - 2 \sqrt{30} Y^{(2)}_{xy}\\- \sqrt{6} Y^{(1)}_{zx} - \sqrt{30} Y^{(2)}_{zx} & - \sqrt{6} Y^{(1)}_{yz} + \sqrt{30} Y^{(2)}_{yz} & - 2 \sqrt{2} Y^{(1)}_{u} + 2 \sqrt{30} Y^{(2)}_{v}\\\sqrt{6} Y^{(1)}_{xy} + \sqrt{30} Y^{(2)}_{xy} & - \sqrt{2} Y^{(1)}_{u} + 3 \sqrt{10} Y^{(2)}_{u} - \sqrt{6} Y^{(1)}_{v} - \sqrt{30} Y^{(2)}_{v} & \sqrt{6} Y^{(1)}_{yz} - \sqrt{30} Y^{(2)}_{yz}\\\sqrt{6} Y^{(1)}_{zx} - \sqrt{30} Y^{(2)}_{zx} & \sqrt{6} Y^{(1)}_{yz} + \sqrt{30} Y^{(2)}_{yz} & 2 \sqrt{2} Y^{(1)}_{u} + 2 \sqrt{30} Y^{(2)}_{v}\\- 2 \sqrt{30} Y^{(2)}_{yz} & 0 & 2 \sqrt{30} Y^{(2)}_{xy}\\\sqrt{2} Y^{(1)}_{u} - 3 \sqrt{10} Y^{(2)}_{u} - \sqrt{6} Y^{(1)}_{v} - \sqrt{30} Y^{(2)}_{v} & - \sqrt{6} Y^{(1)}_{xy} - \sqrt{30} Y^{(2)}_{xy} & - \sqrt{6} Y^{(1)}_{zx} + \sqrt{30} Y^{(2)}_{zx}\\- \sqrt{6} Y^{(1)}_{xy} + \sqrt{30} Y^{(2)}_{xy} & \sqrt{2} Y^{(1)}_{u} + 3 \sqrt{10} Y^{(2)}_{u} + \sqrt{6} Y^{(1)}_{v} - \sqrt{30} Y^{(2)}_{v} & - \sqrt{6} Y^{(1)}_{yz} - \sqrt{30} Y^{(2)}_{yz}\\- \sqrt{2} Y^{(1)}_{u} - 3 \sqrt{10} Y^{(2)}_{u} + \sqrt{6} Y^{(1)}_{v} - \sqrt{30} Y^{(2)}_{v} & \sqrt{6} Y^{(1)}_{xy} - \sqrt{30} Y^{(2)}_{xy} & \sqrt{6} Y^{(1)}_{zx} + \sqrt{30} Y^{(2)}_{zx}\\2 \sqrt{30} Y^{(2)}_{yz} & - 2 \sqrt{30} Y^{(2)}_{zx} & 0\end{pmatrix}$}.
\end{align}

\subsubsection{rank $l=3$}
\begin{align}
  \nonumber
  & \sigma_{ij}^{n\alpha(\mathrm{O/L})} = \\
  & \scalebox{0.8}{$\begin{pmatrix}0 & - 2 \sqrt{30} X^{(2)}_{xyz} & 2 \sqrt{30} X^{(2)}_{xyz}\\- \sqrt{15} X^{\alpha(1)}_{z} - 3 \sqrt{2} X^{\alpha(2)}_{z} + 5 X^{\beta(1)}_{z} + \sqrt{30} X^{\beta(2)}_{z} & - \sqrt{15} X^{\alpha(1)}_{z} + 3 \sqrt{2} X^{\alpha(2)}_{z} - 5 X^{\beta(1)}_{z} + \sqrt{30} X^{\beta(2)}_{z} & 2 \sqrt{15} X^{\alpha(1)}_{z} - 2 \sqrt{30} X^{\beta(2)}_{z}\\\sqrt{15} X^{\alpha(1)}_{y} + 3 \sqrt{2} X^{\alpha(2)}_{y} + 5 X^{\beta(1)}_{y} + \sqrt{30} X^{\beta(2)}_{y} & - 2 \sqrt{15} X^{\alpha(1)}_{y} - 2 \sqrt{30} X^{\beta(2)}_{y} & \sqrt{15} X^{\alpha(1)}_{y} - 3 \sqrt{2} X^{\alpha(2)}_{y} - 5 X^{\beta(1)}_{y} + \sqrt{30} X^{\beta(2)}_{y}\\\sqrt{15} X^{\alpha(1)}_{z} - 3 \sqrt{2} X^{\alpha(2)}_{z} - 5 X^{\beta(1)}_{z} + \sqrt{30} X^{\beta(2)}_{z} & \sqrt{15} X^{\alpha(1)}_{z} + 3 \sqrt{2} X^{\alpha(2)}_{z} + 5 X^{\beta(1)}_{z} + \sqrt{30} X^{\beta(2)}_{z} & - 2 \sqrt{15} X^{\alpha(1)}_{z} - 2 \sqrt{30} X^{\beta(2)}_{z}\\2 \sqrt{30} X^{(2)}_{xyz} & 0 & - 2 \sqrt{30} X^{(2)}_{xyz}\\2 \sqrt{15} X^{\alpha(1)}_{x} - 2 \sqrt{30} X^{\beta(2)}_{x} & - \sqrt{15} X^{\alpha(1)}_{x} - 3 \sqrt{2} X^{\alpha(2)}_{x} + 5 X^{\beta(1)}_{x} + \sqrt{30} X^{\beta(2)}_{x} & - \sqrt{15} X^{\alpha(1)}_{x} + 3 \sqrt{2} X^{\alpha(2)}_{x} - 5 X^{\beta(1)}_{x} + \sqrt{30} X^{\beta(2)}_{x}\\- \sqrt{15} X^{\alpha(1)}_{y} + 3 \sqrt{2} X^{\alpha(2)}_{y} - 5 X^{\beta(1)}_{y} + \sqrt{30} X^{\beta(2)}_{y} & 2 \sqrt{15} X^{\alpha(1)}_{y} - 2 \sqrt{30} X^{\beta(2)}_{y} & - \sqrt{15} X^{\alpha(1)}_{y} - 3 \sqrt{2} X^{\alpha(2)}_{y} + 5 X^{\beta(1)}_{y} + \sqrt{30} X^{\beta(2)}_{y}\\- 2 \sqrt{15} X^{\alpha(1)}_{x} - 2 \sqrt{30} X^{\beta(2)}_{x} & \sqrt{15} X^{\alpha(1)}_{x} - 3 \sqrt{2} X^{\alpha(2)}_{x} - 5 X^{\beta(1)}_{x} + \sqrt{30} X^{\beta(2)}_{x} & \sqrt{15} X^{\alpha(1)}_{x} + 3 \sqrt{2} X^{\alpha(2)}_{x} + 5 X^{\beta(1)}_{x} + \sqrt{30} X^{\beta(2)}_{x}\\- 2 \sqrt{30} X^{(2)}_{xyz} & 2 \sqrt{30} X^{(2)}_{xyz} & 0\end{pmatrix}$},
\end{align}
\begin{align}
  \nonumber
  & \sigma_{ij}^{n\alpha(\mathrm{O/T})} = \\
  & \scalebox{1}{$\begin{pmatrix}2 \sqrt{30} X^{\beta(2)}_{x} & - 3 \sqrt{2} X^{\alpha(2)}_{y} - \sqrt{30} X^{\beta(2)}_{y} & 3 \sqrt{2} X^{\alpha(2)}_{z} - \sqrt{30} X^{\beta(2)}_{z}\\- \sqrt{15} X^{\alpha(1)}_{y} + 6 \sqrt{2} X^{\alpha(2)}_{y} + 5 X^{\beta(1)}_{y} & - \sqrt{15} X^{\alpha(1)}_{x} - 6 \sqrt{2} X^{\alpha(2)}_{x} - 5 X^{\beta(1)}_{x} & 5 X^{(1)}_{xyz}\\\sqrt{15} X^{\alpha(1)}_{z} - 6 \sqrt{2} X^{\alpha(2)}_{z} + 5 X^{\beta(1)}_{z} & - 5 X^{(1)}_{xyz} & \sqrt{15} X^{\alpha(1)}_{x} + 6 \sqrt{2} X^{\alpha(2)}_{x} - 5 X^{\beta(1)}_{x}\\\sqrt{15} X^{\alpha(1)}_{y} + 6 \sqrt{2} X^{\alpha(2)}_{y} - 5 X^{\beta(1)}_{y} & \sqrt{15} X^{\alpha(1)}_{x} - 6 \sqrt{2} X^{\alpha(2)}_{x} + 5 X^{\beta(1)}_{x} & - 5 X^{(1)}_{xyz}\\3 \sqrt{2} X^{\alpha(2)}_{x} - \sqrt{30} X^{\beta(2)}_{x} & 2 \sqrt{30} X^{\beta(2)}_{y} & - 3 \sqrt{2} X^{\alpha(2)}_{z} - \sqrt{30} X^{\beta(2)}_{z}\\5 X^{(1)}_{xyz} & - \sqrt{15} X^{\alpha(1)}_{z} + 6 \sqrt{2} X^{\alpha(2)}_{z} + 5 X^{\beta(1)}_{z} & - \sqrt{15} X^{\alpha(1)}_{y} - 6 \sqrt{2} X^{\alpha(2)}_{y} - 5 X^{\beta(1)}_{y}\\- \sqrt{15} X^{\alpha(1)}_{z} - 6 \sqrt{2} X^{\alpha(2)}_{z} - 5 X^{\beta(1)}_{z} & 5 X^{(1)}_{xyz} & - \sqrt{15} X^{\alpha(1)}_{x} + 6 \sqrt{2} X^{\alpha(2)}_{x} + 5 X^{\beta(1)}_{x}\\- 5 X^{(1)}_{xyz} & \sqrt{15} X^{\alpha(1)}_{z} + 6 \sqrt{2} X^{\alpha(2)}_{z} - 5 X^{\beta(1)}_{z} & \sqrt{15} X^{\alpha(1)}_{y} - 6 \sqrt{2} X^{\alpha(2)}_{y} + 5 X^{\beta(1)}_{y}\\- 3 \sqrt{2} X^{\alpha(2)}_{x} - \sqrt{30} X^{\beta(2)}_{x} & 3 \sqrt{2} X^{\alpha(2)}_{y} - \sqrt{30} X^{\beta(2)}_{y} & 2 \sqrt{30} X^{\beta(2)}_{z}\end{pmatrix}$},
\end{align}
\begin{align}
  \nonumber
  & \sigma_{ij}^{n\alpha(\mathrm{H})} = \\
  & \scalebox{1}{$\begin{pmatrix}6 X^{\alpha(1)}_{x} & - 3 X^{\alpha(1)}_{y} - \sqrt{15} X^{\beta(1)}_{y} & - 3 X^{\alpha(1)}_{z} + \sqrt{15} X^{\beta(1)}_{z}\\- 3 X^{\alpha(1)}_{y} - \sqrt{15} X^{\beta(1)}_{y} & - 3 X^{\alpha(1)}_{x} + \sqrt{15} X^{\beta(1)}_{x} & \sqrt{15} X^{(1)}_{xyz}\\- 3 X^{\alpha(1)}_{z} + \sqrt{15} X^{\beta(1)}_{z} & \sqrt{15} X^{(1)}_{xyz} & - 3 X^{\alpha(1)}_{x} - \sqrt{15} X^{\beta(1)}_{x}\\- 3 X^{\alpha(1)}_{y} - \sqrt{15} X^{\beta(1)}_{y} & - 3 X^{\alpha(1)}_{x} + \sqrt{15} X^{\beta(1)}_{x} & \sqrt{15} X^{(1)}_{xyz}\\- 3 X^{\alpha(1)}_{x} + \sqrt{15} X^{\beta(1)}_{x} & 6 X^{\alpha(1)}_{y} & - 3 X^{\alpha(1)}_{z} - \sqrt{15} X^{\beta(1)}_{z}\\\sqrt{15} X^{(1)}_{xyz} & - 3 X^{\alpha(1)}_{z} - \sqrt{15} X^{\beta(1)}_{z} & - 3 X^{\alpha(1)}_{y} + \sqrt{15} X^{\beta(1)}_{y}\\- 3 X^{\alpha(1)}_{z} + \sqrt{15} X^{\beta(1)}_{z} & \sqrt{15} X^{(1)}_{xyz} & - 3 X^{\alpha(1)}_{x} - \sqrt{15} X^{\beta(1)}_{x}\\\sqrt{15} X^{(1)}_{xyz} & - 3 X^{\alpha(1)}_{z} - \sqrt{15} X^{\beta(1)}_{z} & - 3 X^{\alpha(1)}_{y} + \sqrt{15} X^{\beta(1)}_{y}\\- 3 X^{\alpha(1)}_{x} - \sqrt{15} X^{\beta(1)}_{x} & - 3 X^{\alpha(1)}_{y} + \sqrt{15} X^{\beta(1)}_{y} & 6 X^{\alpha(1)}_{z}\end{pmatrix}$}.
\end{align}

\subsubsection{rank $l=4$}
\begin{align}
  \nonumber
  & \sigma_{ij}^{n\alpha(\mathrm{O/L})} = \\
  & \scalebox{1}{$\begin{pmatrix}28 \sqrt{6} Y^{}_{4} + 2 \sqrt{210} Y^{}_{4u} - 6 \sqrt{70} Y^{}_{4v} & - 14 \sqrt{6} Y^{}_{4} - 4 \sqrt{210} Y^{}_{4u} & - 14 \sqrt{6} Y^{}_{4} + 2 \sqrt{210} Y^{}_{4u} + 6 \sqrt{70} Y^{}_{4v}\\21 \sqrt{10} Y^{\alpha}_{4z} - 3 \sqrt{70} Y^{\beta}_{4z} & - 21 \sqrt{10} Y^{\alpha}_{4z} - 3 \sqrt{70} Y^{\beta}_{4z} & 6 \sqrt{70} Y^{\beta}_{4z}\\- 21 \sqrt{10} Y^{\alpha}_{4y} - 3 \sqrt{70} Y^{\beta}_{4y} & 6 \sqrt{70} Y^{\beta}_{4y} & 21 \sqrt{10} Y^{\alpha}_{4y} - 3 \sqrt{70} Y^{\beta}_{4y}\\21 \sqrt{10} Y^{\alpha}_{4z} - 3 \sqrt{70} Y^{\beta}_{4z} & - 21 \sqrt{10} Y^{\alpha}_{4z} - 3 \sqrt{70} Y^{\beta}_{4z} & 6 \sqrt{70} Y^{\beta}_{4z}\\- 14 \sqrt{6} Y^{}_{4} - 4 \sqrt{210} Y^{}_{4u} & 28 \sqrt{6} Y^{}_{4} + 2 \sqrt{210} Y^{}_{4u} + 6 \sqrt{70} Y^{}_{4v} & - 14 \sqrt{6} Y^{}_{4} + 2 \sqrt{210} Y^{}_{4u} - 6 \sqrt{70} Y^{}_{4v}\\6 \sqrt{70} Y^{\beta}_{4x} & 21 \sqrt{10} Y^{\alpha}_{4x} - 3 \sqrt{70} Y^{\beta}_{4x} & - 21 \sqrt{10} Y^{\alpha}_{4x} - 3 \sqrt{70} Y^{\beta}_{4x}\\- 21 \sqrt{10} Y^{\alpha}_{4y} - 3 \sqrt{70} Y^{\beta}_{4y} & 6 \sqrt{70} Y^{\beta}_{4y} & 21 \sqrt{10} Y^{\alpha}_{4y} - 3 \sqrt{70} Y^{\beta}_{4y}\\6 \sqrt{70} Y^{\beta}_{4x} & 21 \sqrt{10} Y^{\alpha}_{4x} - 3 \sqrt{70} Y^{\beta}_{4x} & - 21 \sqrt{10} Y^{\alpha}_{4x} - 3 \sqrt{70} Y^{\beta}_{4x}\\- 14 \sqrt{6} Y^{}_{4} + 2 \sqrt{210} Y^{}_{4u} + 6 \sqrt{70} Y^{}_{4v} & - 14 \sqrt{6} Y^{}_{4} + 2 \sqrt{210} Y^{}_{4u} - 6 \sqrt{70} Y^{}_{4v} & 28 \sqrt{6} Y^{}_{4} - 4 \sqrt{210} Y^{}_{4u}\end{pmatrix}$},
\end{align}
\begin{align}
  \nonumber
  & \sigma_{ij}^{n\alpha(\mathrm{O/T})} = \\
  & \scalebox{1}{$\begin{pmatrix}6 \sqrt{70} Y^{\beta}_{4x} & - 21 \sqrt{10} Y^{\alpha}_{4y} - 3 \sqrt{70} Y^{\beta}_{4y} & 21 \sqrt{10} Y^{\alpha}_{4z} - 3 \sqrt{70} Y^{\beta}_{4z}\\6 \sqrt{70} Y^{\beta}_{4y} & 6 \sqrt{70} Y^{\beta}_{4x} & - 14 \sqrt{6} Y^{}_{4} - 4 \sqrt{210} Y^{}_{4u}\\6 \sqrt{70} Y^{\beta}_{4z} & - 14 \sqrt{6} Y^{}_{4} + 2 \sqrt{210} Y^{}_{4u} + 6 \sqrt{70} Y^{}_{4v} & 6 \sqrt{70} Y^{\beta}_{4x}\\6 \sqrt{70} Y^{\beta}_{4y} & 6 \sqrt{70} Y^{\beta}_{4x} & - 14 \sqrt{6} Y^{}_{4} - 4 \sqrt{210} Y^{}_{4u}\\21 \sqrt{10} Y^{\alpha}_{4x} - 3 \sqrt{70} Y^{\beta}_{4x} & 6 \sqrt{70} Y^{\beta}_{4y} & - 21 \sqrt{10} Y^{\alpha}_{4z} - 3 \sqrt{70} Y^{\beta}_{4z}\\- 14 \sqrt{6} Y^{}_{4} + 2 \sqrt{210} Y^{}_{4u} - 6 \sqrt{70} Y^{}_{4v} & 6 \sqrt{70} Y^{\beta}_{4z} & 6 \sqrt{70} Y^{\beta}_{4y}\\6 \sqrt{70} Y^{\beta}_{4z} & - 14 \sqrt{6} Y^{}_{4} + 2 \sqrt{210} Y^{}_{4u} + 6 \sqrt{70} Y^{}_{4v} & 6 \sqrt{70} Y^{\beta}_{4x}\\- 14 \sqrt{6} Y^{}_{4} + 2 \sqrt{210} Y^{}_{4u} - 6 \sqrt{70} Y^{}_{4v} & 6 \sqrt{70} Y^{\beta}_{4z} & 6 \sqrt{70} Y^{\beta}_{4y}\\- 21 \sqrt{10} Y^{\alpha}_{4x} - 3 \sqrt{70} Y^{\beta}_{4x} & 21 \sqrt{10} Y^{\alpha}_{4y} - 3 \sqrt{70} Y^{\beta}_{4y} & 6 \sqrt{70} Y^{\beta}_{4z}\end{pmatrix}$}.
\end{align}

\subsection{$t$-multipole}
\subsubsection{rank $l=3$}
\begin{align}
  \nonumber
  & \sigma_{ij}^{n\alpha(\mathrm{O/L})} = \\
  & \scalebox{0.8}{$\begin{pmatrix}0 & - 2 \sqrt{30} X^{(2)}_{xyz} & 2 \sqrt{30} X^{(2)}_{xyz}\\- 2 \sqrt{15} X^{\alpha(1)}_{z} - 3 \sqrt{2} X^{\alpha(2)}_{z} + 10 X^{\beta(1)}_{z} + \sqrt{30} X^{\beta(2)}_{z} & - 2 \sqrt{15} X^{\alpha(1)}_{z} + 3 \sqrt{2} X^{\alpha(2)}_{z} - 10 X^{\beta(1)}_{z} + \sqrt{30} X^{\beta(2)}_{z} & 4 \sqrt{15} X^{\alpha(1)}_{z} - 2 \sqrt{30} X^{\beta(2)}_{z}\\- 5 \sqrt{6} X^{(1)}_{3b} - 3 \sqrt{5} X^{(2)}_{3b} + \sqrt{10} X^{(1)}_{3v} + \sqrt{3} X^{(2)}_{3v} & 5 \sqrt{6} X^{(1)}_{3b} + 3 \sqrt{5} X^{(2)}_{3b} + 3 \sqrt{10} X^{(1)}_{3v} - 5 \sqrt{3} X^{(2)}_{3v} & - 4 \sqrt{10} X^{(1)}_{3v} + 4 \sqrt{3} X^{(2)}_{3v}\\2 \sqrt{15} X^{\alpha(1)}_{z} - 3 \sqrt{2} X^{\alpha(2)}_{z} - 10 X^{\beta(1)}_{z} + \sqrt{30} X^{\beta(2)}_{z} & 2 \sqrt{15} X^{\alpha(1)}_{z} + 3 \sqrt{2} X^{\alpha(2)}_{z} + 10 X^{\beta(1)}_{z} + \sqrt{30} X^{\beta(2)}_{z} & - 4 \sqrt{15} X^{\alpha(1)}_{z} - 2 \sqrt{30} X^{\beta(2)}_{z}\\2 \sqrt{30} X^{(2)}_{xyz} & 0 & - 2 \sqrt{30} X^{(2)}_{xyz}\\5 \sqrt{6} X^{(1)}_{3a} + 3 \sqrt{5} X^{(2)}_{3a} - 3 \sqrt{10} X^{(1)}_{3u} + 5 \sqrt{3} X^{(2)}_{3u} & - 5 \sqrt{6} X^{(1)}_{3a} - 3 \sqrt{5} X^{(2)}_{3a} - \sqrt{10} X^{(1)}_{3u} - \sqrt{3} X^{(2)}_{3u} & 4 \sqrt{10} X^{(1)}_{3u} - 4 \sqrt{3} X^{(2)}_{3u}\\5 \sqrt{6} X^{(1)}_{3b} - 3 \sqrt{5} X^{(2)}_{3b} - \sqrt{10} X^{(1)}_{3v} + \sqrt{3} X^{(2)}_{3v} & - 5 \sqrt{6} X^{(1)}_{3b} + 3 \sqrt{5} X^{(2)}_{3b} - 3 \sqrt{10} X^{(1)}_{3v} - 5 \sqrt{3} X^{(2)}_{3v} & 4 \sqrt{10} X^{(1)}_{3v} + 4 \sqrt{3} X^{(2)}_{3v}\\- 5 \sqrt{6} X^{(1)}_{3a} + 3 \sqrt{5} X^{(2)}_{3a} + 3 \sqrt{10} X^{(1)}_{3u} + 5 \sqrt{3} X^{(2)}_{3u} & 5 \sqrt{6} X^{(1)}_{3a} - 3 \sqrt{5} X^{(2)}_{3a} + \sqrt{10} X^{(1)}_{3u} - \sqrt{3} X^{(2)}_{3u} & - 4 \sqrt{10} X^{(1)}_{3u} - 4 \sqrt{3} X^{(2)}_{3u}\\- 2 \sqrt{30} X^{(2)}_{xyz} & 2 \sqrt{30} X^{(2)}_{xyz} & 0\end{pmatrix}$},
\end{align}
\begin{align}
  \nonumber
  & \sigma_{ij}^{n\alpha(\mathrm{O/T})} = \\
  & \scalebox{0.9}{$\begin{pmatrix}- 3 \sqrt{5} X^{(2)}_{3a} - 5 \sqrt{3} X^{(2)}_{3u} & 3 \sqrt{5} X^{(2)}_{3b} - \sqrt{3} X^{(2)}_{3v} & 3 \sqrt{2} X^{\alpha(2)}_{z} - \sqrt{30} X^{\beta(2)}_{z}\\- 3 \sqrt{5} X^{(2)}_{3b} + 4 \sqrt{10} X^{(1)}_{3v} - 3 \sqrt{3} X^{(2)}_{3v} & - 3 \sqrt{5} X^{(2)}_{3a} + 4 \sqrt{10} X^{(1)}_{3u} + 3 \sqrt{3} X^{(2)}_{3u} & 10 X^{(1)}_{xyz}\\2 \sqrt{15} X^{\alpha(1)}_{z} - 6 \sqrt{2} X^{\alpha(2)}_{z} + 10 X^{\beta(1)}_{z} & - 10 X^{(1)}_{xyz} & 5 \sqrt{6} X^{(1)}_{3a} + 3 \sqrt{5} X^{(2)}_{3a} + \sqrt{10} X^{(1)}_{3u} - 3 \sqrt{3} X^{(2)}_{3u}\\- 3 \sqrt{5} X^{(2)}_{3b} - 4 \sqrt{10} X^{(1)}_{3v} - 3 \sqrt{3} X^{(2)}_{3v} & - 3 \sqrt{5} X^{(2)}_{3a} - 4 \sqrt{10} X^{(1)}_{3u} + 3 \sqrt{3} X^{(2)}_{3u} & - 10 X^{(1)}_{xyz}\\3 \sqrt{5} X^{(2)}_{3a} + \sqrt{3} X^{(2)}_{3u} & - 3 \sqrt{5} X^{(2)}_{3b} + 5 \sqrt{3} X^{(2)}_{3v} & - 3 \sqrt{2} X^{\alpha(2)}_{z} - \sqrt{30} X^{\beta(2)}_{z}\\10 X^{(1)}_{xyz} & - 2 \sqrt{15} X^{\alpha(1)}_{z} + 6 \sqrt{2} X^{\alpha(2)}_{z} + 10 X^{\beta(1)}_{z} & 5 \sqrt{6} X^{(1)}_{3b} + 3 \sqrt{5} X^{(2)}_{3b} - \sqrt{10} X^{(1)}_{3v} + 3 \sqrt{3} X^{(2)}_{3v}\\- 2 \sqrt{15} X^{\alpha(1)}_{z} - 6 \sqrt{2} X^{\alpha(2)}_{z} - 10 X^{\beta(1)}_{z} & 10 X^{(1)}_{xyz} & - 5 \sqrt{6} X^{(1)}_{3a} + 3 \sqrt{5} X^{(2)}_{3a} - \sqrt{10} X^{(1)}_{3u} - 3 \sqrt{3} X^{(2)}_{3u}\\- 10 X^{(1)}_{xyz} & 2 \sqrt{15} X^{\alpha(1)}_{z} + 6 \sqrt{2} X^{\alpha(2)}_{z} - 10 X^{\beta(1)}_{z} & - 5 \sqrt{6} X^{(1)}_{3b} + 3 \sqrt{5} X^{(2)}_{3b} + \sqrt{10} X^{(1)}_{3v} + 3 \sqrt{3} X^{(2)}_{3v}\\4 \sqrt{3} X^{(2)}_{3u} & - 4 \sqrt{3} X^{(2)}_{3v} & 2 \sqrt{30} X^{\beta(2)}_{z}\end{pmatrix}$},
\end{align}
\begin{align}
  \nonumber
  & \sigma_{ij}^{n\alpha(\mathrm{H})} = \\
  & \scalebox{1}{$\begin{pmatrix}3 \sqrt{10} X^{(1)}_{3a} - 3 \sqrt{6} X^{(1)}_{3u} & 3 \sqrt{10} X^{(1)}_{3b} - \sqrt{6} X^{(1)}_{3v} & - 6 X^{\alpha(1)}_{z} + 2 \sqrt{15} X^{\beta(1)}_{z}\\3 \sqrt{10} X^{(1)}_{3b} - \sqrt{6} X^{(1)}_{3v} & - 3 \sqrt{10} X^{(1)}_{3a} - \sqrt{6} X^{(1)}_{3u} & 2 \sqrt{15} X^{(1)}_{xyz}\\- 6 X^{\alpha(1)}_{z} + 2 \sqrt{15} X^{\beta(1)}_{z} & 2 \sqrt{15} X^{(1)}_{xyz} & 4 \sqrt{6} X^{(1)}_{3u}\\3 \sqrt{10} X^{(1)}_{3b} - \sqrt{6} X^{(1)}_{3v} & - 3 \sqrt{10} X^{(1)}_{3a} - \sqrt{6} X^{(1)}_{3u} & 2 \sqrt{15} X^{(1)}_{xyz}\\- 3 \sqrt{10} X^{(1)}_{3a} - \sqrt{6} X^{(1)}_{3u} & - 3 \sqrt{10} X^{(1)}_{3b} - 3 \sqrt{6} X^{(1)}_{3v} & - 6 X^{\alpha(1)}_{z} - 2 \sqrt{15} X^{\beta(1)}_{z}\\2 \sqrt{15} X^{(1)}_{xyz} & - 6 X^{\alpha(1)}_{z} - 2 \sqrt{15} X^{\beta(1)}_{z} & 4 \sqrt{6} X^{(1)}_{3v}\\- 6 X^{\alpha(1)}_{z} + 2 \sqrt{15} X^{\beta(1)}_{z} & 2 \sqrt{15} X^{(1)}_{xyz} & 4 \sqrt{6} X^{(1)}_{3u}\\2 \sqrt{15} X^{(1)}_{xyz} & - 6 X^{\alpha(1)}_{z} - 2 \sqrt{15} X^{\beta(1)}_{z} & 4 \sqrt{6} X^{(1)}_{3v}\\4 \sqrt{6} X^{(1)}_{3u} & 4 \sqrt{6} X^{(1)}_{3v} & 12 X^{\alpha(1)}_{z}\end{pmatrix}$}.
\end{align}

\subsubsection{rank $l=4$}
\begin{align}
  \nonumber
  & \sigma_{ij}^{n\alpha(\mathrm{O/L})} = \\
  & \scalebox{1}{$\begin{pmatrix}3 \sqrt{14} Y^{}_{40} + 7 \sqrt{10} Y^{\beta1}_{4u} - 2 \sqrt{70} Y^{\beta2}_{4u} & \sqrt{14} Y^{}_{40} - 7 \sqrt{10} Y^{\beta1}_{4u} & - 4 \sqrt{14} Y^{}_{40} + 2 \sqrt{70} Y^{\beta2}_{4u}\\7 \sqrt{10} Y^{\beta1}_{4v} - \sqrt{70} Y^{\beta2}_{4v} & - 7 \sqrt{10} Y^{\beta1}_{4v} - \sqrt{70} Y^{\beta2}_{4v} & 2 \sqrt{70} Y^{\beta2}_{4v}\\7 \sqrt{5} Y^{}_{4b} - 3 \sqrt{35} Y^{\alpha}_{4u} & - 7 \sqrt{5} Y^{}_{4b} - \sqrt{35} Y^{\alpha}_{4u} & 4 \sqrt{35} Y^{\alpha}_{4u}\\7 \sqrt{10} Y^{\beta1}_{4v} - \sqrt{70} Y^{\beta2}_{4v} & - 7 \sqrt{10} Y^{\beta1}_{4v} - \sqrt{70} Y^{\beta2}_{4v} & 2 \sqrt{70} Y^{\beta2}_{4v}\\\sqrt{14} Y^{}_{40} - 7 \sqrt{10} Y^{\beta1}_{4u} & 3 \sqrt{14} Y^{}_{40} + 7 \sqrt{10} Y^{\beta1}_{4u} + 2 \sqrt{70} Y^{\beta2}_{4u} & - 4 \sqrt{14} Y^{}_{40} - 2 \sqrt{70} Y^{\beta2}_{4u}\\7 \sqrt{5} Y^{}_{4a} - \sqrt{35} Y^{\alpha}_{4v} & - 7 \sqrt{5} Y^{}_{4a} - 3 \sqrt{35} Y^{\alpha}_{4v} & 4 \sqrt{35} Y^{\alpha}_{4v}\\7 \sqrt{5} Y^{}_{4b} - 3 \sqrt{35} Y^{\alpha}_{4u} & - 7 \sqrt{5} Y^{}_{4b} - \sqrt{35} Y^{\alpha}_{4u} & 4 \sqrt{35} Y^{\alpha}_{4u}\\7 \sqrt{5} Y^{}_{4a} - \sqrt{35} Y^{\alpha}_{4v} & - 7 \sqrt{5} Y^{}_{4a} - 3 \sqrt{35} Y^{\alpha}_{4v} & 4 \sqrt{35} Y^{\alpha}_{4v}\\- 4 \sqrt{14} Y^{}_{40} + 2 \sqrt{70} Y^{\beta2}_{4u} & - 4 \sqrt{14} Y^{}_{40} - 2 \sqrt{70} Y^{\beta2}_{4u} & 8 \sqrt{14} Y^{}_{40}\end{pmatrix}$},
\end{align}
\begin{align}
  \nonumber
  & \sigma_{ij}^{n\alpha(\mathrm{O/T})} = \\
  & \scalebox{1}{$\begin{pmatrix}7 \sqrt{5} Y^{}_{4a} - \sqrt{35} Y^{\alpha}_{4v} & 7 \sqrt{5} Y^{}_{4b} - 3 \sqrt{35} Y^{\alpha}_{4u} & 7 \sqrt{10} Y^{\beta1}_{4v} - \sqrt{70} Y^{\beta2}_{4v}\\- 7 \sqrt{5} Y^{}_{4b} - \sqrt{35} Y^{\alpha}_{4u} & 7 \sqrt{5} Y^{}_{4a} - \sqrt{35} Y^{\alpha}_{4v} & \sqrt{14} Y^{}_{40} - 7 \sqrt{10} Y^{\beta1}_{4u}\\2 \sqrt{70} Y^{\beta2}_{4v} & - 4 \sqrt{14} Y^{}_{40} + 2 \sqrt{70} Y^{\beta2}_{4u} & 7 \sqrt{5} Y^{}_{4a} - \sqrt{35} Y^{\alpha}_{4v}\\- 7 \sqrt{5} Y^{}_{4b} - \sqrt{35} Y^{\alpha}_{4u} & 7 \sqrt{5} Y^{}_{4a} - \sqrt{35} Y^{\alpha}_{4v} & \sqrt{14} Y^{}_{40} - 7 \sqrt{10} Y^{\beta1}_{4u}\\- 7 \sqrt{5} Y^{}_{4a} - 3 \sqrt{35} Y^{\alpha}_{4v} & - 7 \sqrt{5} Y^{}_{4b} - \sqrt{35} Y^{\alpha}_{4u} & - 7 \sqrt{10} Y^{\beta1}_{4v} - \sqrt{70} Y^{\beta2}_{4v}\\- 4 \sqrt{14} Y^{}_{40} - 2 \sqrt{70} Y^{\beta2}_{4u} & 2 \sqrt{70} Y^{\beta2}_{4v} & - 7 \sqrt{5} Y^{}_{4b} - \sqrt{35} Y^{\alpha}_{4u}\\2 \sqrt{70} Y^{\beta2}_{4v} & - 4 \sqrt{14} Y^{}_{40} + 2 \sqrt{70} Y^{\beta2}_{4u} & 7 \sqrt{5} Y^{}_{4a} - \sqrt{35} Y^{\alpha}_{4v}\\- 4 \sqrt{14} Y^{}_{40} - 2 \sqrt{70} Y^{\beta2}_{4u} & 2 \sqrt{70} Y^{\beta2}_{4v} & - 7 \sqrt{5} Y^{}_{4b} - \sqrt{35} Y^{\alpha}_{4u}\\4 \sqrt{35} Y^{\alpha}_{4v} & 4 \sqrt{35} Y^{\alpha}_{4u} & 2 \sqrt{70} Y^{\beta2}_{4v}\end{pmatrix}$}.
\end{align}

  \subsection{$c$-multipole with normalization}
\subsubsection{rank $l = 0$}
\begin{align}
  \nonumber
  & \tilde{\sigma}_{ij}^{n\alpha(\mathrm{O/L})} = \\
  & \scalebox{1}{$\begin{pmatrix}Y^{(1)}_{0} + 4 Y^{(2)}_{0} & Y^{(1)}_{0} - 2 Y^{(2)}_{0} & Y^{(1)}_{0} - 2 Y^{(2)}_{0}\\0 & 0 & 0\\0 & 0 & 0\\0 & 0 & 0\\Y^{(1)}_{0} - 2 Y^{(2)}_{0} & Y^{(1)}_{0} + 4 Y^{(2)}_{0} & Y^{(1)}_{0} - 2 Y^{(2)}_{0}\\0 & 0 & 0\\0 & 0 & 0\\0 & 0 & 0\\Y^{(1)}_{0} - 2 Y^{(2)}_{0} & Y^{(1)}_{0} - 2 Y^{(2)}_{0} & Y^{(1)}_{0} + 4 Y^{(2)}_{0}\end{pmatrix}$},
\end{align}
\begin{align}
  \nonumber
  & \tilde{\sigma}_{ij}^{n\alpha(\mathrm{O/T})} = \\
  & \scalebox{1}{$\begin{pmatrix}0 & 0 & 0\\0 & 0 & 3 Y^{(2)}_{0}\\0 & 3 Y^{(2)}_{0} & 0\\0 & 0 & 3 Y^{(2)}_{0}\\0 & 0 & 0\\3 Y^{(2)}_{0} & 0 & 0\\0 & 3 Y^{(2)}_{0} & 0\\3 Y^{(2)}_{0} & 0 & 0\\0 & 0 & 0\end{pmatrix}$},
\end{align}
\begin{align}
  \nonumber
  & \tilde{\sigma}_{ij}^{n\alpha(\mathrm{H})} = \\
  & \scalebox{1}{$\begin{pmatrix}0 & 0 & 0\\0 & 0 & Y^{(1)}_{0}\\0 & - Y^{(1)}_{0} & 0\\0 & 0 & - Y^{(1)}_{0}\\0 & 0 & 0\\Y^{(1)}_{0} & 0 & 0\\0 & Y^{(1)}_{0} & 0\\- Y^{(1)}_{0} & 0 & 0\\0 & 0 & 0\end{pmatrix}$}.
\end{align}

\subsubsection{rank $l = 1$}
\begin{align}
  \nonumber
  & \tilde{\sigma}_{ij}^{n\alpha(\mathrm{O/L})} = \\
  & \scalebox{1}{$\begin{pmatrix}0 & 0 & 0\\X^{(1)}_{z} - 2 X^{(2)}_{z} + 2 X^{(3)}_{z} & X^{(1)}_{z} - 2 X^{(2)}_{z} - 2 X^{(3)}_{z} & X^{(1)}_{z} + 4 X^{(2)}_{z}\\- X^{(1)}_{y} + 2 X^{(2)}_{y} - 2 X^{(3)}_{y} & - X^{(1)}_{y} - 4 X^{(2)}_{y} & - X^{(1)}_{y} + 2 X^{(2)}_{y} + 2 X^{(3)}_{y}\\- X^{(1)}_{z} + 2 X^{(2)}_{z} + 2 X^{(3)}_{z} & - X^{(1)}_{z} + 2 X^{(2)}_{z} - 2 X^{(3)}_{z} & - X^{(1)}_{z} - 4 X^{(2)}_{z}\\0 & 0 & 0\\X^{(1)}_{x} + 4 X^{(2)}_{x} & X^{(1)}_{x} - 2 X^{(2)}_{x} + 2 X^{(3)}_{x} & X^{(1)}_{x} - 2 X^{(2)}_{x} - 2 X^{(3)}_{x}\\X^{(1)}_{y} - 2 X^{(2)}_{y} - 2 X^{(3)}_{y} & X^{(1)}_{y} + 4 X^{(2)}_{y} & X^{(1)}_{y} - 2 X^{(2)}_{y} + 2 X^{(3)}_{y}\\- X^{(1)}_{x} - 4 X^{(2)}_{x} & - X^{(1)}_{x} + 2 X^{(2)}_{x} + 2 X^{(3)}_{x} & - X^{(1)}_{x} + 2 X^{(2)}_{x} - 2 X^{(3)}_{x}\\0 & 0 & 0\end{pmatrix}$},
\end{align}
\begin{align}
  \nonumber
  & \tilde{\sigma}_{ij}^{n\alpha(\mathrm{O/T})} = \\
  & \scalebox{1}{$\begin{pmatrix}0 & 2 X^{(3)}_{y} & - 2 X^{(3)}_{z}\\3 X^{(2)}_{y} + X^{(3)}_{y} & 3 X^{(2)}_{x} - X^{(3)}_{x} & 0\\- 3 X^{(2)}_{z} - X^{(3)}_{z} & 0 & - 3 X^{(2)}_{x} + X^{(3)}_{x}\\- 3 X^{(2)}_{y} + X^{(3)}_{y} & - 3 X^{(2)}_{x} - X^{(3)}_{x} & 0\\- 2 X^{(3)}_{x} & 0 & 2 X^{(3)}_{z}\\0 & 3 X^{(2)}_{z} + X^{(3)}_{z} & 3 X^{(2)}_{y} - X^{(3)}_{y}\\3 X^{(2)}_{z} - X^{(3)}_{z} & 0 & 3 X^{(2)}_{x} + X^{(3)}_{x}\\0 & - 3 X^{(2)}_{z} + X^{(3)}_{z} & - 3 X^{(2)}_{y} - X^{(3)}_{y}\\2 X^{(3)}_{x} & - 2 X^{(3)}_{y} & 0\end{pmatrix}$},
\end{align}
\begin{align}
  \nonumber
  & \tilde{\sigma}_{ij}^{n\alpha(\mathrm{H})} = \\
  & \scalebox{1}{$\begin{pmatrix}X^{(1)}_{x} + 4 X^{(3)}_{x} & X^{(1)}_{y} - 2 X^{(3)}_{y} & X^{(1)}_{z} - 2 X^{(3)}_{z}\\- X^{(2)}_{y} + 3 X^{(3)}_{y} & X^{(2)}_{x} + 3 X^{(3)}_{x} & 0\\- X^{(2)}_{z} + 3 X^{(3)}_{z} & 0 & X^{(2)}_{x} + 3 X^{(3)}_{x}\\X^{(2)}_{y} + 3 X^{(3)}_{y} & - X^{(2)}_{x} + 3 X^{(3)}_{x} & 0\\X^{(1)}_{x} - 2 X^{(3)}_{x} & X^{(1)}_{y} + 4 X^{(3)}_{y} & X^{(1)}_{z} - 2 X^{(3)}_{z}\\0 & - X^{(2)}_{z} + 3 X^{(3)}_{z} & X^{(2)}_{y} + 3 X^{(3)}_{y}\\X^{(2)}_{z} + 3 X^{(3)}_{z} & 0 & - X^{(2)}_{x} + 3 X^{(3)}_{x}\\0 & X^{(2)}_{z} + 3 X^{(3)}_{z} & - X^{(2)}_{y} + 3 X^{(3)}_{y}\\X^{(1)}_{x} - 2 X^{(3)}_{x} & X^{(1)}_{y} - 2 X^{(3)}_{y} & X^{(1)}_{z} + 4 X^{(3)}_{z}\end{pmatrix}$}.
\end{align}

\subsubsection{rank $l = 2$}
\begin{align}
  \nonumber
  & \tilde{\sigma}_{ij}^{n\alpha(\mathrm{O/L})} = \\
  & \scalebox{0.9}{$\begin{pmatrix}Y^{(1)}_{u} + Y^{(3)}_{u} + 4 Y^{(4)}_{u} - Y^{(1)}_{v} - Y^{(3)}_{v} - 4 Y^{(4)}_{v} & Y^{(1)}_{u} + Y^{(3)}_{u} - 8 Y^{(4)}_{u} + Y^{(1)}_{v} - Y^{(3)}_{v} & - 2 Y^{(1)}_{u} + Y^{(3)}_{u} + 4 Y^{(4)}_{u} - Y^{(3)}_{v} + 4 Y^{(4)}_{v}\\2 Y^{(2)}_{xy} - Y^{(3)}_{xy} - 2 Y^{(4)}_{xy} & - 2 Y^{(2)}_{xy} - Y^{(3)}_{xy} - 2 Y^{(4)}_{xy} & - Y^{(3)}_{xy} + 4 Y^{(4)}_{xy}\\2 Y^{(2)}_{zx} - Y^{(3)}_{zx} - 2 Y^{(4)}_{zx} & - Y^{(3)}_{zx} + 4 Y^{(4)}_{zx} & - 2 Y^{(2)}_{zx} - Y^{(3)}_{zx} - 2 Y^{(4)}_{zx}\\- 2 Y^{(2)}_{xy} - Y^{(3)}_{xy} - 2 Y^{(4)}_{xy} & 2 Y^{(2)}_{xy} - Y^{(3)}_{xy} - 2 Y^{(4)}_{xy} & - Y^{(3)}_{xy} + 4 Y^{(4)}_{xy}\\Y^{(1)}_{u} + Y^{(3)}_{u} - 8 Y^{(4)}_{u} - Y^{(1)}_{v} + Y^{(3)}_{v} & Y^{(1)}_{u} + Y^{(3)}_{u} + 4 Y^{(4)}_{u} + Y^{(1)}_{v} + Y^{(3)}_{v} + 4 Y^{(4)}_{v} & - 2 Y^{(1)}_{u} + Y^{(3)}_{u} + 4 Y^{(4)}_{u} + Y^{(3)}_{v} - 4 Y^{(4)}_{v}\\- Y^{(3)}_{yz} + 4 Y^{(4)}_{yz} & 2 Y^{(2)}_{yz} - Y^{(3)}_{yz} - 2 Y^{(4)}_{yz} & - 2 Y^{(2)}_{yz} - Y^{(3)}_{yz} - 2 Y^{(4)}_{yz}\\- 2 Y^{(2)}_{zx} - Y^{(3)}_{zx} - 2 Y^{(4)}_{zx} & - Y^{(3)}_{zx} + 4 Y^{(4)}_{zx} & 2 Y^{(2)}_{zx} - Y^{(3)}_{zx} - 2 Y^{(4)}_{zx}\\- Y^{(3)}_{yz} + 4 Y^{(4)}_{yz} & - 2 Y^{(2)}_{yz} - Y^{(3)}_{yz} - 2 Y^{(4)}_{yz} & 2 Y^{(2)}_{yz} - Y^{(3)}_{yz} - 2 Y^{(4)}_{yz}\\Y^{(1)}_{u} - 2 Y^{(3)}_{u} + 4 Y^{(4)}_{u} - Y^{(1)}_{v} + 4 Y^{(4)}_{v} & Y^{(1)}_{u} - 2 Y^{(3)}_{u} + 4 Y^{(4)}_{u} + Y^{(1)}_{v} - 4 Y^{(4)}_{v} & - 2 Y^{(1)}_{u} - 2 Y^{(3)}_{u} - 8 Y^{(4)}_{u}\end{pmatrix}$},
\end{align}
\begin{align}
  \nonumber
  & \tilde{\sigma}_{ij}^{n\alpha(\mathrm{O/T})} = \\
  & \scalebox{1}{$\begin{pmatrix}- Y^{(1)}_{yz} + 4 Y^{(4)}_{yz} & - Y^{(1)}_{zx} - 2 Y^{(4)}_{zx} & - Y^{(1)}_{xy} - 2 Y^{(4)}_{xy}\\- Y^{(2)}_{zx} - 3 Y^{(4)}_{zx} & Y^{(2)}_{yz} - 3 Y^{(4)}_{yz} & 6 Y^{(4)}_{u} - 2 Y^{(2)}_{v}\\- Y^{(2)}_{xy} - 3 Y^{(4)}_{xy} & Y^{(2)}_{u} - 3 Y^{(4)}_{u} - Y^{(2)}_{v} - 3 Y^{(4)}_{v} & Y^{(2)}_{yz} - 3 Y^{(4)}_{yz}\\Y^{(2)}_{zx} - 3 Y^{(4)}_{zx} & - Y^{(2)}_{yz} - 3 Y^{(4)}_{yz} & 6 Y^{(4)}_{u} + 2 Y^{(2)}_{v}\\- Y^{(1)}_{yz} - 2 Y^{(4)}_{yz} & - Y^{(1)}_{zx} + 4 Y^{(4)}_{zx} & - Y^{(1)}_{xy} - 2 Y^{(4)}_{xy}\\Y^{(2)}_{u} - 3 Y^{(4)}_{u} + Y^{(2)}_{v} + 3 Y^{(4)}_{v} & - Y^{(2)}_{xy} - 3 Y^{(4)}_{xy} & Y^{(2)}_{zx} - 3 Y^{(4)}_{zx}\\Y^{(2)}_{xy} - 3 Y^{(4)}_{xy} & - Y^{(2)}_{u} - 3 Y^{(4)}_{u} + Y^{(2)}_{v} - 3 Y^{(4)}_{v} & - Y^{(2)}_{yz} - 3 Y^{(4)}_{yz}\\- Y^{(2)}_{u} - 3 Y^{(4)}_{u} - Y^{(2)}_{v} + 3 Y^{(4)}_{v} & Y^{(2)}_{xy} - 3 Y^{(4)}_{xy} & - Y^{(2)}_{zx} - 3 Y^{(4)}_{zx}\\- Y^{(1)}_{yz} - 2 Y^{(4)}_{yz} & - Y^{(1)}_{zx} - 2 Y^{(4)}_{zx} & - Y^{(1)}_{xy} + 4 Y^{(4)}_{xy}\end{pmatrix}$},
\end{align}
\begin{align}
  \nonumber
  & \tilde{\sigma}_{ij}^{n\alpha(\mathrm{H})} = \\
  & \scalebox{1}{$\begin{pmatrix}0 & 2 Y^{(2)}_{zx} & - 2 Y^{(2)}_{xy}\\- Y^{(1)}_{zx} - Y^{(2)}_{zx} & - Y^{(1)}_{yz} + Y^{(2)}_{yz} & - 2 Y^{(1)}_{u} + 2 Y^{(2)}_{v}\\Y^{(1)}_{xy} + Y^{(2)}_{xy} & - Y^{(1)}_{u} + Y^{(2)}_{u} - Y^{(1)}_{v} - Y^{(2)}_{v} & Y^{(1)}_{yz} - Y^{(2)}_{yz}\\Y^{(1)}_{zx} - Y^{(2)}_{zx} & Y^{(1)}_{yz} + Y^{(2)}_{yz} & 2 Y^{(1)}_{u} + 2 Y^{(2)}_{v}\\- 2 Y^{(2)}_{yz} & 0 & 2 Y^{(2)}_{xy}\\Y^{(1)}_{u} - Y^{(2)}_{u} - Y^{(1)}_{v} - Y^{(2)}_{v} & - Y^{(1)}_{xy} - Y^{(2)}_{xy} & - Y^{(1)}_{zx} + Y^{(2)}_{zx}\\- Y^{(1)}_{xy} + Y^{(2)}_{xy} & Y^{(1)}_{u} + Y^{(2)}_{u} + Y^{(1)}_{v} - Y^{(2)}_{v} & - Y^{(1)}_{yz} - Y^{(2)}_{yz}\\- Y^{(1)}_{u} - Y^{(2)}_{u} + Y^{(1)}_{v} - Y^{(2)}_{v} & Y^{(1)}_{xy} - Y^{(2)}_{xy} & Y^{(1)}_{zx} + Y^{(2)}_{zx}\\2 Y^{(2)}_{yz} & - 2 Y^{(2)}_{zx} & 0\end{pmatrix}$}.
\end{align}

\subsubsection{rank $l = 3$}
\begin{align}
  \nonumber
  & \tilde{\sigma}_{ij}^{n\alpha(\mathrm{O/L})} = \\
  & \scalebox{1}{$\begin{pmatrix}0 & - X^{(2)}_{xyz} & X^{(2)}_{xyz}\\- X^{\alpha(1)}_{z} - X^{\alpha(2)}_{z} + X^{\beta(1)}_{z} + X^{\beta(2)}_{z} & - X^{\alpha(1)}_{z} + X^{\alpha(2)}_{z} - X^{\beta(1)}_{z} + X^{\beta(2)}_{z} & 2 X^{\alpha(1)}_{z} - 2 X^{\beta(2)}_{z}\\X^{\alpha(1)}_{y} + X^{\alpha(2)}_{y} + X^{\beta(1)}_{y} + X^{\beta(2)}_{y} & - 2 X^{\alpha(1)}_{y} - 2 X^{\beta(2)}_{y} & X^{\alpha(1)}_{y} - X^{\alpha(2)}_{y} - X^{\beta(1)}_{y} + X^{\beta(2)}_{y}\\X^{\alpha(1)}_{z} - X^{\alpha(2)}_{z} - X^{\beta(1)}_{z} + X^{\beta(2)}_{z} & X^{\alpha(1)}_{z} + X^{\alpha(2)}_{z} + X^{\beta(1)}_{z} + X^{\beta(2)}_{z} & - 2 X^{\alpha(1)}_{z} - 2 X^{\beta(2)}_{z}\\X^{(2)}_{xyz} & 0 & - X^{(2)}_{xyz}\\2 X^{\alpha(1)}_{x} - 2 X^{\beta(2)}_{x} & - X^{\alpha(1)}_{x} - X^{\alpha(2)}_{x} + X^{\beta(1)}_{x} + X^{\beta(2)}_{x} & - X^{\alpha(1)}_{x} + X^{\alpha(2)}_{x} - X^{\beta(1)}_{x} + X^{\beta(2)}_{x}\\- X^{\alpha(1)}_{y} + X^{\alpha(2)}_{y} - X^{\beta(1)}_{y} + X^{\beta(2)}_{y} & 2 X^{\alpha(1)}_{y} - 2 X^{\beta(2)}_{y} & - X^{\alpha(1)}_{y} - X^{\alpha(2)}_{y} + X^{\beta(1)}_{y} + X^{\beta(2)}_{y}\\- 2 X^{\alpha(1)}_{x} - 2 X^{\beta(2)}_{x} & X^{\alpha(1)}_{x} - X^{\alpha(2)}_{x} - X^{\beta(1)}_{x} + X^{\beta(2)}_{x} & X^{\alpha(1)}_{x} + X^{\alpha(2)}_{x} + X^{\beta(1)}_{x} + X^{\beta(2)}_{x}\\- X^{(2)}_{xyz} & X^{(2)}_{xyz} & 0\end{pmatrix}$},
\end{align}
\begin{align}
  \nonumber
  & \tilde{\sigma}_{ij}^{n\alpha(\mathrm{O/T})} = \\
  & \scalebox{1}{$\begin{pmatrix}2 X^{\beta(2)}_{x} & - X^{\alpha(2)}_{y} - X^{\beta(2)}_{y} & X^{\alpha(2)}_{z} - X^{\beta(2)}_{z}\\- X^{\alpha(1)}_{y} + 2 X^{\alpha(2)}_{y} + X^{\beta(1)}_{y} & - X^{\alpha(1)}_{x} - 2 X^{\alpha(2)}_{x} - X^{\beta(1)}_{x} & X^{(1)}_{xyz}\\X^{\alpha(1)}_{z} - 2 X^{\alpha(2)}_{z} + X^{\beta(1)}_{z} & - X^{(1)}_{xyz} & X^{\alpha(1)}_{x} + 2 X^{\alpha(2)}_{x} - X^{\beta(1)}_{x}\\X^{\alpha(1)}_{y} + 2 X^{\alpha(2)}_{y} - X^{\beta(1)}_{y} & X^{\alpha(1)}_{x} - 2 X^{\alpha(2)}_{x} + X^{\beta(1)}_{x} & - X^{(1)}_{xyz}\\X^{\alpha(2)}_{x} - X^{\beta(2)}_{x} & 2 X^{\beta(2)}_{y} & - X^{\alpha(2)}_{z} - X^{\beta(2)}_{z}\\X^{(1)}_{xyz} & - X^{\alpha(1)}_{z} + 2 X^{\alpha(2)}_{z} + X^{\beta(1)}_{z} & - X^{\alpha(1)}_{y} - 2 X^{\alpha(2)}_{y} - X^{\beta(1)}_{y}\\- X^{\alpha(1)}_{z} - 2 X^{\alpha(2)}_{z} - X^{\beta(1)}_{z} & X^{(1)}_{xyz} & - X^{\alpha(1)}_{x} + 2 X^{\alpha(2)}_{x} + X^{\beta(1)}_{x}\\- X^{(1)}_{xyz} & X^{\alpha(1)}_{z} + 2 X^{\alpha(2)}_{z} - X^{\beta(1)}_{z} & X^{\alpha(1)}_{y} - 2 X^{\alpha(2)}_{y} + X^{\beta(1)}_{y}\\- X^{\alpha(2)}_{x} - X^{\beta(2)}_{x} & X^{\alpha(2)}_{y} - X^{\beta(2)}_{y} & 2 X^{\beta(2)}_{z}\end{pmatrix}$},
\end{align}
\begin{align}
  \nonumber
  & \tilde{\sigma}_{ij}^{n\alpha(\mathrm{H})} = \\
  & \scalebox{1}{$\begin{pmatrix}2 X^{\alpha(1)}_{x} & - X^{\alpha(1)}_{y} - X^{\beta(1)}_{y} & - X^{\alpha(1)}_{z} + X^{\beta(1)}_{z}\\- X^{\alpha(1)}_{y} - X^{\beta(1)}_{y} & - X^{\alpha(1)}_{x} + X^{\beta(1)}_{x} & X^{(1)}_{xyz}\\- X^{\alpha(1)}_{z} + X^{\beta(1)}_{z} & X^{(1)}_{xyz} & - X^{\alpha(1)}_{x} - X^{\beta(1)}_{x}\\- X^{\alpha(1)}_{y} - X^{\beta(1)}_{y} & - X^{\alpha(1)}_{x} + X^{\beta(1)}_{x} & X^{(1)}_{xyz}\\- X^{\alpha(1)}_{x} + X^{\beta(1)}_{x} & 2 X^{\alpha(1)}_{y} & - X^{\alpha(1)}_{z} - X^{\beta(1)}_{z}\\X^{(1)}_{xyz} & - X^{\alpha(1)}_{z} - X^{\beta(1)}_{z} & - X^{\alpha(1)}_{y} + X^{\beta(1)}_{y}\\- X^{\alpha(1)}_{z} + X^{\beta(1)}_{z} & X^{(1)}_{xyz} & - X^{\alpha(1)}_{x} - X^{\beta(1)}_{x}\\X^{(1)}_{xyz} & - X^{\alpha(1)}_{z} - X^{\beta(1)}_{z} & - X^{\alpha(1)}_{y} + X^{\beta(1)}_{y}\\- X^{\alpha(1)}_{x} - X^{\beta(1)}_{x} & - X^{\alpha(1)}_{y} + X^{\beta(1)}_{y} & 2 X^{\alpha(1)}_{z}\end{pmatrix}$}.
\end{align}

\subsubsection{rank $l = 4$}
\begin{align}
  \nonumber
  & \tilde{\sigma}_{ij}^{n\alpha(\mathrm{O/L})} = \\
  & \scalebox{1}{$\begin{pmatrix}2 Y^{}_{4} + Y^{}_{4u} - Y^{}_{4v} & - Y^{}_{4} - 2 Y^{}_{4u} & - Y^{}_{4} + Y^{}_{4u} + Y^{}_{4v}\\Y^{\alpha}_{4z} - Y^{\beta}_{4z} & - Y^{\alpha}_{4z} - Y^{\beta}_{4z} & 2 Y^{\beta}_{4z}\\- Y^{\alpha}_{4y} - Y^{\beta}_{4y} & 2 Y^{\beta}_{4y} & Y^{\alpha}_{4y} - Y^{\beta}_{4y}\\Y^{\alpha}_{4z} - Y^{\beta}_{4z} & - Y^{\alpha}_{4z} - Y^{\beta}_{4z} & 2 Y^{\beta}_{4z}\\- Y^{}_{4} - 2 Y^{}_{4u} & 2 Y^{}_{4} + Y^{}_{4u} + Y^{}_{4v} & - Y^{}_{4} + Y^{}_{4u} - Y^{}_{4v}\\2 Y^{\beta}_{4x} & Y^{\alpha}_{4x} - Y^{\beta}_{4x} & - Y^{\alpha}_{4x} - Y^{\beta}_{4x}\\- Y^{\alpha}_{4y} - Y^{\beta}_{4y} & 2 Y^{\beta}_{4y} & Y^{\alpha}_{4y} - Y^{\beta}_{4y}\\2 Y^{\beta}_{4x} & Y^{\alpha}_{4x} - Y^{\beta}_{4x} & - Y^{\alpha}_{4x} - Y^{\beta}_{4x}\\- Y^{}_{4} + Y^{}_{4u} + Y^{}_{4v} & - Y^{}_{4} + Y^{}_{4u} - Y^{}_{4v} & 2 Y^{}_{4} - 2 Y^{}_{4u}\end{pmatrix}$},
\end{align}
\begin{align}
  \nonumber
  & \tilde{\sigma}_{ij}^{n\alpha(\mathrm{O/T})} = \\
  & \scalebox{1}{$\begin{pmatrix}2 Y^{\beta}_{4x} & - Y^{\alpha}_{4y} - Y^{\beta}_{4y} & Y^{\alpha}_{4z} - Y^{\beta}_{4z}\\2 Y^{\beta}_{4y} & 2 Y^{\beta}_{4x} & - Y^{}_{4} - 2 Y^{}_{4u}\\2 Y^{\beta}_{4z} & - Y^{}_{4} + Y^{}_{4u} + Y^{}_{4v} & 2 Y^{\beta}_{4x}\\2 Y^{\beta}_{4y} & 2 Y^{\beta}_{4x} & - Y^{}_{4} - 2 Y^{}_{4u}\\Y^{\alpha}_{4x} - Y^{\beta}_{4x} & 2 Y^{\beta}_{4y} & - Y^{\alpha}_{4z} - Y^{\beta}_{4z}\\- Y^{}_{4} + Y^{}_{4u} - Y^{}_{4v} & 2 Y^{\beta}_{4z} & 2 Y^{\beta}_{4y}\\2 Y^{\beta}_{4z} & - Y^{}_{4} + Y^{}_{4u} + Y^{}_{4v} & 2 Y^{\beta}_{4x}\\- Y^{}_{4} + Y^{}_{4u} - Y^{}_{4v} & 2 Y^{\beta}_{4z} & 2 Y^{\beta}_{4y}\\- Y^{\alpha}_{4x} - Y^{\beta}_{4x} & Y^{\alpha}_{4y} - Y^{\beta}_{4y} & 2 Y^{\beta}_{4z}\end{pmatrix}$}.
\end{align}

\subsection{$t$-multipole with normalization}
\subsubsection{rank $l = 3$}
\begin{align}
  \nonumber
  & \tilde{\sigma}_{ij}^{n\alpha(\mathrm{O/L})} = \\
  & \scalebox{1}{$\begin{pmatrix}0 & - X^{(2)}_{xyz} & X^{(2)}_{xyz}\\- X^{\alpha(1)}_{z} - X^{\alpha(2)}_{z} + X^{\beta(1)}_{z} + X^{\beta(2)}_{z} & - X^{\alpha(1)}_{z} + X^{\alpha(2)}_{z} - X^{\beta(1)}_{z} + X^{\beta(2)}_{z} & 2 X^{\alpha(1)}_{z} - 2 X^{\beta(2)}_{z}\\- X^{(1)}_{3b} - X^{(2)}_{3b} + X^{(1)}_{3v} + X^{(2)}_{3v} & X^{(1)}_{3b} + X^{(2)}_{3b} + 3 X^{(1)}_{3v} - 5 X^{(2)}_{3v} & - 4 X^{(1)}_{3v} + 4 X^{(2)}_{3v}\\X^{\alpha(1)}_{z} - X^{\alpha(2)}_{z} - X^{\beta(1)}_{z} + X^{\beta(2)}_{z} & X^{\alpha(1)}_{z} + X^{\alpha(2)}_{z} + X^{\beta(1)}_{z} + X^{\beta(2)}_{z} & - 2 X^{\alpha(1)}_{z} - 2 X^{\beta(2)}_{z}\\X^{(2)}_{xyz} & 0 & - X^{(2)}_{xyz}\\X^{(1)}_{3a} + X^{(2)}_{3a} - 3 X^{(1)}_{3u} + 5 X^{(2)}_{3u} & - X^{(1)}_{3a} - X^{(2)}_{3a} - X^{(1)}_{3u} - X^{(2)}_{3u} & 4 X^{(1)}_{3u} - 4 X^{(2)}_{3u}\\X^{(1)}_{3b} - X^{(2)}_{3b} - X^{(1)}_{3v} + X^{(2)}_{3v} & - X^{(1)}_{3b} + X^{(2)}_{3b} - 3 X^{(1)}_{3v} - 5 X^{(2)}_{3v} & 4 X^{(1)}_{3v} + 4 X^{(2)}_{3v}\\- X^{(1)}_{3a} + X^{(2)}_{3a} + 3 X^{(1)}_{3u} + 5 X^{(2)}_{3u} & X^{(1)}_{3a} - X^{(2)}_{3a} + X^{(1)}_{3u} - X^{(2)}_{3u} & - 4 X^{(1)}_{3u} - 4 X^{(2)}_{3u}\\- X^{(2)}_{xyz} & X^{(2)}_{xyz} & 0\end{pmatrix}$},
\end{align}
\begin{align}
  \nonumber
  & \tilde{\sigma}_{ij}^{n\alpha(\mathrm{O/T})} = \\
  & \scalebox{1}{$\begin{pmatrix}- X^{(2)}_{3a} - 5 X^{(2)}_{3u} & X^{(2)}_{3b} - X^{(2)}_{3v} & X^{\alpha(2)}_{z} - X^{\beta(2)}_{z}\\- X^{(2)}_{3b} + 4 X^{(1)}_{3v} - 3 X^{(2)}_{3v} & - X^{(2)}_{3a} + 4 X^{(1)}_{3u} + 3 X^{(2)}_{3u} & X^{(1)}_{xyz}\\X^{\alpha(1)}_{z} - 2 X^{\alpha(2)}_{z} + X^{\beta(1)}_{z} & - X^{(1)}_{xyz} & X^{(1)}_{3a} + X^{(2)}_{3a} + X^{(1)}_{3u} - 3 X^{(2)}_{3u}\\- X^{(2)}_{3b} - 4 X^{(1)}_{3v} - 3 X^{(2)}_{3v} & - X^{(2)}_{3a} - 4 X^{(1)}_{3u} + 3 X^{(2)}_{3u} & - X^{(1)}_{xyz}\\X^{(2)}_{3a} + X^{(2)}_{3u} & - X^{(2)}_{3b} + 5 X^{(2)}_{3v} & - X^{\alpha(2)}_{z} - X^{\beta(2)}_{z}\\X^{(1)}_{xyz} & - X^{\alpha(1)}_{z} + 2 X^{\alpha(2)}_{z} + X^{\beta(1)}_{z} & X^{(1)}_{3b} + X^{(2)}_{3b} - X^{(1)}_{3v} + 3 X^{(2)}_{3v}\\- X^{\alpha(1)}_{z} - 2 X^{\alpha(2)}_{z} - X^{\beta(1)}_{z} & X^{(1)}_{xyz} & - X^{(1)}_{3a} + X^{(2)}_{3a} - X^{(1)}_{3u} - 3 X^{(2)}_{3u}\\- X^{(1)}_{xyz} & X^{\alpha(1)}_{z} + 2 X^{\alpha(2)}_{z} - X^{\beta(1)}_{z} & - X^{(1)}_{3b} + X^{(2)}_{3b} + X^{(1)}_{3v} + 3 X^{(2)}_{3v}\\4 X^{(2)}_{3u} & - 4 X^{(2)}_{3v} & 2 X^{\beta(2)}_{z}\end{pmatrix}$},
\end{align}
\begin{align}
  \nonumber
  & \tilde{\sigma}_{ij}^{n\alpha(\mathrm{H})} = \\
  & \scalebox{1}{$\begin{pmatrix}X^{(1)}_{3a} - 3 X^{(1)}_{3u} & X^{(1)}_{3b} - X^{(1)}_{3v} & - X^{\alpha(1)}_{z} + X^{\beta(1)}_{z}\\X^{(1)}_{3b} - X^{(1)}_{3v} & - X^{(1)}_{3a} - X^{(1)}_{3u} & X^{(1)}_{xyz}\\- X^{\alpha(1)}_{z} + X^{\beta(1)}_{z} & X^{(1)}_{xyz} & 4 X^{(1)}_{3u}\\X^{(1)}_{3b} - X^{(1)}_{3v} & - X^{(1)}_{3a} - X^{(1)}_{3u} & X^{(1)}_{xyz}\\- X^{(1)}_{3a} - X^{(1)}_{3u} & - X^{(1)}_{3b} - 3 X^{(1)}_{3v} & - X^{\alpha(1)}_{z} - X^{\beta(1)}_{z}\\X^{(1)}_{xyz} & - X^{\alpha(1)}_{z} - X^{\beta(1)}_{z} & 4 X^{(1)}_{3v}\\- X^{\alpha(1)}_{z} + X^{\beta(1)}_{z} & X^{(1)}_{xyz} & 4 X^{(1)}_{3u}\\X^{(1)}_{xyz} & - X^{\alpha(1)}_{z} - X^{\beta(1)}_{z} & 4 X^{(1)}_{3v}\\4 X^{(1)}_{3u} & 4 X^{(1)}_{3v} & 2 X^{\alpha(1)}_{z}\end{pmatrix}$}.
\end{align}

\subsubsection{rank $l = 4$}
\begin{align}
  \nonumber
  & \tilde{\sigma}_{ij}^{n\alpha(\mathrm{O/L})} = \\
  & \scalebox{1}{$\begin{pmatrix}3 Y^{}_{40} + Y^{\beta1}_{4u} - Y^{\beta2}_{4u} & Y^{}_{40} - Y^{\beta1}_{4u} & - 4 Y^{}_{40} + Y^{\beta2}_{4u}\\Y^{\beta1}_{4v} - Y^{\beta2}_{4v} & - Y^{\beta1}_{4v} - Y^{\beta2}_{4v} & 2 Y^{\beta2}_{4v}\\Y^{}_{4b} - 3 Y^{\alpha}_{4u} & - Y^{}_{4b} - Y^{\alpha}_{4u} & 4 Y^{\alpha}_{4u}\\Y^{\beta1}_{4v} - Y^{\beta2}_{4v} & - Y^{\beta1}_{4v} - Y^{\beta2}_{4v} & 2 Y^{\beta2}_{4v}\\Y^{}_{40} - Y^{\beta1}_{4u} & 3 Y^{}_{40} + Y^{\beta1}_{4u} + Y^{\beta2}_{4u} & - 4 Y^{}_{40} - Y^{\beta2}_{4u}\\Y^{}_{4a} - Y^{\alpha}_{4v} & - Y^{}_{4a} - 3 Y^{\alpha}_{4v} & 4 Y^{\alpha}_{4v}\\Y^{}_{4b} - 3 Y^{\alpha}_{4u} & - Y^{}_{4b} - Y^{\alpha}_{4u} & 4 Y^{\alpha}_{4u}\\Y^{}_{4a} - Y^{\alpha}_{4v} & - Y^{}_{4a} - 3 Y^{\alpha}_{4v} & 4 Y^{\alpha}_{4v}\\- 4 Y^{}_{40} + Y^{\beta2}_{4u} & - 4 Y^{}_{40} - Y^{\beta2}_{4u} & 8 Y^{}_{40}\end{pmatrix}$},
\end{align}
\begin{align}
  \nonumber
  & \tilde{\sigma}_{ij}^{n\alpha(\mathrm{O/T})} = \\
  & \scalebox{1}{$\begin{pmatrix}Y^{}_{4a} - Y^{\alpha}_{4v} & Y^{}_{4b} - 3 Y^{\alpha}_{4u} & Y^{\beta1}_{4v} - Y^{\beta2}_{4v}\\- Y^{}_{4b} - Y^{\alpha}_{4u} & Y^{}_{4a} - Y^{\alpha}_{4v} & Y^{}_{40} - Y^{\beta1}_{4u}\\2 Y^{\beta2}_{4v} & - 4 Y^{}_{40} + Y^{\beta2}_{4u} & Y^{}_{4a} - Y^{\alpha}_{4v}\\- Y^{}_{4b} - Y^{\alpha}_{4u} & Y^{}_{4a} - Y^{\alpha}_{4v} & Y^{}_{40} - Y^{\beta1}_{4u}\\- Y^{}_{4a} - 3 Y^{\alpha}_{4v} & - Y^{}_{4b} - Y^{\alpha}_{4u} & - Y^{\beta1}_{4v} - Y^{\beta2}_{4v}\\- 4 Y^{}_{40} - Y^{\beta2}_{4u} & 2 Y^{\beta2}_{4v} & - Y^{}_{4b} - Y^{\alpha}_{4u}\\2 Y^{\beta2}_{4v} & - 4 Y^{}_{40} + Y^{\beta2}_{4u} & Y^{}_{4a} - Y^{\alpha}_{4v}\\- 4 Y^{}_{40} - Y^{\beta2}_{4u} & 2 Y^{\beta2}_{4v} & - Y^{}_{4b} - Y^{\alpha}_{4u}\\4 Y^{\alpha}_{4v} & 4 Y^{\alpha}_{4u} & 2 Y^{\beta2}_{4v}\end{pmatrix}$}.
\end{align}

  \section{\label{app:Model analysis of spin accumulation}Model analysis of spin accumulation}
In this appendix, we present the spin accumulation in the $mm2$ phase for comparison with the MQ accumulation discussed in Sec.~\ref{sec:Edge Accumulation of Magnetic Quadrupoles Induced by MQ Hall Currents}.
We evaluate the layer-resolved spin response
\begin{align}
  \sigma_x^{y}(\kappa)
  \equiv
  \sigma_x[\sigma_y(\kappa)],
\end{align}
where the local spin operator $\sigma_y(\kappa)$ is defined by
\begin{align}
  \braket{\kappa, \rho\tau\sigma | \sigma_{y}(\kappa) | \kappa, \rho\tau\sigma'} = \sigma_{y}.
  \label{eq:Sy}
\end{align}

For comparison with the conventional description of spin transport,the corresponding spin Hall conductivity is given by
\begin{align}
  \sigma_{zx}^{y(\mathrm H)}
  =
  \frac{
  \sigma_{z;x}^{y}
  -
  \sigma_{x;z}^{y}
  }{2},
\end{align}
which characterizes the generation of a spin current carrying the spin-$y$ component along the $z$ direction in response to an electric field along the $x$ direction, $J_z^y=\sigma_{z;x}^{y}E_x$.
This response is symmetry allowed in the $mm2$ phase.
The spin current is defined as $J_z^y=\{S_y,v_z\}_{+}$~\cite{sinova2004universal}, and the spin Hall response satisfies the conventional relation $J_i^k=\epsilon_{ijk}\sigma_{i;j}^{k}E_j$~\cite{murakami2003dissipationless}, where $\epsilon_{ijk}$ is the Levi--Civita symbol.

Figure~\ref{fig:S_y under mm2} shows the layer dependence of the induced spin-$y$ component for $q=1$.
The obtained edge profile exhibits the characteristic antisymmetric spin accumulation expected in the $mm2$ phase.
This result serves as a reference for the MQ accumulation discussed in the main text and highlights the close analogy between nonequilibrium spin accumulation and MQ accumulation under open-boundary conditions.

\begin{figure*}[htbp]
  \centering
  \includegraphics[width=0.8\columnwidth]{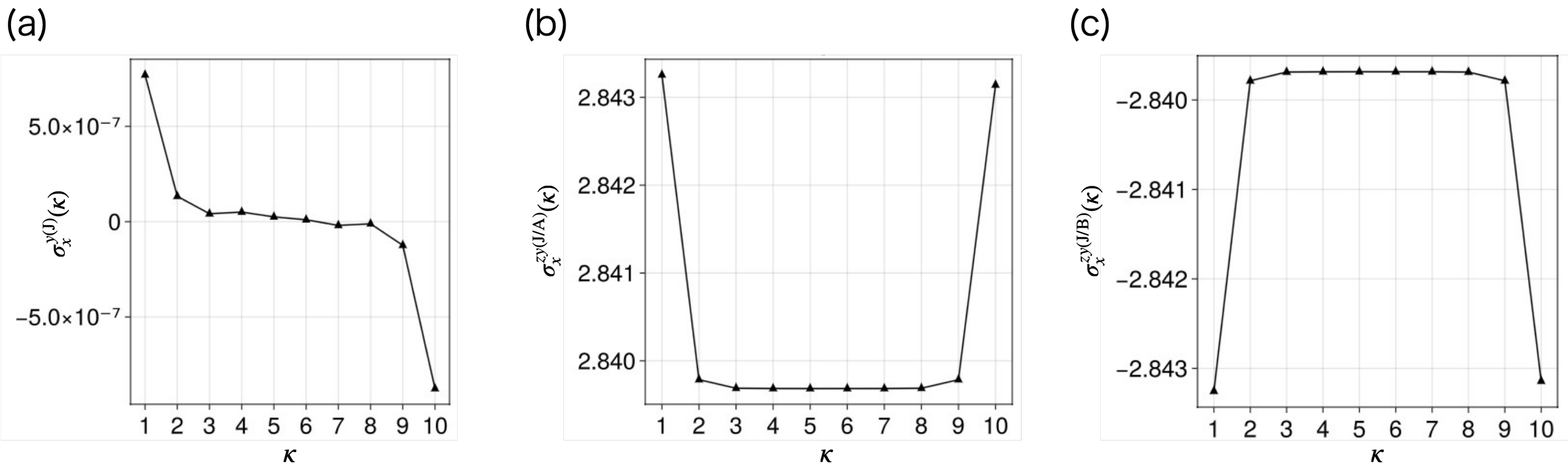}
  \caption{
    (a) Layer dependence of the spin response $\sigma_x^{y(\mathrm J)}(\kappa)$ for $q=1$, $N_z=10$, and $V=2^{15}N_z$.
    (b), (c) Layer dependence of the sublattice-resolved contributions $\sigma_x^{y(\mathrm J/A)}(\kappa)$ and $\sigma_x^{y(\mathrm J/B)}(\kappa)$, obtained from Eq.~(\ref{eq:Sy}) by restricting the sublattice index to $\rho=\mathrm A$ and $\mathrm B$, respectively.
  }
  \label{fig:S_y under mm2}
\end{figure*}

\end{widetext}

\bibliography{ref}

\end{document}